\documentclass[12pt]{article}
\usepackage{a4wide}
\usepackage[centertags]{amsmath}
\usepackage{amssymb}
\usepackage[sort&compress,numbers]{natbib}
\usepackage{ifpdf}
\usepackage{setspace}
\usepackage{amsfonts}
\usepackage{color}
\usepackage{threeparttable}
\usepackage{cancel}
\usepackage{tikz}
\usetikzlibrary{decorations.pathmorphing}
\usepackage{caption}
\usepackage{subcaption}

\usepackage[utf8]{inputenc}
\usepackage{multirow}
\usepackage{hhline}
\usepackage{colortbl}
\usepackage{diagbox} 
\usepackage{makecell}

\usepackage{graphicx,epstopdf}   
\usepackage{sidecap}
\usepackage{caption}
\usepackage{subcaption}

\usepackage{yfonts}

\usetikzlibrary{decorations.markings}
\ifpdf
\usepackage[colorlinks=true,
linkcolor=black,
citecolor=black,
urlcolor=blue,
filecolor=blue,
pdfstartview=FitV,
pdftitle={},
pdfauthor={},
pdfsubject={Hydrodynamics},
pdfkeywords={hydrodynamics, AdS/CFT},
pdfpagemode=None,
bookmarksopen=true
]{hyperref}
\else
\usepackage[hypertex]{hyperref}
\fi
\usepackage{rotating}
\usepackage{rotating}
\makeatletter
\@addtoreset{equation}{section}

\newcommand{\qn}{{\mathfrak{q}}}
\newcommand{\wn}{{\mathfrak{w}}}

\makeatletter
\renewcommand\section{\@startsection {section}{1}{\z@}%
                                   {-3.5ex \@plus -1ex \@minus -.2ex}
                                   {2.3ex \@plus.2ex}%
                                   {\normalfont\large\bfseries}}
\renewcommand\subsection{\@startsection{subsection}{2}{\z@}%
                                     {-3.25ex\@plus -1ex \@minus -.2ex}%
                                     {1.5ex \@plus .2ex}%
                                     {\normalfont\bfseries}}

\def\sec#1{\S \;\ref{#1}}

\title{{Complexified quasinormal modes and the pole-skipping in a holographic system at finite chemical potential }}

\author{Navid Abbasi$^{a}$\footnote{abbasi@lzu.edu.cn},  \ Sara Tahery$^{b}$\footnote{s.tahery@impcas.ac.cn}\\
\small{\emph{}}\\
\small{\emph{$^{a}$School of Nuclear Science and Technology, Lanzhou University,}}\\
\small{\emph	{ 
		222 South Tianshui Road, Lanzhou 730000, China }} \\
\small{\emph	{ $^{b}$
Institute of Modern Physics, Chinese Academy of Sciences, }}\\
\small{\emph	{Lanzhou 730000, China}}
}

\begin{document}

\setlength{\baselineskip}{16pt}
\begin{titlepage}
\maketitle

\vspace{-36pt}

\begin{abstract}
We develop a method  to study coupled dynamics of gauge-invariant variables, constructed out of metric and gauge field  fluctuations on the background of a AdS$_5$ Reissner-Nordstr\"om black brane. Using this method, we compute the numerical spectrum of quasinormal modes associated with fluctuations of spin 0, 1 and 2, non-perturbatively in $\mu/T$. We also analytically compute the spectrum of hydrodynamic excitations in the small chemical potential limit. 
Then, by studying the spectral curve at  complex momenta  in every spin channel,
 we numerically find points at which hydrodynamic and non-hydrodynamic poles collide. We discuss the relation between such collision points and the convergence radius  of the hydrodynamic derivative expansion. 
Specifically in the spin 0 channel, we find that within the range $1.1\lesssim \mu/T\lesssim 2$, the radius of convergence of the hydrodynamic sound mode is set by the absolute value of the complex momentum corresponding to the point at which the sound pole collides with the hydrodynamic diffusion pole.
It shows that in holographic systems at finite chemical potential,  the convergence of the hydrodynamic derivative expansion in the mentioned range is fully controlled by hydrodynamic information.
As the last result, we explicitly show that the relevant information about quantum chaos in our system can be extracted from the pole-skipping points of energy density response function.
We find a threshold value for $\mu/T$,  lower than which the pole-skipping points can be computed perturbatively in a derivative expansion.

  \end{abstract}
\thispagestyle{empty}
\setcounter{page}{0}
\end{titlepage}

\renewcommand{\baselinestretch}{1}  
\tableofcontents
\renewcommand{\baselinestretch}{1.2}  
\section{Introduction}
Low energy dynamics of many body systems near thermal equilibrium can be effectively described in terms of  IR variables.  These variables themselves are of two types.
First are those corresponding to fast thermalizing excitations. They are in correspondence with non-conserved quantities and swallowed by thermal vacuum, very soon after being excited. Second are those corresponding to conserved quantities whose excitations cannot be locally thermalized. They are relaxed just through the transport in long wavelength
compared to local thermalization time scale.

In fact the first-type degrees of freedom are in local equilibrium defined by the second-type ones.
Hydrodynamics is an effective theory which describes the dynamics of the second-type modes, namely slow modes, in a many-body system.  Hydrodynamic equations are then  the conservation equations of energy-momentum tensor $T^{\mu\nu}$ and charge current $J^{\mu}$ in the system. 
The main idea of hydrodynamics is that, in local equilibrium, each of these quantities are given in a perturbative   expansion in terms of the local thermodynamic variables and their derivatives.
Such perturbative expressions are called the hydrodynamic constitutive relations. Using the symmetry considerations, general form of the constitutive relations can be  fixed up to a set of  unknown coefficients, which are referred to as  the transport coefficients.  While hydrodynamics is a universal regime in thermal systems, the associated transport coefficients depend on the underlying microscopic theory by which the system is described.

Considering hydrodynamics as a classical field theory, one can find   simply the response functions of the conserved quantities (see refs. \cite{Kovtun:2012rj,Glorioso:2018wxw}); the corresponding  poles are then the so-called \textit{hydrodynamic modes}.   Hydrodynamic modes are the longest lived modes around thermal equilibrium in the system and are gapless, i.e. $\omega(\textbf{q}\rightarrow 0)=0$. In contrast, the excitations corresponding to the first-type degrees of freedom, mentioned in the first paragraph, are short-lived in the sense that $\omega(\textbf{q}\rightarrow 0)\ne0$. They are called \textit{non-hydrodynamic modes}. 

Hydrodynamic modes can be in general found from the linearized hydrodynamic equations perturbatively, order by order in a derivative expansion (see ref. \cite{Kovtun:2012rj} for a comprehensive review). However, without knowing the microscopic of  the system, the result will be restricted to general perturbative expressions containing unknown transport coefficients.
Moreover, even by having the microscopic theory, computing the transport coefficients in general is limited to  lower orders in the derivative expansion and also to  the perturbative regime of the underlying field theory \cite{Jeon:1995zm}. 

When the microscopic theory has a holographic dual \cite{Maldacena:1997re,Witten:1998qj}, however, extracting information will be relatively straightforward not only in  the hydrodynamic limit,  but also beyond that. In fact holography allows to know the microscopic of certain strongly coupled field theories via studying their gravitational dual in a one-higher-dimensional AdS space time. In the last two decades, by analyzing perturbations around specific AdS black brane solutions, significant information about the strongly coupled $\mathcal{N}=4$ SYM plasma has been found:   transport coefficients of the first order hydrodynamics specifically $\eta/s$  \cite{Policastro:2002se}, conjecturing a lower bound for $\eta/s$ \cite{Kovtun:2004de}, the shear viscosity of R-charged plasma \cite{Son:2006em,Mas:2006dy},  location of the sound poles \cite{Policastro:2002tn}, the leading 't Hooft coupling correction to $\eta/s$ \cite{Buchel:2004di}, the spectrum of  quasinormal modes \cite{Kovtun:2005ev},  the second order transport coefficients \cite{Baier:2007ix,Bhattacharyya:2008jc}, the ’t Hooft coupling correction to the second order transport coefficients \cite{Grozdanov:2014kva}, anomalous transport coefficients \cite{Erdmenger:2008rm,Banerjee:2008th}, the third order transport coefficients for linear   \cite{Bu:2014ena} and full non-linear \cite{Grozdanov:2015kqa} hydrodynamics  and many other related issues. 

At first sight, higher order terms in the hydrodynamic constitutive relations seem just to improve the  accuracy of the derivative expansion. However, as it was found in refs. \cite{Heller:2013fn}, large order behavior of the hydrodynamic expansion might have some information about the non-hydrodynamic modes as well. By numerically computing  the energy density for a holographic boost-invariant flow, up to terms with 240 derivatives, authors of ref.  \cite{Heller:2013fn} found the radius of convergence of the hydrodynamic derivative expansion to be zero. Let us emphasize that this result was associated with an expansion over the inverse proper time in the position space. Then by using  Borel resummation techniques, they found the frequency corresponding with the leading singularity in Borel-transformed hydrodynamic stress tensor to be exactly the same as frequency of the lowest-lying non-hydrodynamic mode, found in refs. \cite{Starinets:2002br,Janik:2006gp} \footnote{The same treatment with large order hydrodynamic gradients in potion space has been applied to FLRW universe in ref. \cite{Buchel:2016cbj}  and to  viscoelastic media in ref. \cite{Baggioli:2018bfa}. }. 

In a recent  paper \cite{Withers:2018srf}, a different approach has been proposed to extract information from the large order behavior of the hydrodynamic derivative expansion. Instead of working with full non-linear hydrodynamics to study a highly symmetric flow in position space \cite{Heller:2013fn}, the  choice of  ref. \cite{Withers:2018srf} is to work with linear hydrodynamics in the complex $q$ plane. The focus of this reference is to study the dispersion relation of the hydrodynamic shear mode at large orders in derivative expansion, in a holographic model in $2+1$ dimensions with finite chemical potential.  Considering the multi-sheet structure of exact $\omega_{\text{shear}}(\textbf{q})$, the closest non-analytic points on the longest-lived sheet are found to be as  $q=\pm i q_{*}$ \cite{Withers:2018srf}, where $q=|\textbf{q}|$. The author of ref. \cite{Withers:2018srf} then argues that $q_{*}$ sets a finite radius of convergence for the hydrodynamic expansion of the shear mode \footnote{This result seems to be in contrast to that of  \cite{Heller:2013fn}. However in a very recent paper \cite{Heller:2020uuy}, it was shown that  hydrodynamic derivative expansion converges in the position space as well,  if initial data have
	support in momentum space not exceeding a critical value.}. As explicitly mentioned in ref. \cite{Withers:2018srf}, the physical origin of $q_*$ is  the collision between a hydrodynamic and non-hydrodynamic mode on the imaginary $q$ axis.

Much more recently, it was shown that  the same result could be found from  analytic structure of spectral curves in classical hydrodynamics\footnote{In the present discussion we do not consider hydrodynamic fluctuations. See refs. \cite{Kovtun:2012rj} and  \cite{Glorioso:2018wxw} for  inclusion of fluctuations in classical hydrodynamics.}. In ref. \cite{Grozdanov:2019kge}, the above-mentioned  collision points were found from  the associated  spectral curves in a holographic neutral fluid in 3+1 dimensions, for both sound and shear hydrodynamic modes.  
In the mentioned reference, such collision points are called level-crossing points. In fact the theory of refs. \cite{Grozdanov:2019kge,Grozdanov:2019uhi} is the general theory for studying spectral curves and level-crossing.

Let us denote that in \cite{Withers:2018srf}, the  relation between $q_{*}$ and the convergence radius of the shear mode has been numerically confirmed at a specific fixed value of $\mu/T$. 
To investigate how actually $q_{*}$ may vary with $\mu/T$ and also to study the convergence of other modes, in this paper we follow the issue in a holographic model at finite chemical potential in $3+1$ dimensions. In the gravity side we consider a AdS$_5$ Reissner-Nordstr\"om black brane. In the bulk of AdS$_5$, such black brane  is identified with the temperature $T$ as well as   a  parameter $Q$, both related to  $\mathcal{N}=4$ SYM boundary theory, where $Q$ is a monotonically increasing function of $\mu/T$.

The main difficulty comes from the fact that in our model parturbations  of gravity and gauge field in the bulk are coupled. One may think of finding  master fields and then deriving decoupled equations governing their dynamics \cite{Kodama:2003kk}\footnote{The master equations for AdS$_5$ RN black brane have been found in \cite{Ge:2008ak,Matsuo:2009yu}. See also \cite{Jansen:2019wag} for the case of Einstein-Maxwell-Dilaton black branes.}. But we choose to work with coupled equations! The reason is related to the numerical method that we want to adopt.
We construct the generalized version of Frobenius expansion used in  ref. \cite{Kovtun:2005ev} to find the quasinormal modes associated with \textit{coupled} differential equations. The advantage of working with coupled  equations then is that it lets  each of our results at $Q=0$ be comparable with some counterpart result within refs. \cite{Kovtun:2005ev} or \cite{Grozdanov:2019kge}. Let us denote that our numerical method works very well in the range $0\le Q\lesssim0.88$, or equivalently for $0\le \frac{\mu}{T} \lesssim 4$. However, in our plots we will demonstrate the results within the range $0\le Q\le0.85$.

We first construct  gauge-invariant variables out of the bulk field perturbations. 
We classify them according to their transformation
properties under $SO(2)$ group corresponding to the isotropy  of the transverse plane perpendicular to $\textbf{q}$. This is actually the subject of \sec{Gauge_section}. In \sec{method} we develop  a new method to find the quasinormal modes of coupled differential equations in the bulk. Our method is based on the combination of analytic and numerical computations. 
In \sec{RN_equations}    we will  derive  coupled equations governing dynamics of gauge-invariant variables in each of the spin channels, on the AdS$_5$ RN black brane background. Our equations can be regarded as  non-trivial generalizations of decoupled equations associated with an AdS$_5$ Schwarzschild black brane accompanied by a probe gauge field  \cite{Kovtun:2005ev}. 

Then in each of the spin channels, we use the method developed in \sec{method} to solve the (coupled) equations and find the corresponding quasinormal modes. We find the spectrum of quasinormal modes by considering variations of both $\qn=q/(2\pi T)$ and $Q$. Spin 0 and spin 1 spectra, each  turns out to be a superposition of two types of poles. In the spin 0 channel, poles are in correspondence with fluctuations of two scalar \textit{master} operators in the boundary theory. The latter operators reduce to energy density and charge density in the vanishing $Q$ limit. The corresponding hydrodynamic modes  are the two sounds  together with the diffusion mode.
In the spin 1 channel,  poles correspond to fluctuations of two vector \textit{master} operators in the boundary. In fact these operators become the transverse momentum density and transverse component of charge current  when $Q$ vanishes. In this case, only one hydrodynamic mode does exist which is actually the shear mode. Finally, from the single dynamical equation of the  spin 2 channel,  we find poles corresponding to fluctuations of the transverse \textit{master} stress tensor. The latter reduces to the transverse stress tensor at $Q=0$. As expected, there is no any hydrodynamic mode in this channel.

 It should be noted that in our present case with $Q\ne0$, the perturbations of $T^{\mu\nu}$ and $J^{\mu}$ on the boundary are coupled; for instance in the spin 0 channel $\langle T J\rangle$ is non-vanishing.  By the \textit{master} quantities in this channel we then mean  $\bar{T}^L$ and $\bar{J}^L$ with vanishing  cross correlators: $\langle\bar{T}^L\bar{J}^L\rangle=0$ \footnote{Here the superscript "$L$" denotes the "longitudinal" direction. In the spin 1 channel we use the superscript "$T$" as denoting the transverse direction.}. The latter quantities are sourced by bulk master fields. We neither explicitly write down the expressions of bulk master field nor need to work with them. 

The spectrum of quasinormal modes at finite chemical potential have been already computed  in AdS$_4$ RN model \cite{Brattan:2010pq,Edalati:2010hk,Edalati:2010pn}. But in holographic models in 3+1 dimensions, well-known results are just limited to the spin 1 and spin 2 channels \cite{Maeda:2006by,Janiszewski:2015ura}.  We find the spectrum of quasinormal modes in all three channels, non-perturbatively in $\mu/T$.
 To the best of our knowledge our study is the first computation of quasinormal modes in the spin 0 channel of  AdS$_5$ RN model. In addition, by analytically solving the coupled dynamical equations in the hydrodynamic limit, we find the spectrum of hydrodynamic modes as well. Again, this is the first analytic computation of the hydrodynamic modes from AdS$_5$ RN black brane. Although, we will find analytic solutions in the bulk perturbatively in $\mu/T$. We then use fluid/gravity \cite{Erdmenger:2008rm,Banerjee:2008th} to confirm our results by explicit computation of the hydrodynamic modes at finite $\mu/T$.

In the second part of the paper, in \sec{complex_life}, we study quasinormal modes at complex momenta. We find radii of convergence of the derivative expansions corresponding to dispersion relations of all four hydrodynamic modes, separately,  within  the  range $0\le Q\le 0.85$. To this end, at each value of $Q$ we gradually increase $|\qn|$ from $0$ and look for the first collision of hydrodynamic poles with the other poles.

For  the shear mode and at small values of $Q$, we find the level crossing between hydrodynamic and non-hydrodynamic poles associated with the transverse \textit{master} current spectrum, namely the poles of $\langle
\bar{J}^T\bar{J}^T\rangle$,  to  set the convergence radius of the dispersion relation. It turns out that the radius of convergence monotonically increases when $Q$ increases from $0$ to a certain threshold. When $Q$ exceeds the latter, the collision between diffusion pole with a non-hydrodynamic pole which belongs to the transverse \textit{master} momentum density spectrum becomes important. The latter in fact is the pole of $\langle
\bar{T}^T\bar{T}^T\rangle$.
In this regime, radius of convergence has a sub-branch decreasing in $Q$. This  might be reminiscent of  the analytic formula found in ref. \cite{Withers:2018srf} in AdS$_4$ RN model.

Our surprising results are mostly related to the spin 0 channel. We motivate that for each of the sound and diffusion modes, one has to find the critical points of spectral curve lying on their own branch of Puiseux series \footnote{Every branch of the spectral curve at the vicinity of origin corresponds to a specific hydrodynamic mode $\wn(\qn\rightarrow 0)=0$. The small-$\qn$ expansion of each mode is given by a Puiseux series. }.
Doing so, for the diffusion case we show that within the whole range of $Q$ that we consider, the level crossing happens between  the diffusion pole and gapped poles associated with the \textit{master} charge density spectrum. In other words, all collisions happen among poles of $\langle
\bar{J}^L\bar{J}^L\rangle$.  But  sound modes are found to collide  with various types of poles, depending on the value of $\mu/T$. Their first collisions are actually with either gapped poles of the \textit{master} energy density spectrum or those of the \textit{master} charge density spectrum or even with the diffusion pole!  The latter is a novel aspect of level crossing phenomenon, specific to holographic systems at finite chemical potential \footnote{In a  different context, collision between hydrodynamic sound mode and the hydrodynamic diffusion mode was already observed in \cite{Novak:2018pnv}.}. It can be regraded as a counterexample for the statement that finite radius of convergence of hydrodynamic derivative expansion is determined by the interplay between hydrodynamic and non-hydrodynamic modes. In other words,  we find that in a specific range of $\mu/T$, the convergence radius of derivative expansion associated with the sound mode is fully determined by hydrodynamic information.

In the last part of the paper, in \sec{chaos}, we study one another aspect of quasinormal modes.
Following recent studies on the relation between hydrodynamics and quantum chaos in maximally chaotic systems \cite{Grozdanov:2017ajz,Blake:2017ris,Blake:2018leo}, we will show that the pole-skipping points of energy density response function in our system precisely coincide with the chaos points in the system. This result is found without making any assumption about the value of $\mu/T$. Thus it provides a new support for the hydrodynamic origin of quantum chaos \cite{Blake:2017ris}.  Then we focus on the spin 0 channel and numerically find the dispersion relation of sound and diffusion poles at purely imaginary momenta. For a typical value of $\mu/T$, we show that the above-mentioned pole-skipping points lie actually on the sound curve.  

In the very last part, we discuss the possibility of finding the chaos point by using the derivative expansion. Using the spin 0 channel results of \sec{complex_life}, we will find that the chaos point does not always lie within the domain of convergence of the hydrodynamic derivative expansion. We find a critical value $Q_c$, beyond which the chaos points lie outside the mentioned domain. It simply means that for this range of $Q$, the pole-skipping point of energy density response function has to be found non-perturbatively.

Finally in \sec{conclusion}, we briefly review our results and discuss possible follow-up directions.

\section{Gauge invariant variables}
\label{Gauge_section}
The background solution on which we would like to find the quasinormal modes can be written in general as 
\begin{equation}\label{}
\begin{split}
ds^2&=\, a(r)\left(-f(r)dt^2+\frac{}{}\sum_{i=1}^{p}dx_i^2\right)+b(r)dr^2,\\
A&=\,\tilde{q} \,c(r)\, dt,
\end{split}
\end{equation}
where $\tilde{q}=q_0/r_h^p$ is the electric charge density on the horizon. $r_h$ is the radial location of the horizon: $f(r_h)=0$.
We take the fluctuations of metric and gauge field to be $h_{\mu \nu}(r)e^{- i \omega t+ i q z}$ and $a_{\mu}(r)e^{- i \omega t+ i q z}$, respectively, where $z=x^p$. We focus on $p=3$ case. We then find the specific combinations of perturbations which are invariant under the simultaneous general diffeomorphism, denoted by $\xi^{\mu}$, and gauge transformations $\phi$ in the bulk
\begin{equation}\label{diff_gauge}
\begin{split}
h_{\mu\nu}&\rightarrow	h_{\mu\nu}-\nabla_{\mu}\xi_{\nu}-\nabla_{\nu}\xi_{\mu},\\
a_{\mu}&	\rightarrow a_{\mu}-\partial_{\mu} \phi-\xi^{\lambda}\nabla_{\lambda}a_{\mu}-a_{\lambda}\nabla_{\mu}\xi^{\lambda}.
\end{split}
\end{equation}
It is convenient to arrange perturbations $h_{\mu\nu}$ and $a_{\mu}$ according to representations of $SO(2)$ group associated with the two dimensional plane perpendicular to $q$ (see  Appendix \ref{diff_gauge_App}). We now start to find gauge invariant variables associated with different spin channels.

\subsection*{Spin $0$ channel}
\label{}
In this channel six perturbations $G_{i}\in\{a_t, a_z, h_{tt}, h_{tz}, h_{zz}, h\}$, where $h=\frac{1}{2}(h_{xx}+h_{yy})$, are coupled to each other.  Any first order gauge invariant variable, denoted by $G_{\text{inv}}$, is naturally a linear combination of $G_i$'s:
\begin{equation}\label{G_inv}
G_{\text{inv}}=\sum_{i=1} ^{6}\alpha_i G_i.
\end{equation}
On the other hand, there are in general four diffeomorphism and gauge parameter functions in this channel: $\zeta_{i}\in\{\phi, \xi_t, \xi_z, \xi_r\}$. 
The task is to find $a_i$'s such that $G_{\text{inv}}$ to be independent of $\zeta_i$'s. Under the transformations \eqref{diff_gauge}, $G_i$'s are transformed as
\begin{equation}\label{}
G_{\text{inv}} \rightarrow G_{\text{inv}}+\sum_{i=1} ^{4}\beta_i \zeta_i,
\end{equation}
with $\beta_i$ being a linear combination of $\alpha_i$'s. Gauge invariance demands $\beta_i$'s must vanish. So we find four equations among six parameters $\alpha_i$.  As a result, two of the parameters $\alpha_i$ remain free; it simply shows that in the sound channel we deal with two gauge invariant variables.
Let us now find them in details.

From \eqref{gauge_trans_Spin_0}, the above mentioned two gauge invariant variables can be found either by \newline
$\bullet$ finding an appropriate combination of its first four lines; or by \newline
$\bullet$  finding an appropriate combination of its last three lines.

In the first case, we take  $h_{zz}+ \ell_1 h_{tt}+ \ell_2 h_{tz}+ \ell_3 h$
and then demand its diff-gauge transformed  be independent of $\xi_{t}$, $\xi_{z}$ and $\xi_{r}$. We find 
\begin{equation}\label{}
\ell_1=\frac{q^2}{\omega^2}\,\,\,\,\,\ell_2=\frac{2 q }{\omega},\,\,\,\,\,\,\ell_3=\frac{1}{2}\left(-1+\frac{q^2 f(r)}{\omega^2}+\frac{q^2 a(r)f'(r)}{\omega^2 a'(r)}\right).
\end{equation}
In the second case, we take $a_{z}+ e_1 a_{t}+ e_2 h$ combination.
Again, demanding its diff-gauge transformed be independent of $\xi_r$ and $\xi_t$ fixes the coefficients $e_1$ and $e_2$:
\begin{equation}\label{}
e_1=\frac{q}{\omega},\,\,\,\,\,e_2=-\frac{q}{ 2\omega}\,\frac{ \tilde{q}\,c'(r)}{a'(r)}.
\end{equation}
In summary, the gauge-invariant variables in the spin $0$ channel can be written as
\begin{equation}\label{Z_2}\boxed{
	\begin{split}
	Z_0&=\qn^2 f H_{tt}+2\, \wn \,\qn H_{tz}+\wn^2 H_{zz}+\qn^2 f\left(1+\frac{a f'}{a' f}-\frac{\wn^2}{\qn^2 f}\right)H\\ 
	E_{z}&=\qn A_t+\wn A_z-\qn\left(\tilde{q}\,\frac{a \,c'}{a'\,c_h}\right)\,H
	\end{split}}
\end{equation}
where $A_{t}=a_{t}/c_h$, $A_{z}=a_{z}/c_h$, $H_{tt}=h_{tt}/a f$, $H_{tz}=h_{tz}/a$, $H_{ij}=h_{ij}/a \,\,(i,j\ne t)$ and $H=h/(p-1) a$. We have defined $c_h=c(r_h)$. In the above equations, $0$ denotes the spin.  At this point one could continue to work with $h_{\mu\nu}$'s and construct $Z_0$ in terms of them. But as we will see later, $H_{\mu\nu}=h^{\mu}_{\,\,\nu}$'s behave like scalar field perturbations in the bulk and  asymptote to  finite values on the boundary.  Let us emphasize that $a_{\mu}$ fields have already finite limit at the boundary and the rescaling factor $c_{h}$ is nothing but a normalization constant.

The dynamical equations governing dynamics of $Z_0$ and $E_z$ on the AdS$_5$ RN background will be given by \eqref{Sound_channel_diff}.
\subsection*{ Spin $1$ channel}
\label{}
From \eqref{gauge_trans_Spin_1}, it is obvious that $E_{\alpha}=i \omega A_{\alpha}$ ($\alpha= x, y $) are gauge invariant. In addition, two another gauge-invariant variables can be constructed by demanding $h_{z\alpha}+ \kappa  h_{t \alpha}$ be independent of diff-gauge parameter functions. Therefore there are four gauge invariant variables in this channel that due to the $SO(2)$ symmetry, we need to work only with two of them
\begin{equation}\label{Z_1}\boxed{
	\begin{split}
	Z_{1}&= \wn\, H_{z x}\,+\,\qn \,H_{t x}\\ 
	E_{x}&= \wn\,A_{x}
	\end{split}}
\end{equation}
where $A_{\alpha}=a_{\alpha}/c_h$ and $H_{t\alpha}=h_{t\alpha}/a$. 
The dynamical equations governing dynamics of $Z_1$ and $E_x$ on the AdS$_5$ RN background will be given by \eqref{Shear_channel_diff}.
\subsection*{ Spin $2$ channel}
\label{}
From \eqref{gauge_trans_Spin_2} one easily finds that there are two gauge invariant variables in this channel
\begin{equation}\label{gauge_inv_Spin_2}\boxed{
	\begin{split}
	Z_{2}&= H_{xy}\\ 
	W_{2}&=H_{xx}-H_{yy}
	\end{split}}
\end{equation}
We will find that on the AdS$_5$ RN background, $Z_2$ and $W_2$ obey a common equation (see \eqref{Spin_2_dynamics}).
\section{Quasinormal modes from coupled differential equations }
\label{method}
As we will see later, in each channel except for the spin 2 one, dynamical variables constitute of two gauge invariant quantities  whose dynamics is governed by a specific set of coupled differential equations. In this section we construct a method to find, in general,  the quasinormal modes associated with such coupled equations. Let us take the two gauge invariant variables in these channels as $Z(x^{\mu},u)$ and $E(x^{\mu},u)$ (see  \eqref{Sound_channel_diff} for spin 0 and \eqref{Shear_channel_diff} for spin 1 case).  The radial coordinate $u$ is related to $r$ through $r\sim1/\sqrt{u}$ with the horizon located at $u=1$. We then make Fourier transformation in the boundary directions
\begin{equation}\label{Fourier_Z_E}
(Z(x^{\mu},u), \,E(x^{\mu},u))=\,\int\frac{d^dk}{(2\pi)^d}\,e^{i k x}(Z(u),E(u)).
\end{equation}
In each channel, upon imposing the ingoing boundary condition at the horizon, the solutions to Fourier component fields take the following form
\begin{equation}\label{Our_solution}
\begin{split}
Z(u)&= (1-u^2)^{-i\wn/2}G(u),\\ 
E(u)&=(1-u^2)^{-i\wn/2}Y(u),
\end{split}
\end{equation}
which are fixed up to two normalization consonants:
\begin{equation}\label{G_Y_bdy}
G(1)=\,C_Z,\,\,\,\,\,\,\,\,\,Y(1)=\,C_E.
\end{equation}
In fact the space of solutions to $Z$ and $E$ corresponds to the two-parameter space defined by $C_Z$ and $C_E$.
 When $Q=0$, $Z$ and $E$ decouple from each other. Then for a given point in the parameter space, i.e. $(C_Z,C_E)$, one expects $C_Z$ solely to specify  $G(u)$ and $C_E$ to do the same separately with $Y(u)$. In the language of linear algebra, the functions $G(u)$ and $Y(u)$ are two vectors directed along the basis vectors in the parameter space, in $Q=0$ case. 
 
Now let us investigate how a point $(C_Z,C_E)$ in the parameter space specifies $Z(u)$ and $E(u)$, when $Q\ne0$.
 In the latter case, $Z$ and $E$ are coupled. The natural expectation is that the vectors $G(u)$ and $Y(u)$ are no longer parallel to the basis vectors. In other words, 
  $C_Z$ cannot solely specify $G(u)$ as neither can  $C_Z$ specify $Y(u)$. Thus both $C_Z$ and $C_E$ should 
 be present in the expressions of $G(u)$ and $Y(u)$.
 Considering all the above explanations, we are led to consider the general form of the solutions as following
 \begin{equation}\label{ingoing_coupled_sol}
 \begin{split}
 Z(u)&= (1-u^2)^{-i\wn/2}\big(C_{Z}\,g_{Z}(u)+\,C_{E}\,y_{Z}(u)\big),\\ 
 E(u)&=(1-u^2)^{-i\wn/2}\big( C_{Z}\,g_{E}(u)+\,C_{E}\,y_{E}(u)\big).
 \end{split}
 \end{equation}
The coupling between $Z(u)$ and $E(u)$ has been taken into account by  considering functions $y_Z(u)$ and $g_{E}(u)$. Then, equations \eqref{ingoing_coupled_sol} are actually giving two vectors with nonzero components along both two basis directions in the parameter space. Because of \eqref{G_Y_bdy}, it is needed that $g_{Z}(1)=\,y_{E}(1)=1$ and $y_Z(1)=\,g_{E}(1)=0$. The latter simply means that $Z$ and $E$ decouple at the horizon $u=1$. Moreover, decoupling at $Q=0$ forces  $y_Z(u)$ and $g_{E}(u)$ to vanish at this limit as well. 
 
In order to find associated quasinormal modes, we need to know the near boundary expansion of bulk fields. Considering the conformal dimensions of $Z$ and $E$ (see Appendix \ref{bdy_bahavior}), one finds
\begin{equation}\label{near_bdy_exp}
\begin{split}
u\rightarrow 0:\,\,\,\,\,\,\,\,\,\,\,\,\,Z(u)&= \left(c_1+\frac{}{}\cdots\right)+\,u^2\big(c_2+\cdots\big),\\ 
E(u)&=\left( c_3+\frac{}{}\cdots\right)+\,u\big(c_4+\cdots\big),
\end{split}
\end{equation}
where 
\begin{equation}\label{c_1_c_3}
c_1=\,C_{Z}\,g_{Z}(0)+\,C_{E}\,y_{Z}(0),\,\,\,\,\,\,\,\,c_3=\, C_{Z}\,g_{E}(0)+\,C_{E}\,y_{E}(0).
\end{equation}
Using these solutions, the bilinear boundary bulk action takes the following general form\footnote{At this point we partly adopt the idea of ref. \cite{Kaminski:2009dh} to write the boundary action as the first line of \eqref{bilinear_bdy_action}. In this reference, a framework for calculating holographic Green's functions from this bilinear action has been developed. However, what we are going to do is different from the framework of this reference (see Appendix \ref{Comparison} for a comparison between the method  of ref.  \cite{Kaminski:2009dh} and that of us). See also \cite{Janik:2016btb} for yet another treatment with coupled dynamics in the bulk.}:
\begin{equation}\label{bilinear_bdy_action}
\begin{split}
S \sim &\lim_{u\rightarrow 0}\int_{\omega, \textbf{q}}\begin{pmatrix}
Z'_{k}(u)&E'_{k}(u)
\end{pmatrix}\begin{pmatrix}
A(u)&B_1(u)\\
B_2(u)& D(u)
\end{pmatrix}\begin{pmatrix}
Z_{-k}(u)\\
E_{-k}(u)
\end{pmatrix}+\,\text{contact\,\,terms},\\
\sim& \lim_{u\rightarrow 0}\int_{\omega, \textbf{q}}\begin{pmatrix}
Z_k(u)&E_k(u)
\end{pmatrix}\begin{pmatrix}
2u\frac{c_2}{c_1}A(u)&2u\frac{c_2}{c_1}B_1(u)\\
\frac{c_4}{c_3}B_2(u)& \frac{c_4}{c_3}D(u)
\end{pmatrix}\begin{pmatrix}
Z_{-k}(u)\\
E_{-k}(u)
\end{pmatrix}+\,\text{contact\,\,terms}.
\end{split}
\end{equation}
We temporarily used the subscript $k=(\omega, \textbf{q})$  for Fourier fields $Z$ and $E$ to distinguish between $k$ and $-k$ Fourier components. In the following we drop the subscript and continue to respect the convention made in  \eqref{Fourier_Z_E}.  In  \eqref{bilinear_bdy_action}, "contact terms" are finite parts of the boundary counter-term. Needless to say that $2\,u\,A(u)$, 2\,$u\,B_1(u)$, $B_2(U)$ and $D(u)$ all go to finite values when $u\rightarrow 0$: 
\begin{equation}\label{limits}
\lim_{u\rightarrow 0}2u\,A(u)=\,a,\,\,\lim_{u\rightarrow 0}2u\,B_1(u)=\,b,\,\,\,\lim_{u\rightarrow 0}B_2(u)=\,c,\,\,\,\lim_{u\rightarrow 0}D(u)=\,d.
\end{equation}
By making an appropriate unitary transformation $U$, the middle matrix in above is simply diagonalized:
\begin{equation}\label{diagonalized}
\begin{split}
S\sim  &\int_{\omega, \textbf{q}}\begin{pmatrix}
Z_{k}(0)&E_{k}(0)
\end{pmatrix}U\begin{pmatrix}
\boldsymbol{G}_1&0\\
0& \boldsymbol{G}_2
\end{pmatrix}U^{\dagger}\begin{pmatrix}
Z_{-k}(0)\\
E_{-k}(0)
\end{pmatrix}+\,\text{contact\,\,terms}\\
\sim&  \int_{\omega, \textbf{q}}\begin{pmatrix}
\mathcal{Z}_{k}&\mathcal{E}_{k}
\end{pmatrix}\begin{pmatrix}
\boldsymbol{G}_1&0\\
0& \boldsymbol{G}_2
\end{pmatrix}\begin{pmatrix}
\mathcal{Z}_{-k}\\
\mathcal{E}_{-k}
\end{pmatrix}+\,\text{contact\,\,terms}
\end{split}
\end{equation}
where $\boldsymbol{G}_1$ and $\boldsymbol{G}_2$ are some expressions in terms of $a$, $b$, $c$ and $d$ together with $c_i; i=1,2,3,4$.
The transformed variables at $u=0$ are then given by 
\begin{equation}\label{new_Z_E}
\begin{split}
\mathcal{Z}_{k}=&\,U^{\dagger}_{11}Z_k(0)+\,U^{\dagger}_{12}E_{k}(0),\\
\mathcal{E}_{k}=&\,U^{\dagger}_{21}Z_k(0)+\,U^{\dagger}_{22}E_{k}(0).
\end{split}
\end{equation}
At this point the holographic AdS/CFT duality \cite{Witten:1998qj} implies that  $\mathcal{Z}$ and $\mathcal{E}$ couple to specific operators $\mathcal{O}_1$ and $\mathcal{O}_2$ in the boundary theory. In other words, $\mathcal{Z}_{k}$ and $\mathcal{E}_{k}$ are the decoupled \textit{master} fields in the bulk and $\boldsymbol{G}_1$ and $\boldsymbol{G}_2$ are the Green's functions of the corresponding boundary \textit{master} operators \footnote{When $Q=0$ and in the spin 0 channel, $\boldsymbol{G}_1$ and $\boldsymbol{G}_2$ correspond to $G_{2}(\omega,q)$ and $\Pi^{L}(\omega, q)$ in ref. \cite{Kovtun:2005ev}, respectively. At the same $Q$ and in the spin 1 channel, $\boldsymbol{G}_1$ and $\boldsymbol{G}_2$ correspond to $G_{1}(\omega,q)$ and $\Pi^{T}(\omega, q)$ in the mentioned reference.}. 
 Using  \eqref{bilinear_bdy_action}, \eqref{limits} and \eqref{diagonalized}, one finds
\begin{equation}\label{}
\langle \mathcal{O}_1 \mathcal{O}_1 \rangle_{R}\sim \boldsymbol{G}_1\sim \frac{\mathcal{K}_{+}(c_1,c_2,c_3,c_4)}{c_1 c_3}+\,\text{c.t.} ,\,\,\,\,\,\,\,\,\,\,\,\,\,\,\,
\langle \mathcal{O}_2 \mathcal{O}_2 \rangle_{R}\sim \boldsymbol{G}_2 \sim\frac{\mathcal{K}_{-}(c_1,c_2,c_3,c_4)}{c_1 c_3}+\,\text{c.t.}
\end{equation}
where "c.t." stands for the contact term contributions \cite{Policastro:2002tn} and
\begin{equation}\label{}
\mathcal{K}_{\pm}=\,\frac{1}{2}\left(a\, c_2 c_3+d\, c_1c_4\pm \sqrt{(a\, c_2 c_3 + d\, c_1 c_4)^2-4 c_1 c_2 c_3 c_4(a d - b c )}\right).
\end{equation}
By definition, the quasinormal modes of bulk perturbations correspond to the  poles of retarded Green's functions \cite{Nunez:2003eq,Birmingham:2001pj,Horowitz:1999jd}. One simply finds that $c_1=0$ gives the poles of $\boldsymbol{G}_1$ while poles of $\boldsymbol{G}_2$ are the roots of $c_3=0$. Using \eqref{c_1_c_3}, the corresponding roots are then found from 
\begin{equation}\label{general_Dirichlet}
\begin{split}
C_{Z}\,g_{Z}(0)+\frac{}{}C_{E}\,y_{Z}(0)=&\,0,\\
C_{Z}\,g_{E}(0)+\frac{}{}C_{E}\,y_{E}(0)=&\,0.
\end{split}
\end{equation}
As we will see in next sections, each of $g_{Z}(0)$, $y_Z(0)$, $\cdots$ is a complicated analytic function of $\omega$ and $\textbf{q}$. In order for \eqref{general_Dirichlet} leads to non-trivial relations between $\omega$ and $\textbf{q}$, it is required that
\begin{equation}\label{det_quasi}
\det\begin{pmatrix}
g_{Z}(0)&y_{Z}(0)\\
g_{E}(0)& y_{E}(0)
\end{pmatrix}=\,0.
\end{equation}
This is our first result in this paper; the equation from which, we can numerically find the quasinormal modes of coupled perturbations $Z$ and $E$. We will also show that in the hydrodynamic limit, equation \eqref{det_quasi} can be solved analytically.

Let us summarize the method developed in this section. In order to find the quasinormal modes of coupled perturbations $Z$ and $E$ in  the spin 0 and 1 channels, we first construct  the corresponding coupled differential equations. We should find the (analytic) solutions to them which are ingoing at the horizon. We do the latter by using the Frobenius expansion. The corresponding solutions can be formally written in the form of \eqref{ingoing_coupled_sol}, up to two arbitrary normalization constants $C_Z$ and $C_E$. Having specified the functions $g_Z$, $g_E$, $y_Z$ and $y_E$, then equation \eqref{det_quasi} determines the spectrum of quasinormal modes in the associated channel.

\section{Quasinormal spectrum and hydrodynamic modes in $\mathcal{N}=4$ SYM theory at finite chemical potential}
\label{RN_equations}
Our system of interest is holographically described by dynamics of metric and  a $U(1)$ gauge field in the bulk of AdS. 
The corresponding action is given by
	\begin{equation}
	S = \frac{1}{16 \pi G_5} \int d^5 x \,\,\sqrt{-g} \left(R + \frac{12}{L^2} - F^{M N} F_{M N}\right)+ S_{bdy}.
	\end{equation}
	where $S_{bdy}$ is the boundary counter term. The equations of motion  are given by:
	\begin{equation}\label{EoM}
	\begin{split}
	G_{\mu \nu}-6 g_{\mu \nu}+2\left(F_{\mu \rho} F^{\rho}_{\,\,\nu}+\,\frac{1}{4}F^{\alpha \beta}F_{\alpha \beta}\,\,g_{\mu \nu}\right)=\,0,\,\,\,\,\,\,\,\,\,
	\nabla_{\mu}F^{\mu \nu}=\,0.
		\end{split}
	\end{equation}
		We work in the unite where $L=1$. The solution in the Poincare patch is written as it follows
		\begin{equation}\label{back_ground}
		ds^2=r^2\left(-f(r)dt^2+dx^2+dy^2+\frac{}{}dz^2\right)+\frac{1}{r^2 f(r)}dr^2,\,\,\,\,\,\,\,	A=-\frac{\sqrt{3}q_b}{2 r^2}dt,
		\end{equation}
		with  $	f(r)=1-\frac{m}{r^4}+\frac{q_b^2}{r^6}$. Parameters $m$ and $q$ are two constants.
		It is convenient to re-scale the quantities with the radius of the outer horizon, $R$, namely the largest positive root of $f(r)=0$. We may write 
\begin{equation}\label{rescale}
\rho\equiv\frac{r}{R},\,\,\,\,\,\,\,\,M=\frac{m}{R^4},\,\,\,\,\,\,\,\,Q=\frac{q_b}{R^3},\,\,\,\,\,\,\,\,Q^2=M-1.
\end{equation}
In the rescaled coordinates, the outer horizon locates at $\rho=1$ while the boundary is identified with $\rho\rightarrow \infty$. For further requirements, we need to work in  a system  with finite domain of the radial coordinate; for this purpose,  we make the  change
$\rho=1/\sqrt{u}$. The function $f$ then takes the following form 
\begin{equation}
 f(u)=1-(Q^2+1)u^2+Q^2 u^3.
 \end{equation}
Using this together with  \eqref{rescale}, the Hawking temperature $T$ and the chemical potential $\mu$ of the boundary theory are found to be  
\begin{equation}\label{mu_T}
\begin{split}
T=&\,\frac{1}{4\pi}\left(-2R\, f'_u\big|_{u=1}\right)=\,\frac{R}{2\pi}\,(2-Q^2),\\
\mu=&\,A_{t}\big|_{\infty}-A_{t}\big|_{u=1}=\,\frac{ \sqrt{3} q}{2R^2}=\,\frac{\sqrt{3}}{2}Q R.
\end{split}
\end{equation}
From the  expressions \eqref{mu_T}, one can derive $Q$ as a function of $\mu/T$:
\begin{equation}\label{Q__mu_T}
Q=\,\frac{\sqrt{2}\,a\, (\mu/T)}{1+\sqrt{1+a^2\,(\mu/T)^2}},\,\,\,\,\,\,\,\,\,\,\,\,\,\,\,\,a=\left(\frac{8}{3\pi^2}\right)^{1/2}.
\end{equation}
This relation simply shows that there is a one-to-one map between $\mu/T$ and $Q$. Thus in the following, we take $T$ and $Q$ as the two independent thermodynamic variables. Then  the bulk metric and gauge field can be finally written in $(t,x,y,z, u)$ coordinates as the following:
 \begin{equation}\label{Metric_Gauge_u_coord}
  \begin{split}
 ds^2&=\frac{1}{u}\left(\frac{2\pi T}{2-Q^2}\right)^2\left(-f(u)dt^2+\frac{}{}dx^2+dy^2+dz^2\right)+\frac{1}{4u^2 f(u)}du^2,\\
 A&=-\frac{\sqrt{3}\pi T Q}{2-Q^2}\,u\,dt.
   \end{split}
\end{equation}

\subsection{Spin $0$ Channel}
We turn on the set of perturbations $\{h_{tt}, h_{tz}, h_{zz}, h, A_t, A_z\}$ in the radial gauge. On the RN background solution  \eqref{Metric_Gauge_u_coord}, the gauge invariant variables \eqref{Z_2} take the following form
 \begin{equation}\label{Z_0_E_0_on_shell}
\begin{split}
Z_0&=\qn^2 f H_{tt}+ 2 \qn \,\wn H_{tz}+\wn^2 H_{zz}+\,\qn^2\left(2-Q^2 u^3-\frac{\wn^2}{\qn^2}-f\right)H,\\
E_z&=\qn A_t+ \wn A_z+\qn\,Q\,H.
\end{split}
\end{equation}
Note that on the background solution \eqref{Metric_Gauge_u_coord}, $\tilde{q}=Q$.
To find the coupled dynamical equations of the above two variables, firstly it is required the perturbations $\{h_{tt}, h_{tz}, h_{zz}, h, a_t, a_z\}$ in the spin $0$ equations \eqref{EoM} to be replaced with $\{a f H_{tt}, a H_{tz}, a H_{zz}, 2 a H, c_h A_t, c_h A_z\}$  (see \eqref{Z_2} and explanations given below that). Doing so, we find equations of the latter "six" perturbations.
The difficult task is to eliminate all these perturbations in favor of the two gauge invariant variables $Z_0$ and $E_z$. After long computations, which are not shown here, we arrive at the two following coupled differential equations
\begin{equation}\label{Sound_channel_diff}
\begin{split}
Z_0''&+\,{{\textswab{a}}}_{1}\,Z_0'+\,{{\textswab{a}}}_{2}\,Z_0+\,{{\textswab{b}}}_{3}\,E_z+\,{{\textswab{b}}}_{4}\,E_z'=0,\\
E_z''&+\,{{\textswab{b}}}_{1}\,E_z'+\,{{\textswab{b}}}_{2}\,E_z+\,{{\textswab{a}}}_{3}\,Z_0+\,{{\textswab{a}}}_{4}\,Z_0'=0.
\end{split}
\end{equation}
We have put $\textswab{a}_i$ coefficients in front of $Z_0$ and $Z_0'$ and have done the same for  $\textswab{b}_i$ coefficients with  $E_z$ and $E_z'$. Considering $\wn=\omega/2\pi T$ and $\qn=q/2\pi T$, the coefficients are found to be as 
\begin{equation}\label{Z_0_coef}
\begin{split}
{{\textswab{a}}}_{1}&=\frac{(Q^2u^3-2)\wn^2\mathcal{D}+f(\qn^4(Q^4u^6-4)+\qn^2\wn^2(2-Q^2u^3)+3\wn^4)+ \qn^4(4+2Q^2u^3-3f)f^2}{u f(\wn^2-\qn^2 f)\left(\mathcal{D}-\qn^2 f\right)},\\
{{\textswab{a}}}_{2}&=\frac{\tilde{Q}^2u \,\wn^2\,\mathcal{D}+\qn^2 f\left((Q^2u^3-2)(8-\qn^2\tilde{Q}^2u+8Q^2u^3)-4\tilde{Q}^2u \wn^2+(32+\qn^2\tilde{Q}^2u+8Q^2 u^3-16f)f)\right)}{4 u^2 f^2\, \left(\mathcal{D}-\qn^2 f\right)},\\
{{\textswab{a}}}_{3}&=\,-\,\frac{2\,\qn \,Q\,(1+Q^2 u^3-f)}{u f (\mathcal{D}-\qn^2 f)},\,\,\,\,\,\,\,\,\,\,\,\,\,\,\,\,\,\,\,\,\,\,\,\,\,\,\,\,\,\,\,\,
{{\textswab{a}}}_{4}=\frac{\qn\, Q\,(\qn^2(2+Q^2u^3)-\wn^2-\qn^2 f)}{(\mathcal{D}-\qn^2 f)(\qn^2 f-\wn^2)}
\end{split}
\end{equation}
and
\begin{equation}\label{E_z_coef}
\begin{split}
{{\textswab{b}}}_{1}&=\frac{2  \qn^2Q^2 u^2 f-\wn^2f'}{f(\qn^2 f-\wn^2)},\,\,\,\,\,\,\,\,\,\,\,\,\,\,\,\,\,\,\,\,\,\,\,\,\,\,\,\,\,\,\,\,\,\,\,\,\,\,\,\,\,\,\,\,\,\,\,\,\,\\
\\
{{\textswab{b}}}_{2}&=\frac{-\tilde{Q}^2\wn^4 \mathcal{D}+2\qn^2 \wn^2 f(\qn^2 \tilde{Q}^2+6 Q^2 u^2)(Q^2 u^3-2)+\wn^4f(7 \qn^2 \tilde{Q}^2+36 Q^2 u^2)}{4 u f^2(\mathcal{D}-\qn^2 f)(\qn^2 f -\wn^2)}\\
&\,\,\,\,\,\,\,\,\,\,\,\,\,\,\,\,\,\,\,\,\,\,+\frac{\qn^2 f^2(48 \qn^2 Q^2 u^2-\qn^4\tilde{Q}^2(Q^2 u^3-2)-5(\qn^2\tilde{Q}^2+12Q^2 u^2)\wn^2)+\qn^6\tilde{Q}^2f^3}{4 u f^2(\mathcal{D}-\qn^2 f)(\qn^2 f -\wn^2)},\\
{{\textswab{b}}}_{3}&=\,-\frac{6\,\qn\, Q\bigg((Q^2u^3-2)\wn^2\mathcal{D}+(2\qn^4(Q^2u^3-2)+\qn^2\wn^2(2-Q^2u^3)+6\wn^4)f+(4\qn^4-6\qn^2\wn^2)f^2\bigg)}{f\,(\qn^2 f-\wn^2)\,\left(\mathcal{D}-\,\qn^2 f\right)},\\
{{\textswab{b}}}_{4}&=\frac{2 \,\qn\, Q\,u\,\left(\mathcal{D}-\qn^2 f\right)}{(\wn^2-\qn^2 f)}.
\end{split}
\end{equation}
We have defined $\mathcal{D}=(\qn^2(Q^2u^3-2)+3 \wn^2)$ and $\tilde{Q}=Q^2-2$. 
At this point it should be noted that when $Q=0$, the above equations reduce exactly to the pair of decoupled equations (4.5b) and (4.35) in ref. \cite{Kovtun:2005ev}. 
Finding the quasinormal modes as well as hydrodynamic modes from these equations is the subject of following subsections.

\subsubsection*{Quasinormal modes}
The analytic solution to equations \eqref{Sound_channel_diff} is unknown; so to find the associated spectrum of quasinormal modes, 
To this end,  we combine the  Frobenius expansions of $Z_0$ and $E_z$ in the bulk  (see Appendix \ref{Frobenius} for more details) with  the method developed in \sec{method}. In what follows, the corresponding results will be given.

\begin{figure}
	\centering
	\includegraphics[width=0.42\textwidth]{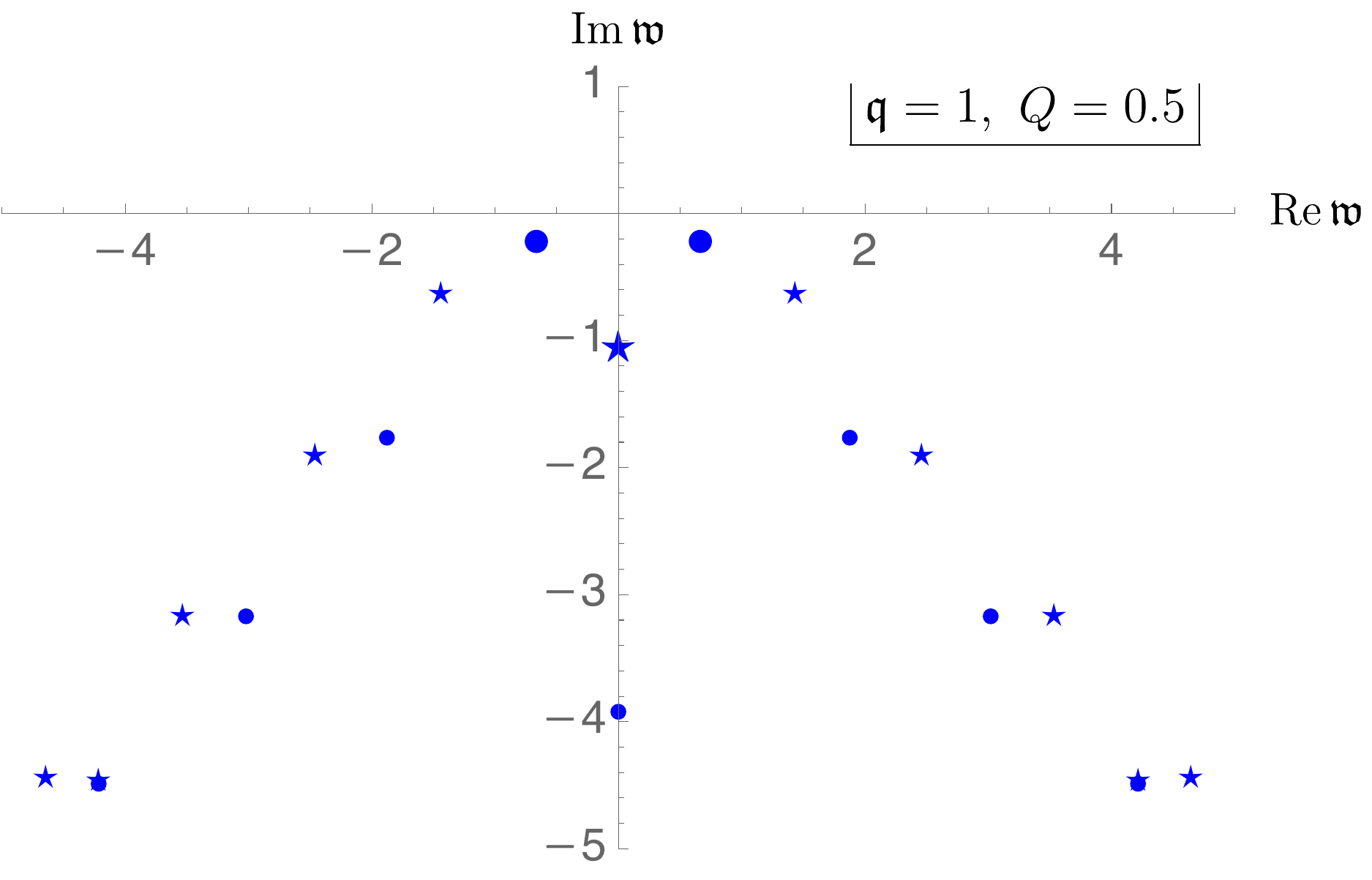}\,\,\,\,\,\,\,\,	\includegraphics[width=0.42\textwidth]{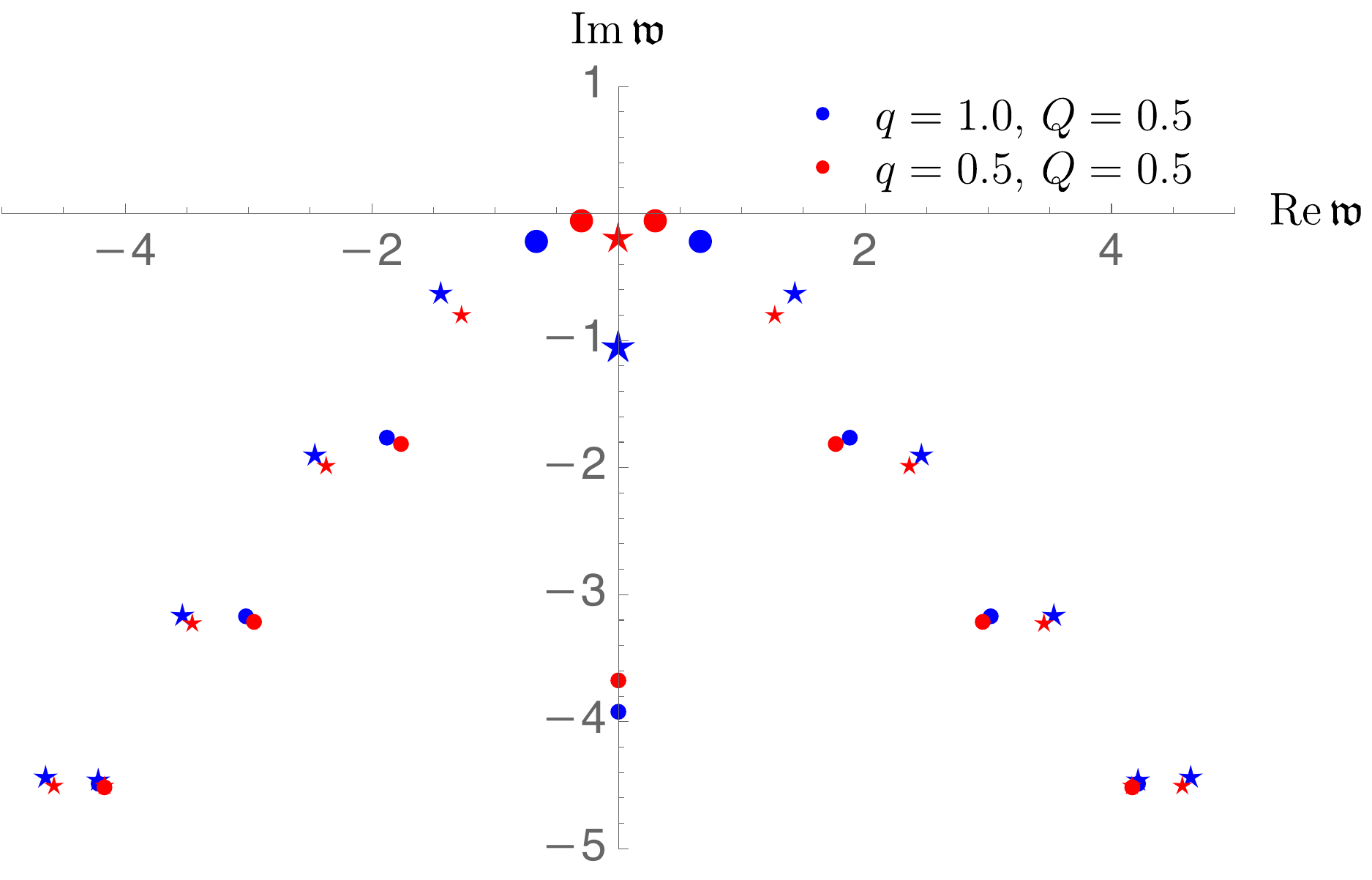}
	\caption{Left panel: Stars represent poles of the \textit{master} charge density Green’s function and dots
		correspond with poles of the \textit{master} energy density Green’s function.  As $\qn$ decreases, all poles stay at a finite distance
		from the real axis, except for the one marked with a large star and the two marked with large dots. Large dots therefore manifest the existence  of two sound modes and the large dot corresponds with the existence of a diffusive $U(1)$ charge mode in the boundary $\mathcal{N}=4$ SYM theory at finite chemical potential.}
		\label{quasi_EzZ0_Q_12_fig}
\end{figure}

In Fig.\ref{quasi_EzZ0_Q_12_fig}, the typical arrangement of poles has been demonstrated  for  two cases in this channel. In the \textit{left panel,} we have shown the quasinormal modes associated with $\qn=1$ at $Q=0.5$. As can be seen, we have splitted them into two sets, denoted by dots and stars. The idea for such spitting comes from the  knowledge about the arrangement of poles in the sound channel as well as in the  diffusion channel on the AdS-Schwarzschild background \cite{Kovtun:2005ev}\footnote{Throughout this paper we follow the terminology of \cite{Kovtun:2005ev}; we refer to the diffusion of $U(1)$ $R-$charge  simply as the diffusion and to the that of momentum as the shear.}. In the latter case the $R-$current fluctuations decouple from the spin 0 fluctuations of energy-momentum tensor; then one can refer to dots (sound channel) and stars (diffusion channel) as the poles of energy density and the charge density Green's functions, respectively. 
In our present case, however, such distinction is no longer  true; all the correlation functions have poles which correspond to all of the quasinormal modes \footnote{We thank anonymous referee for pointing this out to us.}(see the discussion in the section 2.5 of \cite{Kovtun:2012rj}). 
To be consistent with $Q=0$ case, we use the language of \textit{master} fields. In fact, each of the two sets of poles, namely dots and stars, correspond to poles of a specific \textit{master operator} on the boundary. We call the two corresponding operators associated with spin 0 channel as the \textit{master} energy  and \textit{master} charge; we also show them by $\bar{T}^{L}$ and $\bar{J}^{L}$. Thus the sound and diffusion channels correspond to poles of $\langle\bar{T}^{L}\bar{T}^{L}\rangle $ and $\langle\bar{J}^{L}\bar{J}^{L}\rangle$, respectively.  Needless to say, when $Q=0$ the latter two correlates reduce to $G_2(\omega,q)$ and $\Pi^L(\omega,q)$ of the ref. \cite{Kovtun:2005ev}.

In the \textit{ right panel }of Fig.\ref{quasi_EzZ0_Q_12_fig}, we have compared the left panel spectrum with the one corresponding  to the same $Q$ but at a smaller $\qn$. One clearly notices that  as $\qn$ decreases, all complex poles move away from the horizontal  axis, but the two large dots and the large star move towards the origin.
This simply shows that the spectral curve of the spin 0 fluctuations includes three branches of Puiseux series passing through the origin of complex plane. In the next subsection, we explicitly derive the equation of these branches in the vicinity of origin. These three branches correspond to three hydrodynamic modes: \textit{two sound modes} together with \textit{one diffusion mode}.   It should be also noted that the poles other than these three are all gapped.

Let us give a comment about the lowest purely imaginary gapped pole in Fig.\ref{quasi_EzZ0_Q_12_fig} which has been shown with a small dot on the imaginary axis. The latter indicates that this is actually pole of the \textit{master} energy density Green's function. To understand why this is so, one can compare Fig.\ref{quasi_EzZ0_Q_12_fig} with the spectrum of sound  or diffusion poles on the AdS$_5$-Schwarzschild background. To compare this pole with the similar pole in the AdS$_5$-Schwarzschild case, we need to keep track of it when $Q$ goes from $0$ to $0.5$. But it turns out that at small values of $Q$ such pole lies beyond the domain of convergence of our method. Thus the  comparison with the  AdS$_5$-Schwarzschild fails to work. At this point we exploit the results of ref. \cite{Edalati:2010pn} about the large $Q$ limit of quasinormal modes in the spin 0 channel in AdS$_4$ RN case. Comparing Fig.\ref{quasi_EzZ0_Q_12_fig} with figure.1 in ref. \cite{Edalati:2010pn}, we may say that one of the two lowest poles on the imaginary axis is a diffusion pole while the other one should belong to the spectrum of \textit{master} energy density, namely to the sound channel.  Since the lowest pole is a diffusion (star) pole, the next one would be a dot pole.
\begin{SCfigure}
	\centering
	\includegraphics[width=0.45\textwidth]{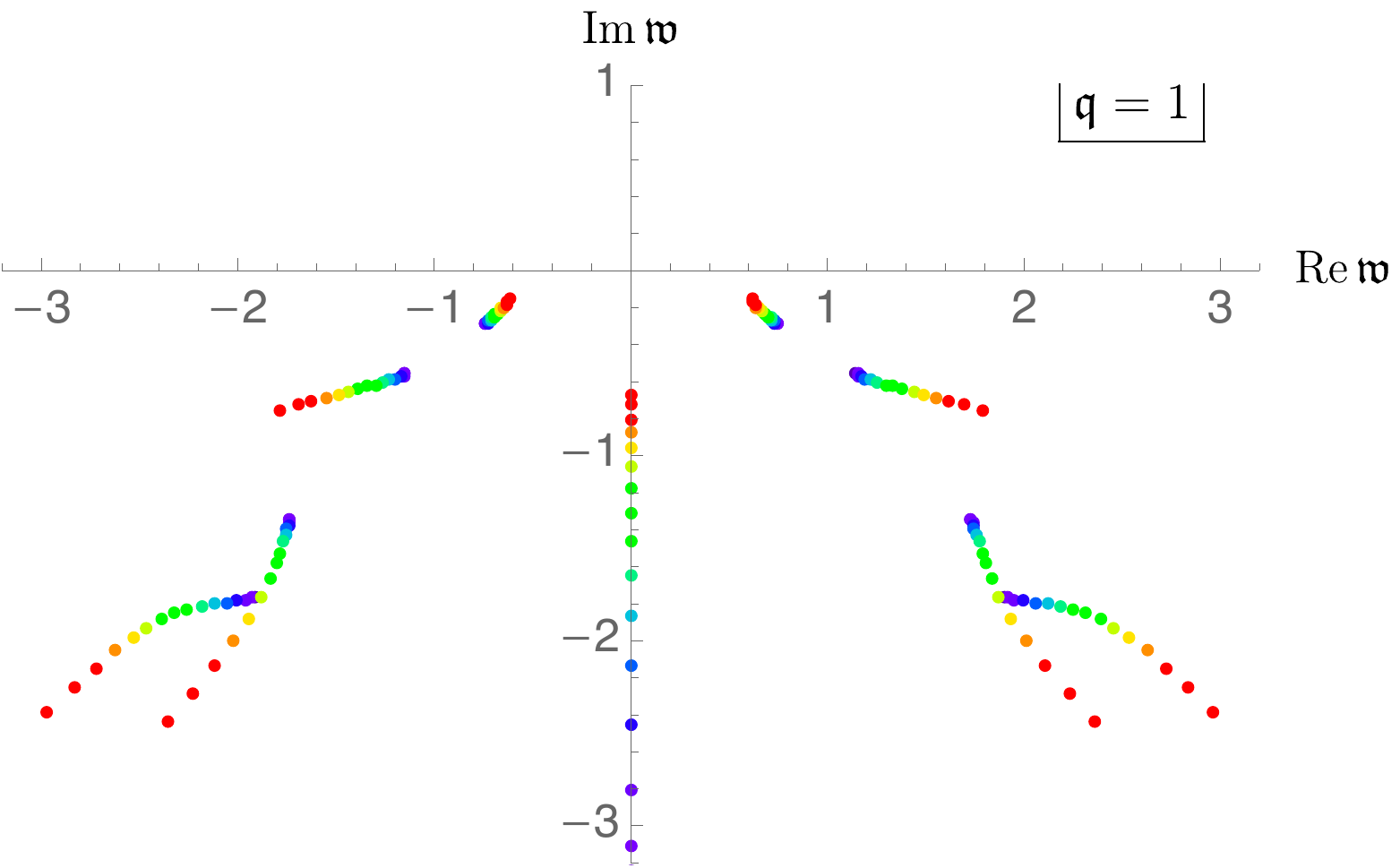}	
	\caption{The eight lowest complex quasinormal modes and the lowest purely imaginary one in the spin 0 channel, at $\mathfrak{q}=1$. Every colorful path-like set of points, starting in purple and ending in red,  shows the  change of one specific mode when $Q$ discretely increases from $0$ to $0.8$. We have considered 16 regular steps by increments of $\Delta Q=0.05$.}
	\label{quasi_EzZ0_Q_12_fig_color}
\end{SCfigure}

So far we have just talked about the quasinormal modes at a fixed finite value of $Q$. In Fig.\ref{quasi_EzZ0_Q_12_fig_color}, we have depicted part of the spectrum of quasinormal modes at the fixed momentum $\qn=1$ for several values of $Q$ within the range $0\le Q \le 0.8$.  One observes that by approaching towards the extremal limit\footnote{The extremal limit is identified with $Q=\sqrt{2}$.}, even at a finite fixed value of momentum, e.g. at $\qn=1$, the non-hydrodynamic  (gapped) poles move away from the origin. As mentioned in the Introduction, our numerical method works well within the range $0\le Q\le0.88$. It would be interesting to try other numerical methods to find the extrapolation of colorful paths depicted in Fig.\ref{quasi_EzZ0_Q_12_fig_color} at larger values of $Q$, specifically when $Q\rightarrow\sqrt{2}$ or equivalently when $T\rightarrow 0$. 

See also Appendix.\ref{refrence_data} for some reference numerical data.
\subsubsection*{Hydrodynamic limit}
Although the complete analytic solution to equations \eqref{Sound_channel_diff} is unknown,  one can analytically solve them in the  hydrodynamics limit. At first sight, equations \eqref{Sound_channel_diff} may seem impossible to become decoupled. But as we will show, at small $\mu/T$ limit, namely when $Q\ll1$, they decouple in the hydrodynamic limit. What we are going to do is to perturbatively solve \eqref{Sound_channel_diff} in the hydrodynamic expansion and also in $Q$ expansion.

Demanding the solutions be ingoing at the horizon, the near horizon behavior of $Z_0$ and $E_z$ are fixed as follows:
\begin{equation}\label{N_H_sound}
u\rightarrow 1\,\,\,\,\,\,\,\,\,\,\,\,\,\,Z_0(u),  E_z(u)\,\sim\,\left(1-(1+Q^2)u^2+\frac{}{}Q^2 u^3\right)^{-i \wn/2}.
\end{equation}
In order to enter the hydrodynamic expansion, we apply the rescaling $\wn\rightarrow \epsilon \,\wn$ and $\qn \rightarrow \epsilon \,\qn$ to the dynamical equations as well as to \eqref{N_H_sound}.  A quick look at the the coefficients \eqref{Z_0_coef} and \eqref{E_z_coef} reveals that
\begin{eqnarray}
\textswab{a}_{1}, \textswab{a}_{2} \sim O(\epsilon^0),& &\,\,\,\,\,\,\,\textswab{b}_{3}, \textswab{b}_{4} \sim O(\epsilon),\\
\textswab{b}_{1}, \textswab{b}_{2} \sim O(\epsilon^0),& &\,\,\,\,\,\,\,\textswab{a}_{3}, \textswab{a}_{4} \sim O(\epsilon^{-1}).
\end{eqnarray}
Such scaling simply means that in each of the equations \eqref{Sound_channel_diff}, the function $Z_0$ must be one order higher in $\epsilon$ expansion, than the function $E_z$.
 Thus  the appropriate ansatz for the functions $Z_0(u)$ and $E_z(u)$ in the double expansion over  $\epsilon$ and $Q$ is given by
\begin{equation}\label{Z_hydro}
\begin{split}
Z_0(u) &=\left(1-(1+Q^2)u^2+\frac{}{}Q^2 u^3\right)^{-i \wn/2}\, \sum_{k=0} \wn^{k+1}\left(Z_{0}^{k,0}(u)+ Q Z_{0}^{k,1}(u)+ Q^2 Z_{0}^{k,2}(u)\right),\\
E_z(u) &=\left(1-(1+Q^2)u^2+\frac{}{}Q^2 u^3\right)^{-i \wn/2}\,  \sum_{k=0} \wn^{k}\left(E_{z}^{k,0}(u)+ Q\frac{}{} E_{z}^{k,1}(u)+ Q^2 E_{z}^{k,2}(u)\right).
\end{split}
\end{equation}
Substituting the above expressions into equations \eqref{Sound_channel_diff} and expanding over $\epsilon$ and $Q$, we obtain a set of second order ordinary differential equations for the functions $Z_{0}^{n,m}$ and $E_z^{n,m}$. 

Starting from the lowest order in $\epsilon$ and $Q$, one firstly finds $Z_{0}^{0,0}(u)$ from the first line in  \eqref{Sound_channel_diff} up to an unknown coefficient. Regularity at $u=1$ together with fixing its value at the same point, namely $Z_{0}^{0,0}(1)=C_1$, picks out a unique regular solution to $Z_{0}^{0,0}(u)$. Using this solution, then the next function that can be found is $E_{z}^{0,0}(u)$; from the second equation in \eqref{Sound_channel_diff}. Again, regularity at $u=1$ and demanding $E_z^{0,0}(1)=C_2$ fix the solution.

  In the Appendix \ref{App_sol_HYDRO_SOUND}, we have listed the corresponding perturbative solutions at higher orders, according to the ordering we have found them through the perturbation theory. It should be noted that at every order of perturbation, one of the two  boundary conditions is regularity at $u=1$ and the second one is the the normalization of the solution at the same point:
  \begin{equation}\label{Dirichlete_sound}
  \begin{split}
 Z^{0,0}(1)=&\,C_1, \,\,\,\,\,\,\,E_z^{0,0}(1)=C_2,\\
 \text{and}\,\,\,\,\,\,\,\,\,\, Z^{m,n}(1)=&\,0, \,\,\,\,\,\,\,\,\,E_z^{m,n}(1)=0:\,\,\,\,\,\,m^2+n^2\ne 0.
  \end{split}
  \end{equation}
  Let us denote that in what follows, for convenience, we replace $C_1$ with $(2-3\textswab{y}^2) C_1$.  
 Quasinormal modes are, by definition \cite{Nunez:2003eq}, the modes obtained upon applying the Dirichlet boundary condition to \eqref{Z_hydro}. One thus writes 
 \begin{eqnarray}
 Z_0(0)=0,\,\,\,\,\,E_z(0)=0.
 \end{eqnarray}
According to our earlier discussions these two equations are coupled. By explicit computations (Appendix \ref{App_sol_HYDRO_SOUND}) we find\footnote{These equations are analogue of the formal equations \eqref{general_Dirichlet}; however, here are specifically associated with spin 0 channel and  with the hydrodynamic limit as well.}
\begin{equation}\label{Coupled_Dirichete}
\begin{split}
 Z_0(0)=0&:\,\,\,\,\,\,\,\,\,m_{11}(\wn, \qn; Q) \,C_1+m_{12}(\wn, \qn; Q)\, C_2=0,\\
E_z(0)=0&:\,\,\,\,\,\,\,\,\,m_{21}(\wn, \qn; Q)\, C_1+m_{22}(\wn, \qn; Q)\, C_2=0,
\end{split}
\end{equation}
where $C_1$ and $C_2$ are the normalization coefficients defined in \eqref{Dirichlete_sound}. The $m_{ij}$ coefficients, up to first order in derivatives, are found to be
  \begin{equation}\label{m_ij_Spin_0}
  \begin{split}
m_{11}&=(1-3\textswab{y}^2)\,\wn-2 i\, \wn^2+\bigg(\frac{5-6\textswab{y}^2}{3\textswab{y}^2-2}\,\wn+\frac{8-2\textswab{y}^2(\log 8+1)+3\textswab{y}^4(1-3\textswab{y}^2)(3-\log 8)}{2\textswab{y}^2(3\textswab{y}^2-2)}\,\wn^2\bigg)\,Q^2,\\
m_{12}&=\bigg(-\frac{4\sqrt{3}}{\textswab{y}}\,\wn+\frac{4i\,(-2+\textswab{y}^2(\log 8-3))}{\sqrt{3}\textswab{y}^3}\,\wn^2\bigg)\,Q,\\
m_{21}&=\bigg(-\frac{\sqrt{3} \,\textswab{y}}{4}-\frac{i\sqrt{3}\,(-1+\textswab{y}^2(\log 2-1))}{4 \textswab{y}}\,\wn\bigg)\,Q,\\
m_{22}&=1+i \left(\frac{1}{\textswab{y}^2}-\log 2\right)\,\wn+\bigg(\frac{3}{2}+\frac{1}{2}i\, \left(\frac{1}{\textswab{y}^2}-\log 2\right)\wn\bigg)\,Q^2
    \end{split}
  \end{equation}
  where $\textswab{y}=\wn/\qn$. Obviously, in the limit $Q=0$, the two non-diagonal coefficients $m_{12}$ and $m_{21}$ vanish and one is left with two decoupled equations in \eqref{Coupled_Dirichete}; $C_1\,m_{11}=0$ and $C_2\,m_{22}=0$. Requiring $C_1, \,C_2\ne0$, these equations then give the well-known dispersion relations, namely equations (4.43) and (4.16) in \cite{Kovtun:2005ev}, respectively. When $Q\ne0$, however, equations \eqref{Coupled_Dirichete} are coupled and have a non-trivial set of solutions if and only if one demands
  \begin{equation}\label{m_ij_Spin_0_matrix}
  \det\begin{pmatrix}
  m_{11}&m_{12}\\
  m_{21}& m_{22}
  \end{pmatrix}=\,0.
  \end{equation}
Solving the recent equation to second order in $\qn$ and $Q$, we find the dispersion of  the spin 0 hydrodynamic excitations in a holographic charged fluid as it follows 
  \begin{equation}\label{hydro_mode_holog_spin_0}
  \begin{split}
\wn^{\pm}_{\text{sound}}=&\,\pm\frac{1}{\sqrt{3}}\qn-\frac{i}{6}\left(2-3 \frac{}{}Q^2\right)\qn^2,\\
\wn_{\text{diffusion}}=&\,-i\left(1- \frac{}{}Q^2\right)\qn^2,\\
\wn_{\text{non-hydro}}=&-\frac{i}{\log 2}(1+4 Q^2)\,+i \qn^2 \left(1-\frac{1+ \log 2}{\log 2}Q^2\right).
  \end{split}
  \end{equation}
  The expression in the first line of \eqref{hydro_mode_holog_spin_0} is the dispersion relation of the two hydrodynamic sound modes in the charged fluid. The second line is showing the hydrodynamic diffusion of the $U(1)$ charge. Finally the expression given  in the third line does obviously correspond to a gapped mode which lies beyond the regime of hydrodynamics. To the best of our knowledge, this is the first computation of hydrodynamic modes in the AdS RN background\footnote{In \cite{Matsuo:2009yu}, using the master field method, only the first term of sound mode, namely $\wn=1/\sqrt3\qn$ has been found. Authors of \cite{Matsuo:2009yu} have also computed the first order hydrodynamic transport coefficients  via using the relevant Kubo formulas.}.

\subsection{Spin $1$ Channel}
We turn on the set of perturbations $ \{h_{tx}, h_{zx}, h, A_x\}$ in the radial gauge. As mentioned in \sec{Gauge_section}, the gauge invariant variables associated with this channel are identified with $Z_1$ and $E_x$ given by \eqref{Z_1}. Just like in the spin 0 channel, the difficult part of the computation here is to find dynamical equations governing dynamics of these variables. Combining all spin 1 components of \eqref{EoM} and performing long computations, which are not shown here, we have eliminated $ \{h_{tx}, h_{zx}, h, A_x\}$ in favor of  $Z_1$ and $E_x$. Eventually we have arrived at the following two coupled dynamical equations
\begin{equation}\label{Shear_channel_diff}
\begin{split}
0=&\,Z_1''+\,\frac{(Q^2 u^3-2)\wn^2+\wn^2f + \qn^2 f^2}{u f(\wn^2-\qn^2 f)}\,Z_1'+\,\frac{\tilde{Q}^2(\wn^2-\qn^2 f)}{4 u f^2}\,Z_1+\,\frac{3 Q \,\qn\, \wn \,\,u\, f'}{(\wn^2-\qn^2 f)\,f}\,E_x+\,\frac{3 Q \qn u }{\wn}\,E_x',\\
0=&\,E_x''+\,\frac{f'}{f}\,E_x'+\,\frac{\tilde{Q}^2(\wn^2-\qn^2 f)^2-12 \,Q^2 \,\wn^2\,u^2  f}{4 \,u f^2\,(\wn^2-\qn^2 f)}\,E_x-\frac{ Q \,\qn\, \wn}{(\wn^2-\qn^2 f)}\,Z_1'.
\end{split}
\end{equation}
As it must be, one can readily check that when $Q=0$, the above equations reduce exactly to the pair of decoupled equations (4.5a) and (4.26) in ref. \cite{Kovtun:2005ev}. 
In the following two subsections, we proceed with numerically and analytically solving the above equations, respectively.
\subsubsection*{Quasinormal modes}
The analytic solution to the coupled equations \eqref{Shear_channel_diff} is unknown. However,  just like what was done in the spin 0 channel, we can use the Frobenius expansions of $Z_1$ and $E_x$ to find the corresponding quasinormal modes via the method developed in \sec{method}.

\begin{figure}
	\centering
	\includegraphics[width=0.42\textwidth]{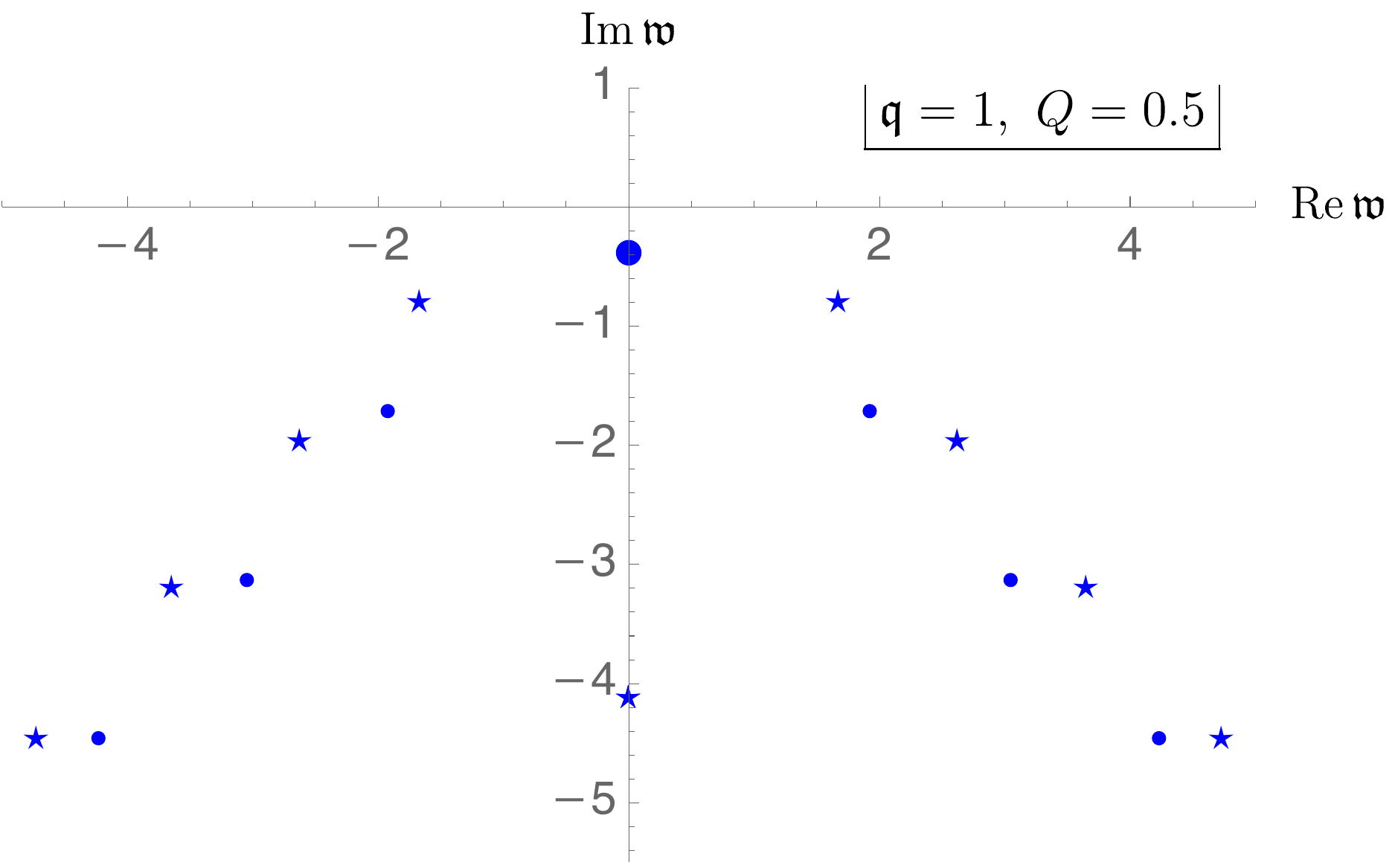}\,\,\,\,\,\,\,\,	\includegraphics[width=0.42\textwidth]{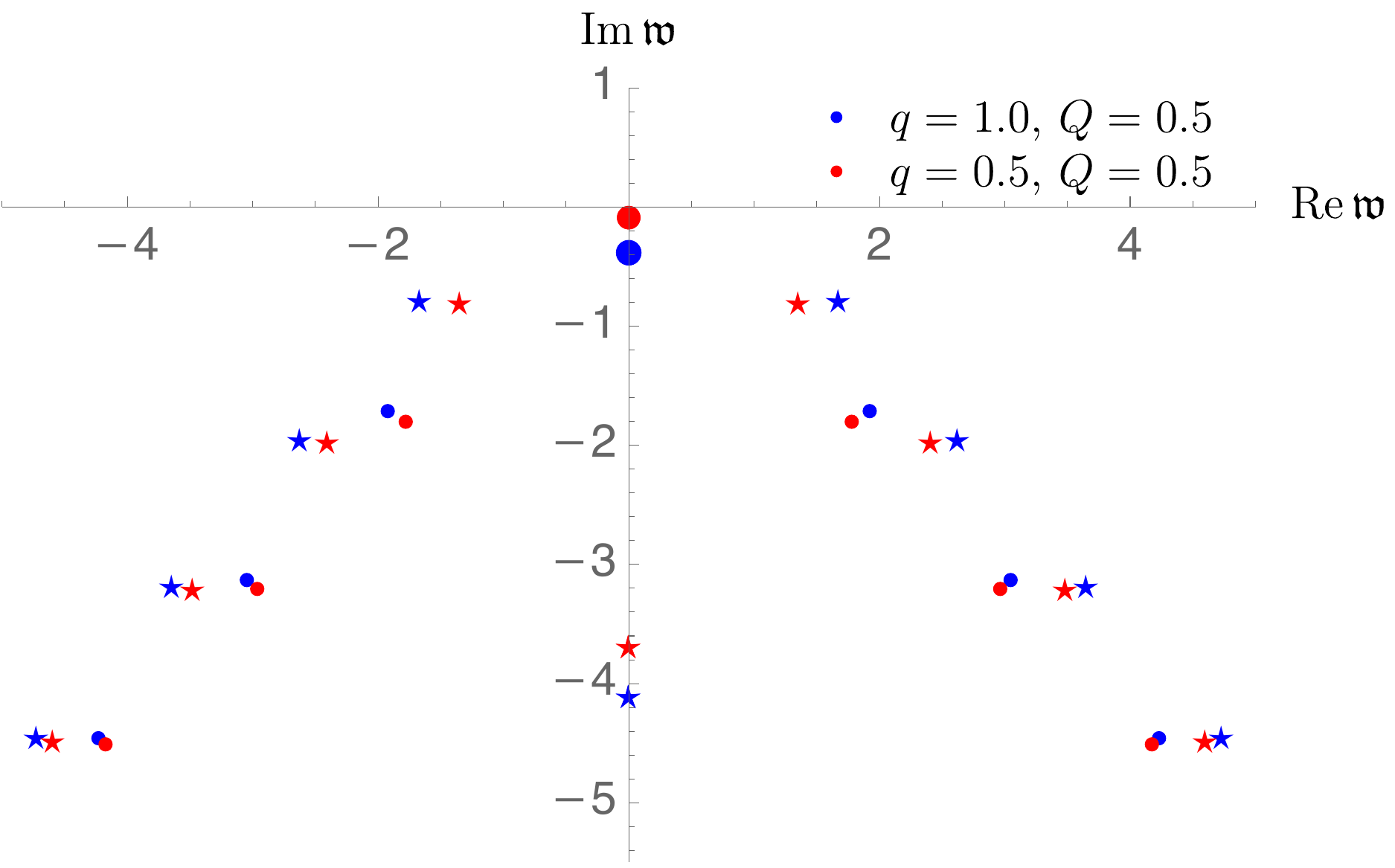}
	\caption{Left panel: Stars are poles of the transverse \textit{master} momentum density Green's function and dots correspond with the poles of the transverse \textit{master} charge current Green's function. In fact,  dots identify the shear poles. As $\qn$ decreases, all poles stay at a finite distance
		from the real axis, except for the one marked with a large dot. The latter manifests the existence of a diffusive shear mode in the boundary $\mathcal{N}=4$ SYM theory at finite chemical potential.  }
	\label{quasi_ExZ1_Q12_fig}
\end{figure}

Firstly in order to get familiar with the typical arrangement of quasinormal modes in the spin 1 channel, in the \textit{left} panel of Fig.\ref{quasi_ExZ1_Q12_fig}   we have shown the spectrum at $\qn=1$ for $Q=0.5$. We have splitted the poles into two sets, denoted by stars and dots. The idea of such splitting originates from the corresponding spectra of quasinormal modes associated with decoupled variables $Z_1$ and $E_x$ on an AdS-Schwarzschild background. One naturally concludes that dots in the above figures correspond to the poles associated with fluctuations of transverse \textit{master} momentum density, i.e. $\bar{T}^{T}$. They are actually the shear poles. Thus the shear channel corresponds to poles of $\langle\bar{T}^{T}\bar{T}^{T}\rangle $.  Needless to say, when $Q=0$ the latter correlator reduces to $G_1(\omega,q)$ of the ref. \cite{Kovtun:2005ev}. It os clear that tars  identify the poles associated with fluctuations of transverse \textit{master } charge current, i.e. $\bar{J}^{T}$. 

In the \textit{right} panel of Fig.\ref{quasi_ExZ1_Q12_fig} we have compared the left panel spectrum with the spectrum associated for the same $Q$ but at a smaller $\qn$. As it can be seen, when $\qn$ decreases, all complex frequency poles move away from the real axis. At the same time, the large dot mode becomes closer and closer to the origin. It simply shows that there exists one branch of Puiseux series associated with the spectral function of spin 1 channel that passes through the origin. In other words, there is only one gapless mode in this channel which is actually the hydrodynamic \textit{shear mode}. In the next subsection we explicitly derive the dispersion relation of this mode.

Let us give a comment about the lowest purely imaginary gapped pole in Fig.\ref{quasi_ExZ1_Q12_fig}.  By similar arguments to what we made in the spin 0 channel, we compare this figure with the figure.1  in ref. \cite{Edalati:2010hk}. Then one immediately concludes that the mentioned pole should be a star one.

\begin{SCfigure}
	\centering
	\includegraphics[width=0.45\textwidth]{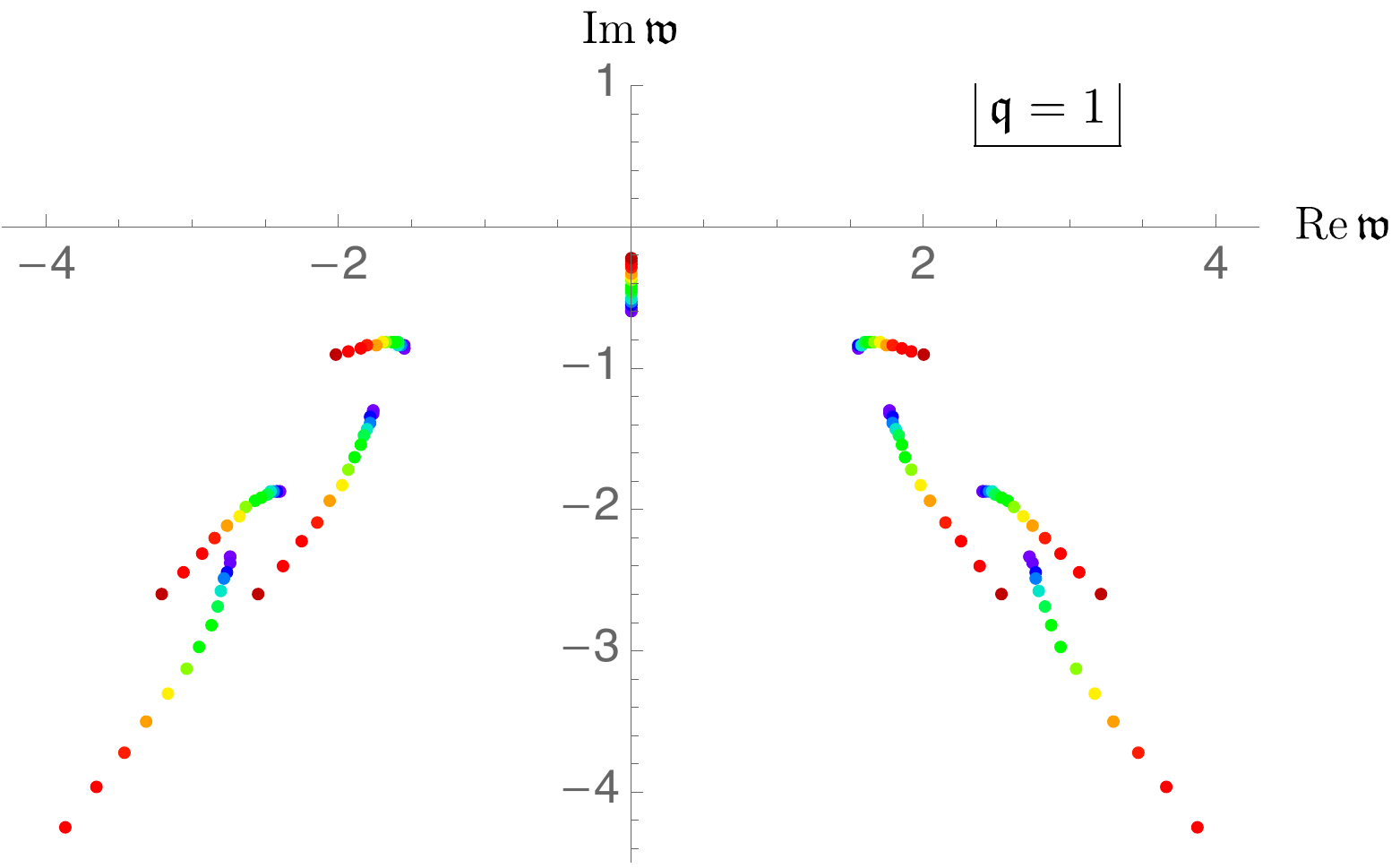}	
	\caption{The  eight lowest complex quasinormal modes and the lowest purely imaginary one in spin 1 channel,  at $\mathfrak{q}=1$. Every colorful path-like set of points, starting in purple and ending in red,  shows the  change of one specific mode when $Q$ discretely increases from $0$ to $0.8$. We have considered 16 regular steps by increments of $\Delta Q=0.05$.}
	\label{quasi_ExZ1_Q12_fig_color}
\end{SCfigure}

Our discussion on quasinormal modes in this channel has been so far limited to the case with a fixed value of $Q$. In Fig.\ref{quasi_ExZ1_Q12_fig_color}, we have demonstrated part of the spectrum of quasinormal modes at the fixed momentum $\qn=1$ for several values of $Q$ within the range $0\le Q \le 0.8$.  One observes that by approaching towards extremality, the gapped poles move away from the real axis, while the gapless mode becomes close to the origin. It would be interesting to study the spectrum in the extrapolation region $0.9 < Q \le \sqrt{2}$ to see what the fate of this mode in the extremal limit will be \cite{navid:2020}.

\subsubsection*{Hydrodynamic limit}
In this subsection we are going to find an analytic solution to the dynamical equations  \eqref{Shear_channel_diff} in the hydrodynamic limit $\wn(\qn\rightarrow0)=\,0$. As mentioned in \cite{Kovtun:2005ev}, it turns out that the appropriate rescaling in this channel is $\wn\rightarrow \epsilon^2 \wn$ and $\qn \rightarrow \epsilon \qn$ with $\epsilon \ll1$. But even in this limit, equations \eqref{Shear_channel_diff} are coupled; 
 then as was the case in the spin 0 channel, we proceed with perturbatively expanding equations over $Q$ as well. This expansion for $Q\ll1$ together with the derivative expansion over $\epsilon$ make it possible to analytically find $Z_1$ and $E_x$.

To specify the general form of the solutions, it is required to investigate the behavior of the coefficients in \eqref{Shear_channel_diff} under the  above-mentioned hydro rescaling.  Doing so we find that
the appropriate ansatz for the functions $Z_1$ and $E_x$ is given by:
\begin{equation}\label{Shear_hydro}
\begin{split}
Z_1(u) &=\left(1-(1+Q^2)u^2+\frac{}{}Q^2 u^3\right)^{-i \wn/2}\, \sum_{k=0} \wn^{k}\left(Z_{1}^{k,0}(u)+ Q Z_{1}^{k,1}(u)+ Q^2 Z_{1}^{k,2}(u)\right),\\
E_x(u) &=\left(1-(1+Q^2)u^2+\frac{}{}Q^2 u^3\right)^{-i \wn/2}\,  \sum_{k=1} \left(\frac{\wn}{\qn}\right)^{k}\left(E_{x}^{k,0}(u)+ Q\frac{}{} E_{x}^{k,1}(u)+ Q^2 E_{x}^{k,2}(u)\right).
\end{split}
\end{equation}
Thus the main task is to find the functions $Z_{1}^{k,m}$ and $E_x^{k,m}$. It should be noted that the ingoing boundary condition at the horizon has been already fulfilled in \eqref{Shear_hydro}. The remaining boundary conditions are then given by
\begin{equation}\label{Dirichlete_shear}
\begin{split}
Z_1^{0,0}(1)=&\,C_3, \,\,\,\,\,\,\,E_x^{0,0}(1)=C_4,\\
\text{and}\,\,\,\,\,\,\,\,\,\, Z_1^{m,n}(1)=&\,0, \,\,\,\,\,\,\,\,\,E_x^{m,n}(1)=0:\,\,\,\,\,\,m^2+n^2\ne 0.
\end{split}
\end{equation}
Here $C_3$ and $C_4$ are the values of $Z_1$ and $E_x$ at the horizon.
The solution functions that obey the above boundary conditions have been given in the Appendix \ref{App_sol_HYDRO_SOUND}. The hydrodynamic modes in this channel are then found by applying the Dirichlet condition to the solutions evaluated at $u=0$ (see \sec{method}). The latter can be formally written as
\begin{equation}\label{Coupled_Dirichete_Shear}
\begin{split}
Z_1(0)=0&:\,\,\,\,\,\,\,\,\,s_{11}(\wn, \qn; Q) \,C_3+s_{12}(\wn, \qn; Q)\, C_4=0,\\
E_x(0)=0&:\,\,\,\,\,\,\,\,\,s_{21}(\wn, \qn; Q)\, C_3+s_{22}(\wn, \qn; Q)\, C_4=0.
\end{split}
\end{equation}
To leading order in $\epsilon$ and second order in $Q$, we find
  \begin{equation}\label{Dirichlete_Spin_1}
 \begin{pmatrix}
  s_{11}&s_{12}\\
  s_{21}& s_{22}
  \end{pmatrix}=\, \begin{pmatrix}
  1+ \frac{i \qn^2}{2 \wn}&\frac{3Q}{2}\\
  \frac{i \qn^2 Q}{2\wn}& \left(1+\frac{3Q^2}{2}\right)
  \end{pmatrix}.
  \end{equation}
It is obvious that at $Q=0$ limit,  equations \eqref{Coupled_Dirichete_Shear} decouple and one finds $\wn=-i \qn^2/2$, the well-known shear mode in a holographic neutral fluid \cite{Kovtun:2005ev}. When $Q\ne0$,  equations are coupled and have non-trivial solutions if and only if $det(s_{ij})=0$; this gives
\begin{equation}\label{hydro_mode_holog_spin_1}
\wn =- \frac{i}{4}\left(2-3 Q^2\right) \qn^2,
\end{equation}
which shows how the density of $U(1)$ charge perturbatively modifies the dispersion of the shear mode in the system.

\subsection{Spin $2$ Channel}
As discussed in \sec{Gauge_section}, by turning on metric perturbations $ \{h_{xy}, h_{xx}, h_{yy}\}$ in the radial gauge, two spin 2 gauge invariant variables $Z_2$ and $W_2$ are excited (see \eqref{gauge_inv_Spin_2}). We find that on the RN background solution  \eqref{Metric_Gauge_u_coord}, these quantities  commonly obey the following equation
\begin{equation}\label{Spin_2_dynamics}
	Z_2''+\,\left(\frac{f'}{f}-\frac{1}{u}\right)\,Z_2'+\,\frac{\tilde{Q}^2\,u (\wn^2- \qn^2 f)+f\big(12(1-f)+6Q^2 u^3+10 u f'- 4 u^2 f''\big)}{4 u^2 f^2}\,Z_2=\,0.
\end{equation}
With no need to follow \sec{method}, we take the following Frobenius expansion 
\begin{equation}
Z_2(u)=(1-(1+Q^2)u^2+Q^2 u^3)^{-i \wn/2}\sum_{n=0}^{\infty} c_n(\wn, \qn)(1-u)^n.
\end{equation}
Substituting it into \eqref{Spin_2_dynamics}, we find coefficients $c_n$  all in terms of $c_0$. Then by applying the Dirichlet boundary condition we arrive at the spectral curve of fluctuations in this channel:
\begin{equation}
Z_2(0)=\sum_{n=0}^{\infty} c_n(\wn, \qn)=\,0.
\end{equation}
By keeping sufficient number of terms in the sum, one can numerically find the spectrum of quasinormal modes. 

\begin{figure}
	\centering
	\includegraphics[width=0.6\textwidth]{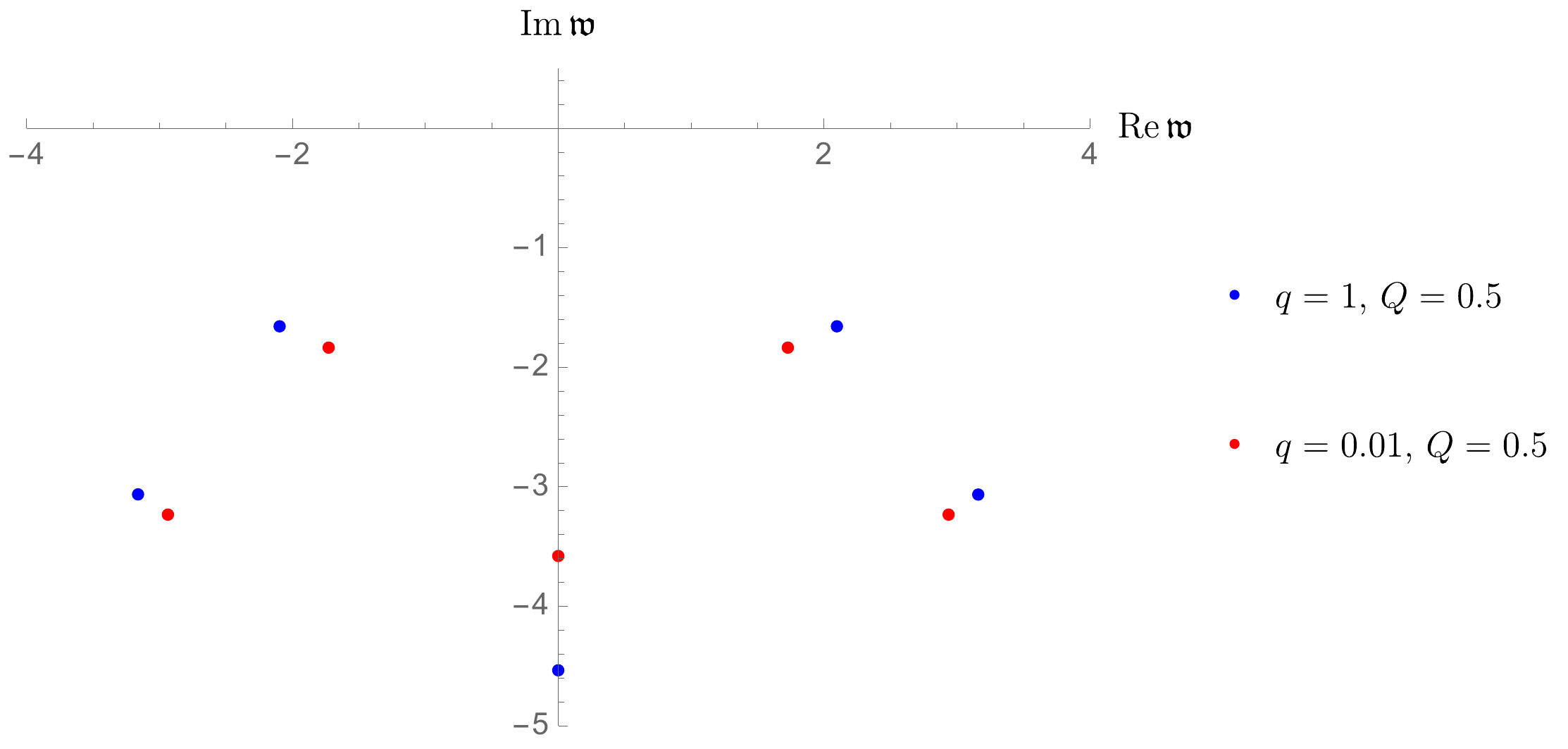}	
	\caption{Quasinormal modes associated with $Q=0.5$ at two different values of $\qn$,  in spin 1 channel.  At each value of $\qn$, we have shown five lowest quasinormal modes. As $\qn$ decreases, all poles stay at finite distances
		from the real axis. The latter manifests the non-existence of any spin 2 hydrodynamic mode in $\mathcal{N}=4$ SYM theory at finite chemical potential.}
	\label{quasi_Exy}
\end{figure}
\begin{SCfigure}
	\centering
	\includegraphics[width=0.45\textwidth]{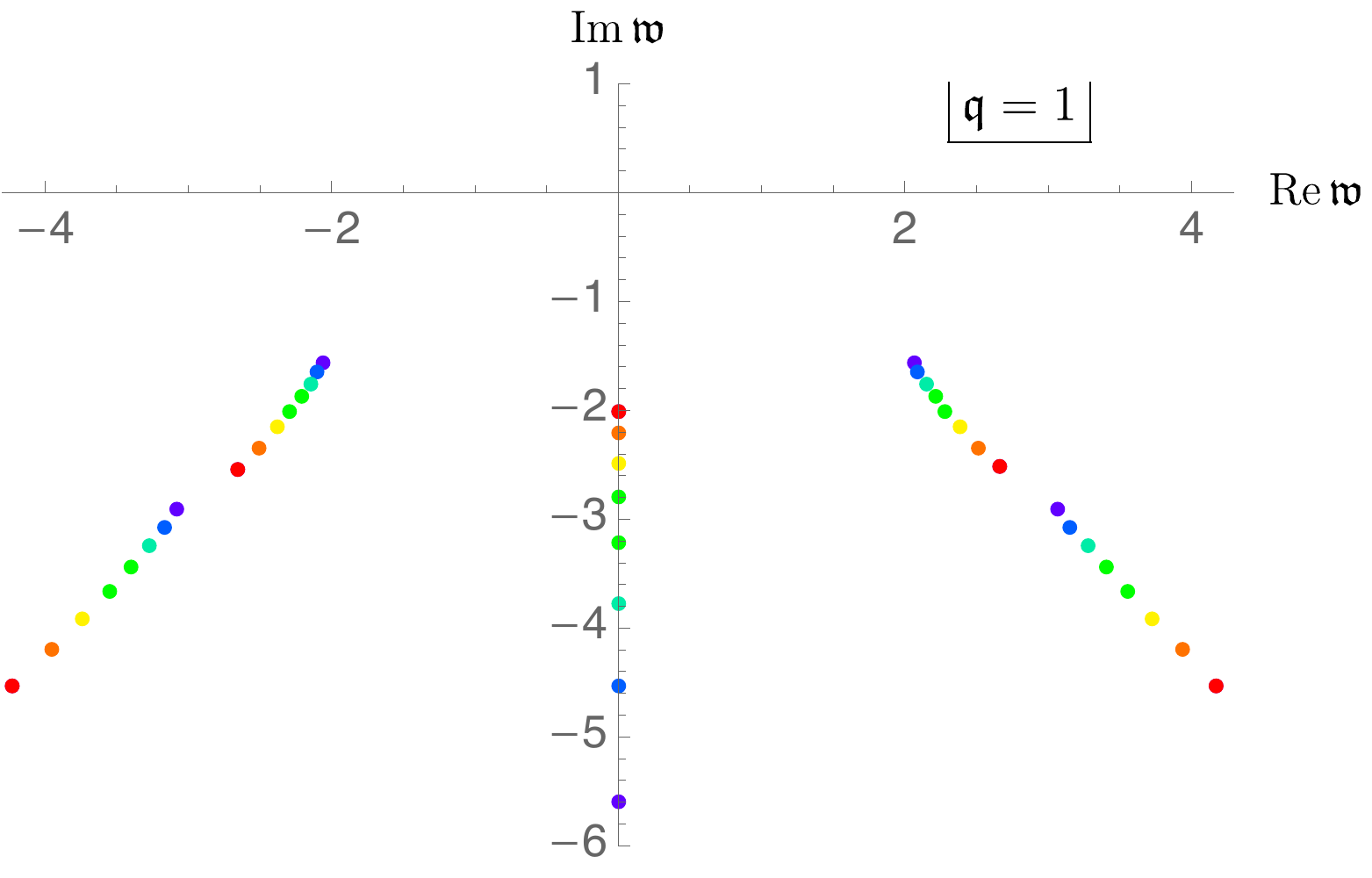}	
	\caption{The  four lowest complex quasinormal modes and the lowest purely imaginary one in the spin 1 channel,  at $\mathfrak{q}=1$. Every colorful path-like set of points, starting in purple and ending in red,  shows the  change of one specific mode when $Q$ discretely increases from $0.45$ to $0.8$. We have considered 8 regular steps by increments of $\Delta Q=0.05$.}
	\label{quasi_Exy_color}
\end{SCfigure}

In Fig.\ref{quasi_Exy}, we have demonstrated the spectra associated with $Q=0.5$ for two values of $\qn$. For both cases, we have shown five lowest-lying quasinormal modes. When $\qn$ decreases, all complex frequency poles move away from the real axis. At the same time, the purely imaginary mode becomes closer to the origin, however, as long as the systems is not extremal, it never reaches the origin. These observations are enough to  conclude that there is no any branch of poles passing through the origin when $\qn\rightarrow 0$. In other words, spin 2 channel of fluctuations contains only gapped excitations.

Our discussion on quasinormal modes in this channel has been so far limited to the case with a fixed value of $Q$. In Fig.\ref{quasi_Exy_color}, we have demonstrated part of the spectrum of quasinormal modes at the fixed momentum $\qn=1$ for various values of $Q$ within the range $0.45\le Q \le 0.8$.  One observes that by approaching towards extremality, the gapped modes move away from the real axis, while the purely imaginary mode becomes close to the origin. It would be interesting to study the spectrum extrapolating region $0.9 < Q \le \sqrt{2}$ to see what the fate of this mode in the extremal limit will be.

\section{Complex life of quasinormal modes and the radius of convergence of the hydrodynamic derivative expansion}
\label{complex_life}
Let us recall that in every channel of our study, the Dirichlet boundary condition at $u=0$ gives an algebraic equation between $\wn$ and $\qn$. Such an equation specifies the spectral curve of the collective excitations in the associated channel. For the spin $n$ channel, it can be formally written as\footnote{It should be noted that we did not explicitly write down the spectral curve equations anywhere in this paper. However it would be useful to note that in the hydrodynamic limit, the perturbative spectral curve of the spin 0 channel, to first order in gradients and second order in $Q$, is obtained by substituting \eqref{m_ij_Spin_0} into \eqref{m_ij_Spin_0_matrix}. In the spin 1 channel, it is simply given by the determinant of \eqref{Dirichlete_Spin_1}.}
\begin{equation}\label{Spectral_Spin_n}
\boldsymbol{F}_n(\qn^2, \wn)=0.
\end{equation}
It is clear that both non-hydrodynamic and possible hydrodynamic modes are encoded in this equation. Hydrodynamic modes,  $\wn(\qn^2\rightarrow 0)= 0$, are expected just to live  in the vicinity of origin $(0,0)$. The small-$\qn$ expansion of $\wn(\qn^2)$ then can be found by using the theorem of Puiseux.
The Puiseux analysis implies that the domain of convergence of Puiseux series centered
at the origin is the circle whose radius is set by the distance from the origin, to the nearest critical point of the associated spectral curve \cite{Grozdanov:2019kge}.
The critical points of the spectral curve, themselves, can be found by solving the following set of equations: 
\begin{equation}\label{Critical}
\boldsymbol{F}_n( \qn_c^2, \wn_c)=0,\,\,\,\,\,\,\,\frac{\partial\boldsymbol{F}_n( \qn_c^2, \wn_c)}{\partial \wn}=\,0.
\end{equation}
In \cite{Grozdanov:2019kge}, it has been shown that the critical points obtained from these equations are exactly the level-crossing points of the complexified quasinormal modes in the associated channel. 

The focus of \cite{Grozdanov:2019kge} is to study the quasinormal modes of a holographic \textit{neutral} fluid. 
However, when fluid carries a conserved charge as well, the arrangement of complexified quasinormal modes may change significantly. Specifically in the spin 0 channel, in addition to the sound branches of  the Puiseux series passing through the origin, the diffusion branch passes through the same point, too. It may possibly give rise to emergence of new critical points, due to crossing between sound and diffusion branches. As we explicitly show in the next subsection,  such critical points will appear when the parameter $Q$ exceeds a specific threshold.

\subsection{Spin 0 channel}
In \sec{Gauge_section} we formally argued how to find the quasinormal modes of coupled variables in the bulk. For the present spin channel, the spectral equation, namely $\boldsymbol{F}_\text{0}( \qn^2, \wn; Q)=0$, was already solved numerically in previous section and
the corresponding spectrum of quasinormal modes typically was shown in Fig.\ref{quasi_EzZ0_Q_12_fig}. From the arrangement of poles in that  figure one notices that $\boldsymbol{F}_\text{0}( \qn^2, \wn; Q)=0$ has three eigen frequencies in the vicinity of $(0,0)$.
 Assuming the analyticity of $\boldsymbol{F}_0$ at $(0,0)$, then the implicit function theorem gives these three branches by three Puiseux series as it follows
\begin{equation}\label{Puiseux}
\begin{split}
\wn^{\pm}_{\text{sound}}=&\,-i\sum_{n=1}^{\infty}a_n e^{\pm \frac{i \pi n}{2}}\qn^n=\,\pm a_1 \qn+\, i a_2\qn^2+\cdots\\
\wn_{\text{diffusion}}=&\,-i\sum_{n=1}^{\infty}c_n \qn^{2n}=\,- i c_1 \qn^2+\cdots .
\end{split}
\end{equation}
Following refs. \cite{Withers:2018srf,Grozdanov:2019kge}, what we refer to as the hydrodynamic derivative expansion is the form of the above expansions in the momentum space.

Let us recall that the two sound branches are located symmetrically with respect to imaginary axis in the complex $\wn$ plane. Thus, in half of the $\wn$ plane, for instance, where $\text{Re} \, \wn>0$ there are exactly two branches passing through the origin; $\wn^{+}_{\text{sound}}$ together with $\wn_{\text{diffusion}}$.
By naively applying the statement of \cite{Grozdanov:2019kge} to the present case, one may  conclude that the distance between $(0,0)$ and the critical point of $\boldsymbol{F}_0$, the nearest to origin, identifies the radius of convergence of the derivative expansion in this channel. But an immediate follow-up question is: to which series given in \eqref{Puiseux} such radius corresponds?

Needless to say that $\wn^{+}_{\text{sound}}$ and $\wn_{\text{diffusion}}$
may have different radii of convergence. In fact \textit{ the radius of convergence  of $\wn^{+}_{\text{sound}}$ ($\wn_{\text{diffusion}}$) is identified with the distance between origin and the nearest critical point  located  on the sound (diffusion) branch of Puiseux series}.
\begin{figure}
	\centering
		\includegraphics[width=0.7\textwidth]{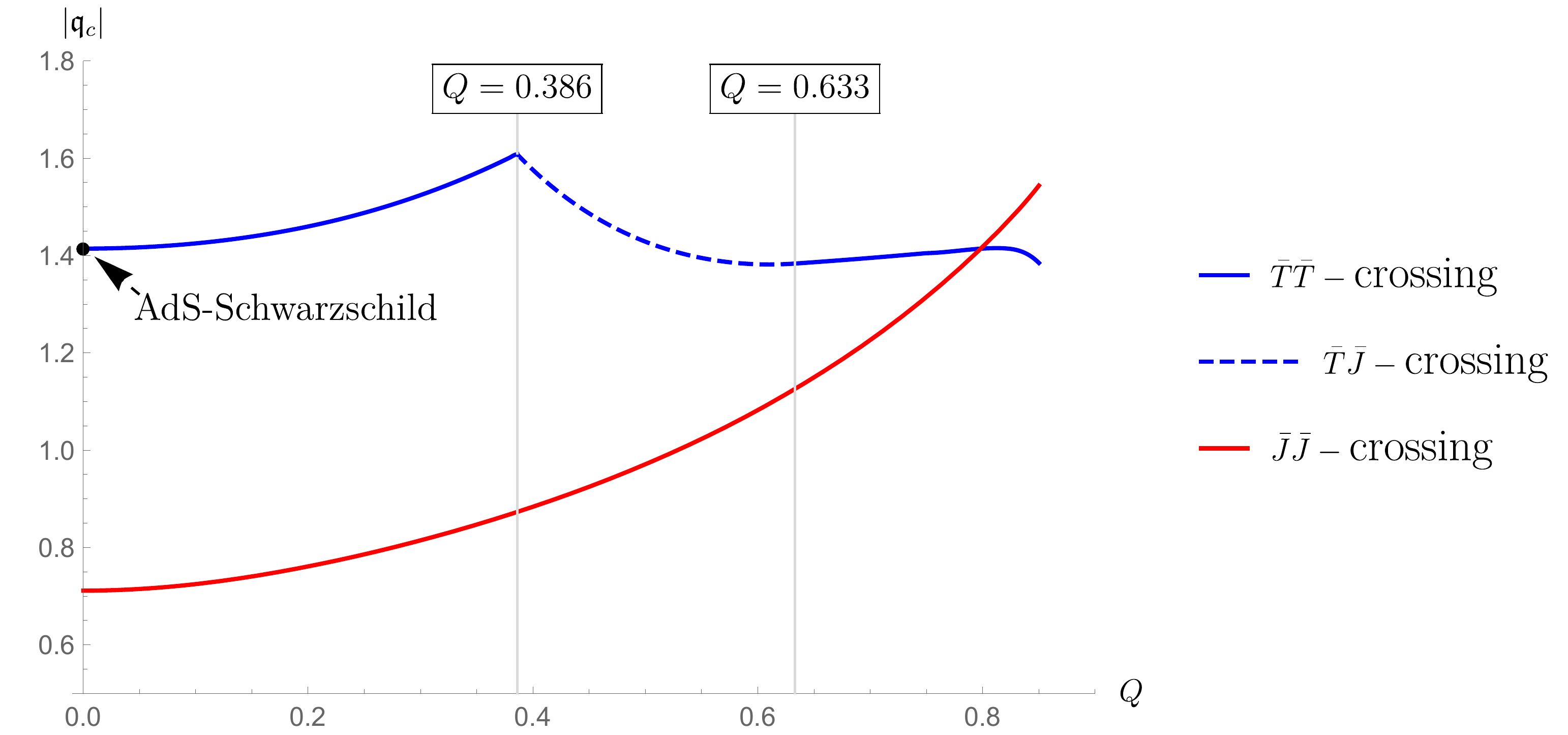}	
\caption{ The radius of convergence of the derivative expansion versus $Q$.  As $Q$ increases, the radius of convergence of $\wn_{\text{diffusion}}$ (red curve) monotonically increases. At the same time, domain of convergence of $\wn_{\text{sound}}$ non-trivially changes. As discussed in the text, it can be studied in three different  intervals: $0\le Q\le 0.386$, $0.386\le Q \le 633$ and $0.633\le Q\le 0.850$. The intersection point of blue curve with the vertical axis, related to the sound mode in the $\mathcal{N}=4$ SYM theory in the vanishing $\mu$ limit, was found in \cite{Grozdanov:2019kge}.}
\label{q_c_Sound_diiusion}
\end{figure}
Thus in addition to find the nearest critical points, via solving \eqref{Critical}, one has to be careful about positioning of them on the considered branch of Puiseux series.

We have numerically found the nearest critical points to the origin, on both sound and diffusion branches of Puiseux series,  for several values of $Q$, within the range $0\le Q\le 0.85$.  The result has been shown in Fig.\ref{q_c_Sound_diiusion}. The typical behavior of the diffusion mode, shown by the red curve,  seems to be qualitatively the same for the whole range of $Q$. The situation for sound mode, i.e. the blue curve,  however, is more complicated. It should be studied in three different intervals; $(i)\, 0\le Q \le 0.386$, $(ii) \,0.386 \le Q \le 0.633$ and $(iii) \,0.633\le Q \le 0.850$. 


To explore more on the relation between radius of convergence of the  derivative expansion and $Q$, we now start to study the complex life of quasinormal modes. We assume  $\qn^2$ to be  a complex number $\qn^2=|\qn^2|e^{i \theta }$. Then  $\theta=0$ simply corresponds to poles with real $\qn^2$, already shown in Fig.\ref{quasi_EzZ0_Q_12_fig}.  As before, we show them by (large) dots and (large) stars in upcoming figures. Then we let the phase $\theta$ change from $0$ to $2\pi$. For a given $|\qn^2|$, such change of $\theta$ corresponds to moving dots and stars along some trajectories in  the complex $\wn$ plane.
The interaction of poles via crossing of their trajectories is the main issue that we will discuss in details for each of the three intervals mentioned above, separately.

\begin{figure}[tb]
	\centering
	\includegraphics[width=0.35\textwidth]{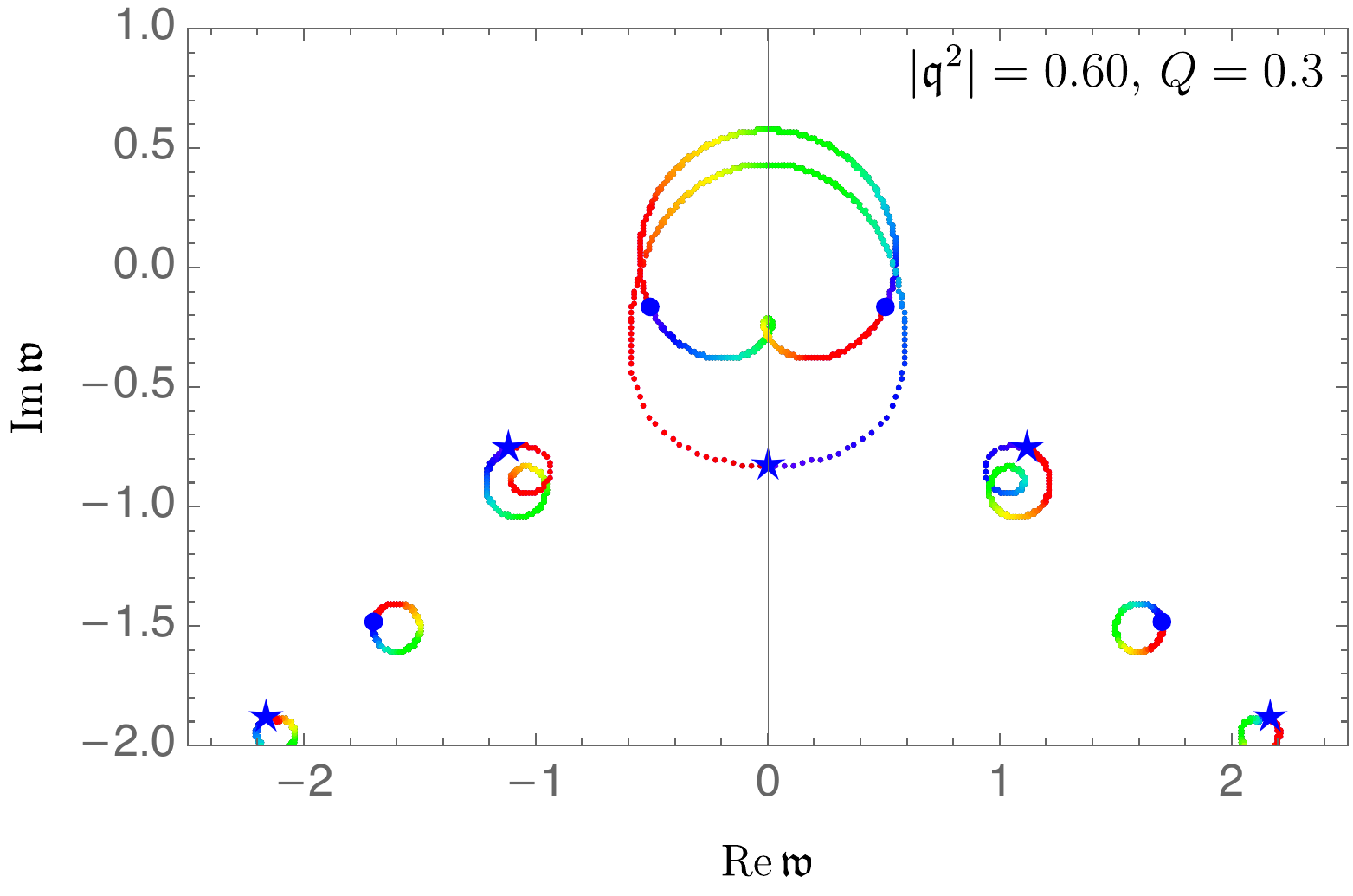}\includegraphics[width=0.35\textwidth]{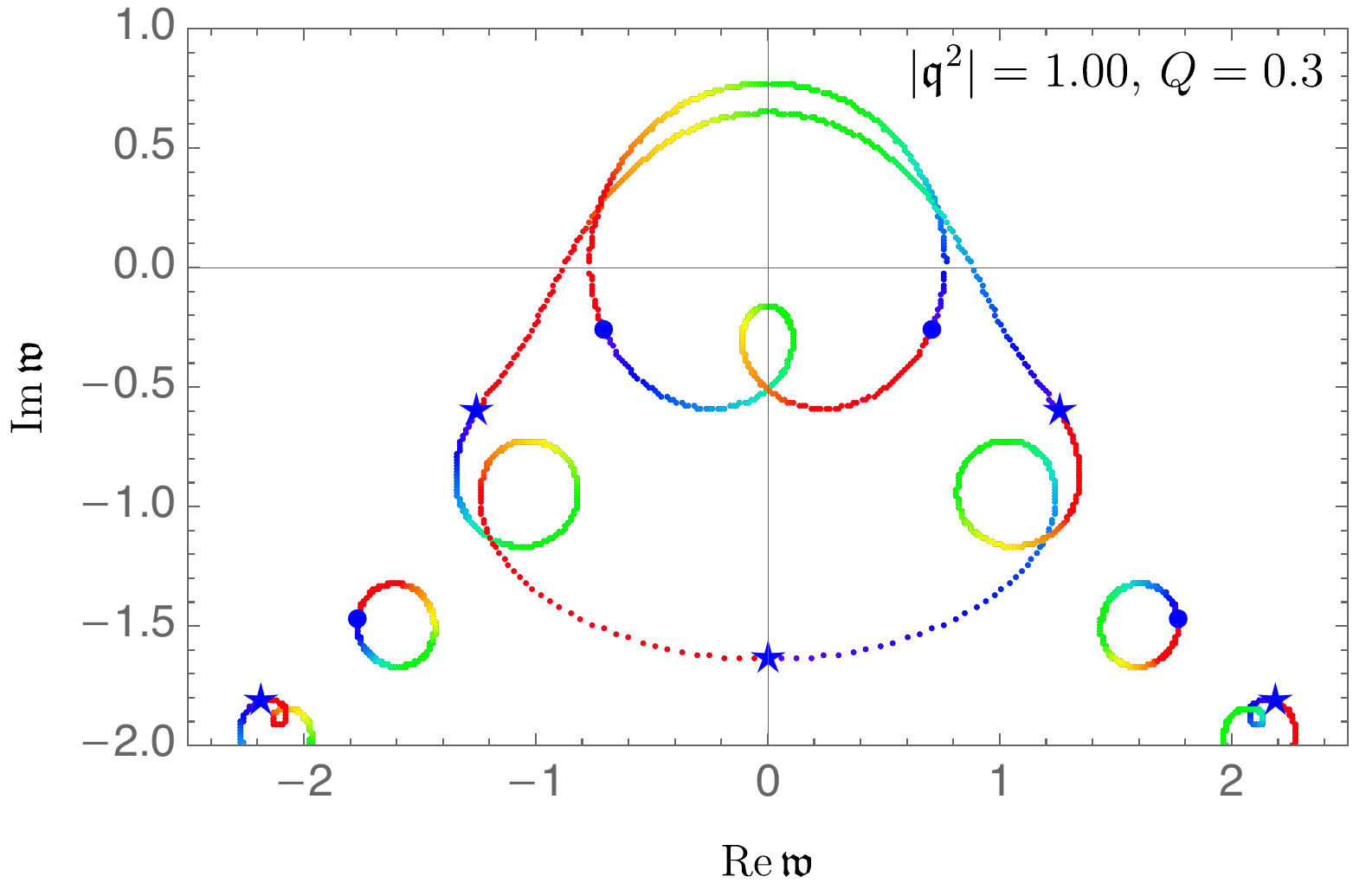}\includegraphics[width=0.35\textwidth]{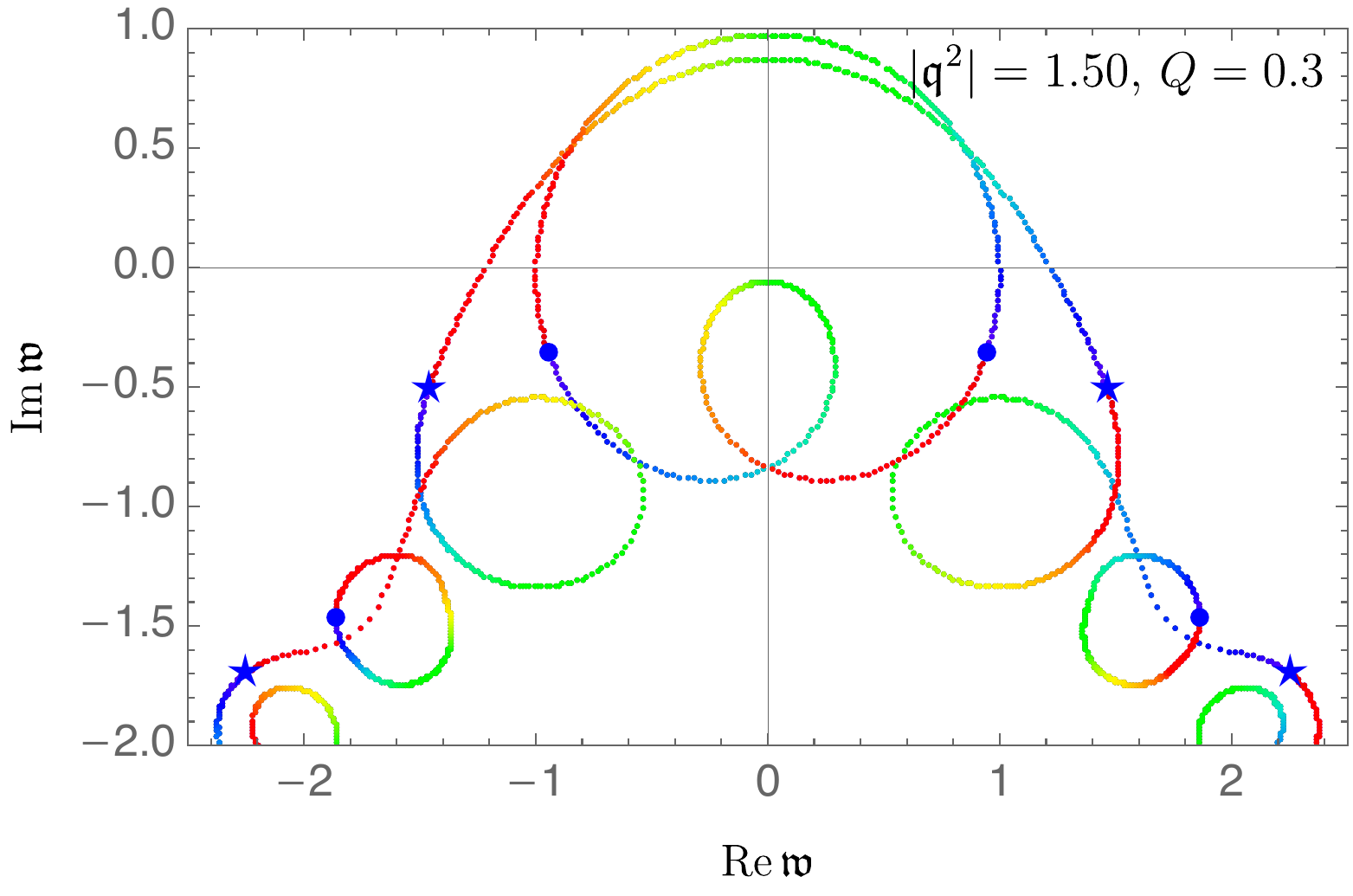}
	\includegraphics[width=0.42\textwidth]{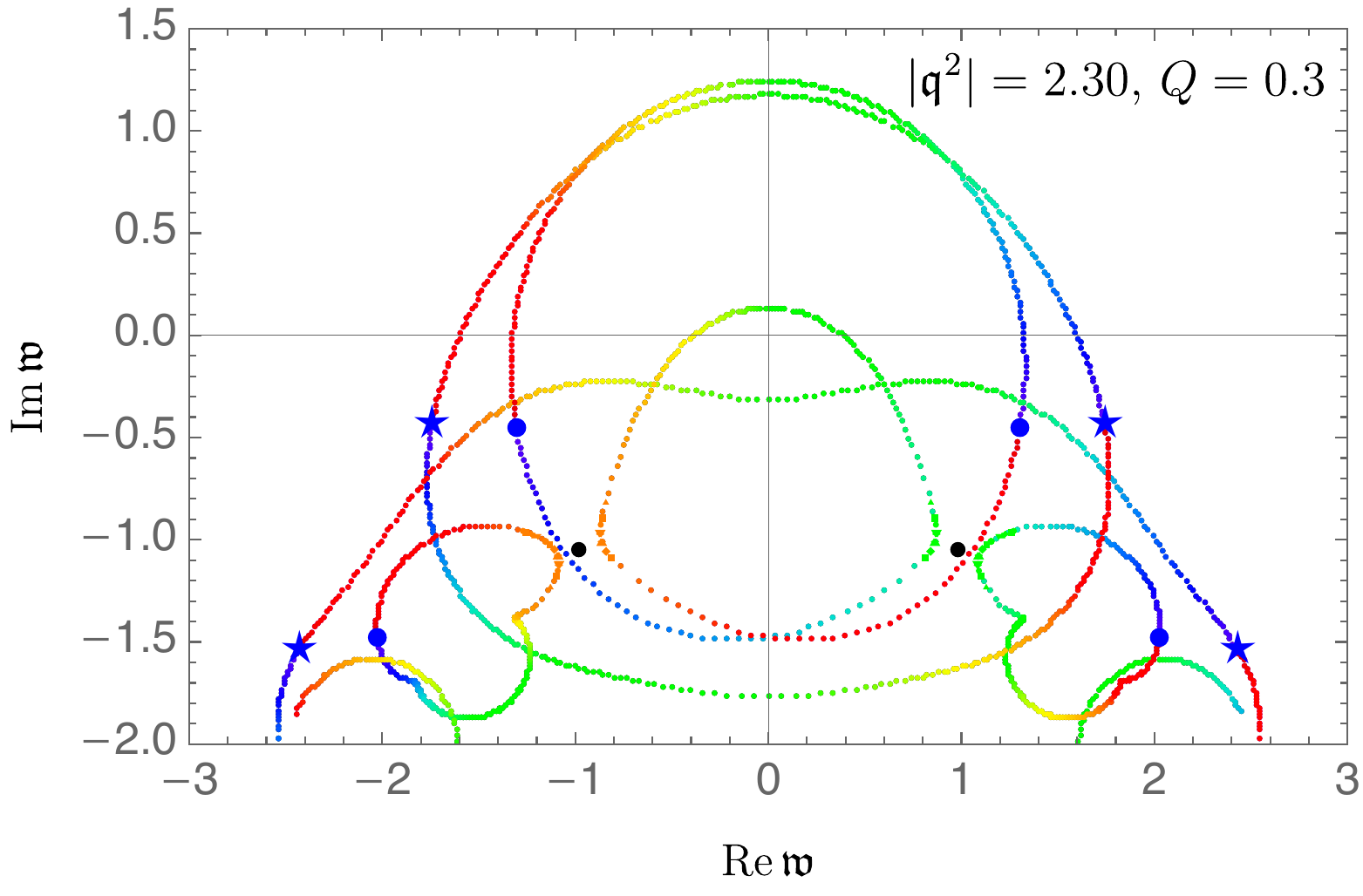}\,\,\,\,\,\,\includegraphics[width=0.42\textwidth]{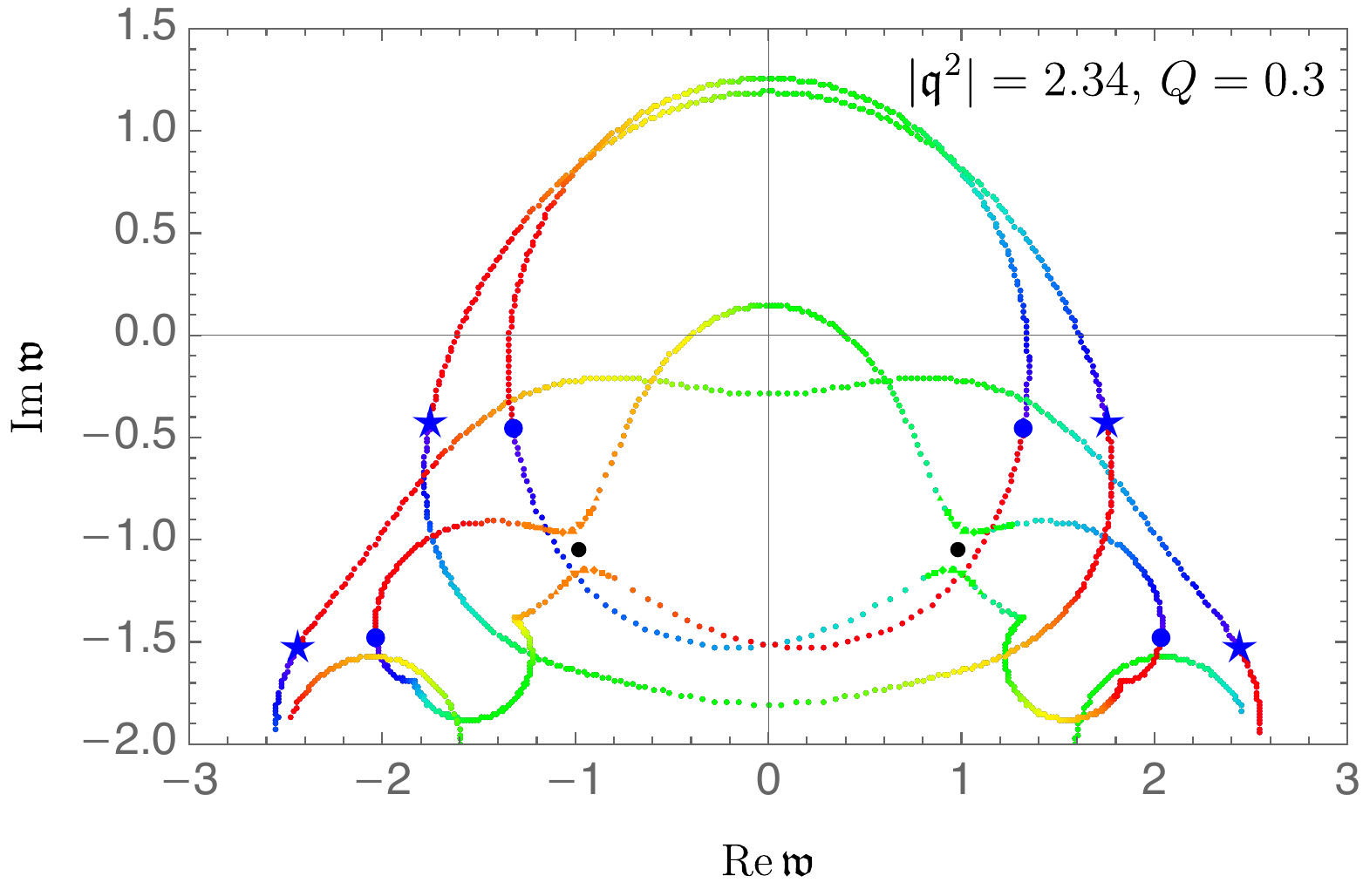}
	\caption{Poles of the retarded two-point function in the spin 0 channel at $Q=0.3$, in the complex $\wn-$plane, at various values of
		the complexified momentum $\qn^2=|\qn^2|e^{i \theta}$. Large dots and large stars correspond to the poles with  purely real momentum (i.e. at $\theta=0$). As $\theta$ increases from $0$ to $2\pi$, each pole moves counter-clockwise	following the trajectory whose color changes continuously from blue to red. While at $|\qn^2|=0.60$ (top left panel) all poles follow closed orbits, at $|\qn^2|=1.00$ (top middle panel), the hydrodynamic diffusion pole and the two lowest (namely nearest to the horizontal axis) star gapped poles follow open orbits. It simply means that the dispersion relation $\wn_{\text{diffusion}}(\qn^2)$ has branch point singularities in the complex momentum squared plane at $(0.60)^{1/2}<|\qn_c|<(1.00)^{1/2}$. It is clear that the point $(Q=0.5, |\qn_c|)$ then lies on the red curve in Fig.\ref{q_c_Sound_diiusion}. At $|\qn^2|=1.50$ (top right panel) the orbits of the hydrodynamic sound pole and the two nearest dot poles are still closed. However, by further increasing $|\qn^2|$, their associated trajectories come close to each other. Finally they collide at the positions marked by black dots in the bottom row plots. The collision points are identified with critical value of momentum $|\qn_c^2|=2.32$.  It is clear that the point $(Q=0.3, (2.32)^{1/2})$  lies on the blue curve in Fig.\ref{q_c_Sound_diiusion}. After the collision, for instance at $|\qn^2|=2.34$, the orbits of sound pole and the two nearest dot gapped poles are no longer closed: four of them exchange their positions as the phase $\theta$ increases from $0$ to $2\pi$. This is the manifestation of the $\boldsymbol{\bar{T}\bar{T}-}$\textbf{crossing}. }
	\label{Complex_Sound_Chennel}
\end{figure}

$\boldsymbol{(i)\,\, 0\le Q \le 0.386}$: In this interval we choose to show the results associated with $Q=0.3$. See Fig.\ref{Complex_Sound_Chennel}. It is clear that by increasing $|\qn^2|$, trajectories of poles become more complicated. Let us firstly consider the highest large dots in the figure. These pols  actually lie on the two sound branches of the Puiseux series near the origin. When $|\qn_c^2|\approx 2.32$, their trajectory collides with that of (dot)  gapped mode poles,  at two points (marked by black dots in the bottom row plots). One then concludes that at $Q=0.3$ the radius of convergence of the derivative expansion for the $\wn^{\pm}_{\text{sound}}$ is $|\qn^{\text{sound}}_c|\approx(2.32)^{1/2}\approx 1.52$. 

\par\bigskip 
\noindent

Since the above-mentioned collision happens between two (dot) poles which both belong to the \textit{master} energy density spectrum,  we call the crossing of the associated trajectories  the $\boldsymbol{\bar{T}\bar{T}}-$\textbf{crossing}\footnote{It might be better to refer to this collision as $\bar{T}^L\bar{T}^L-$crossing. However, for the sake of brevity we omit the superscripts. }. As a result, within the range $0\le Q \le 0.386$, this is $\bar{T}\bar{T}-$crossing which determines the radius of convergence of the derivative expansion for the sound branch. As pointed out in Fig.\ref{q_c_Sound_diiusion}, the AdS-Schwarzschild case with $Q=0$, studied in \cite{Grozdanov:2019kge}, falls into the same range. 

\begin{figure}[tb]
	\centering
	\includegraphics[width=0.35\textwidth]{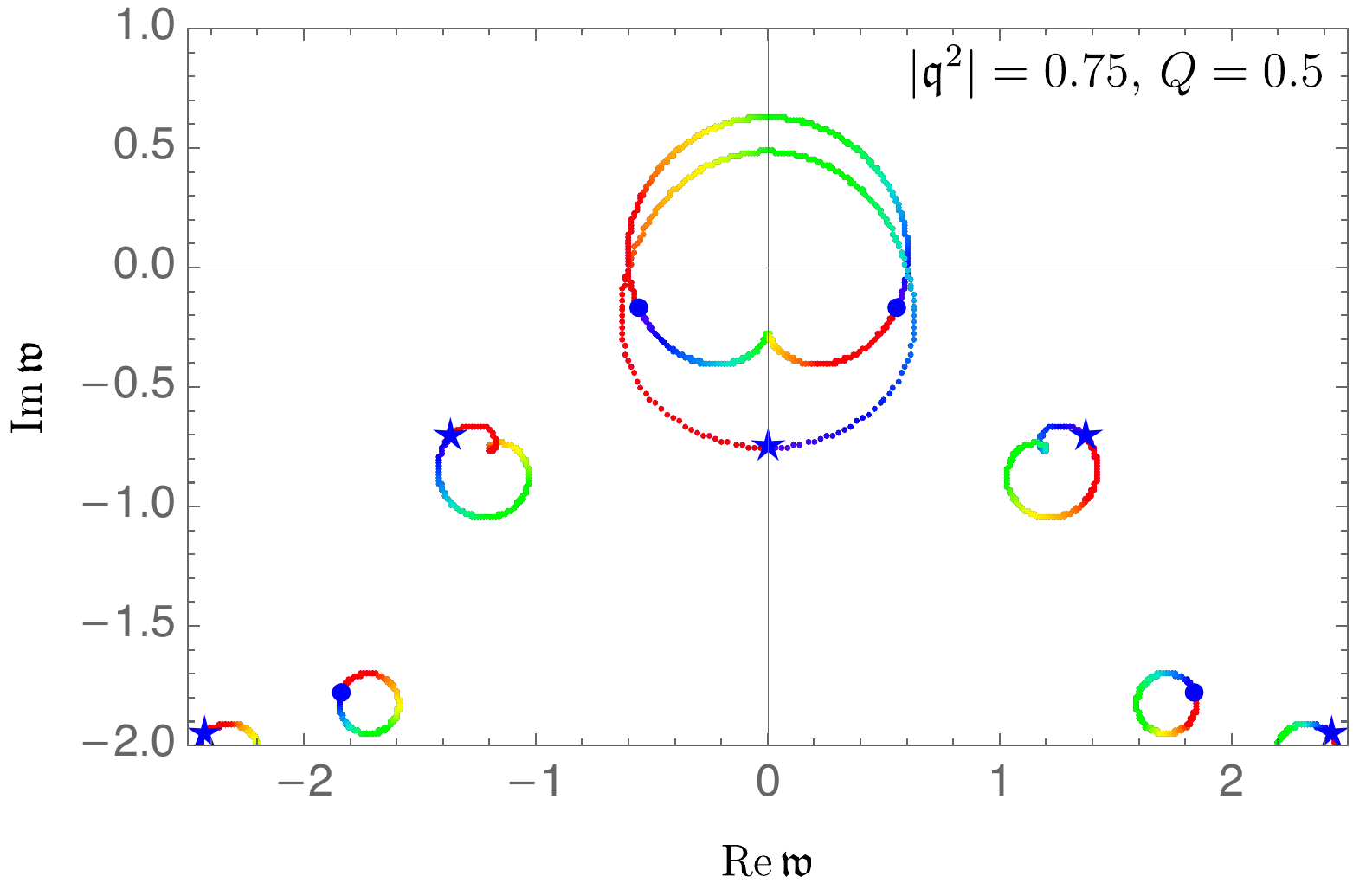}\includegraphics[width=0.35\textwidth]{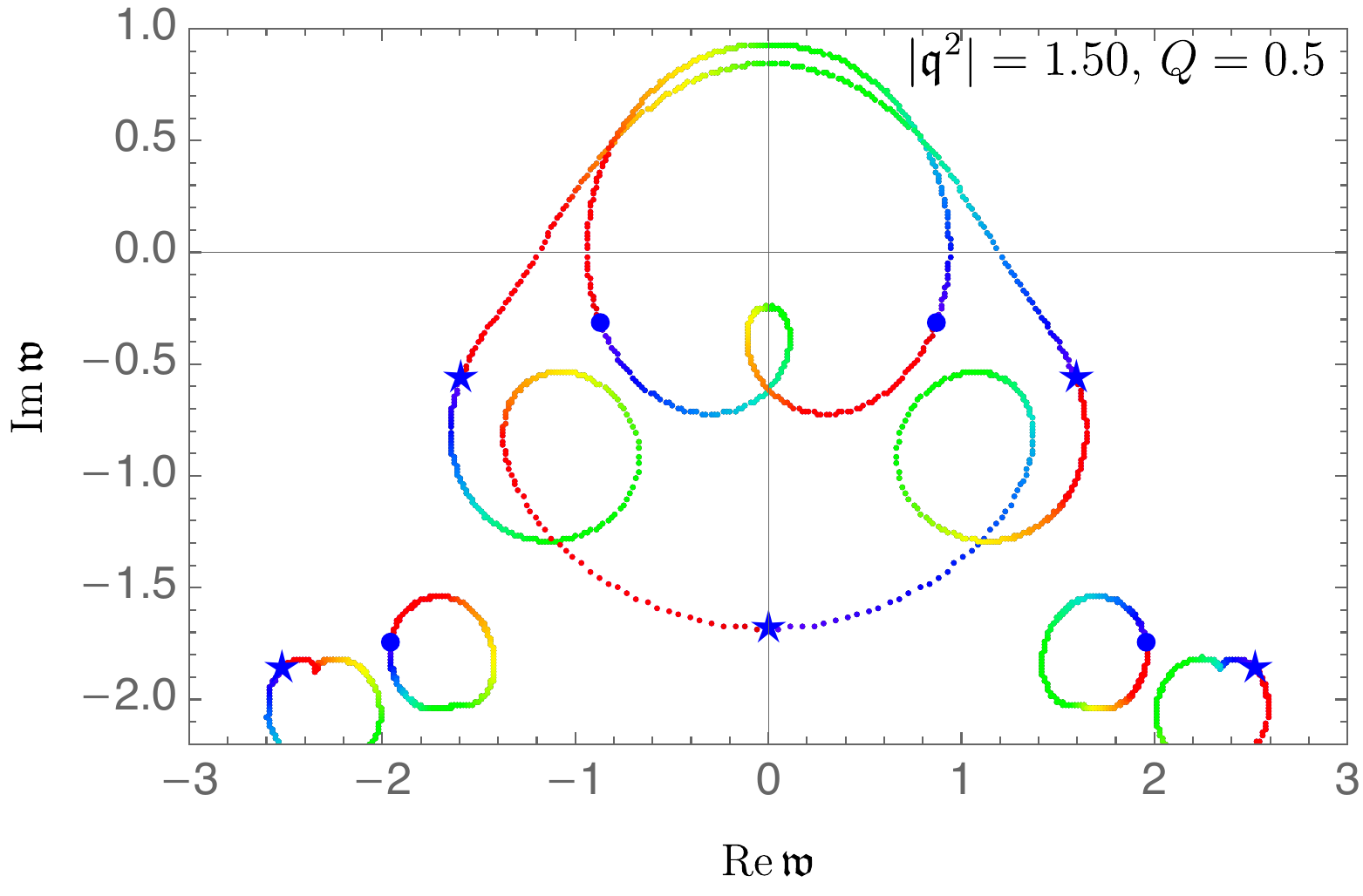}\includegraphics[width=0.35\textwidth]{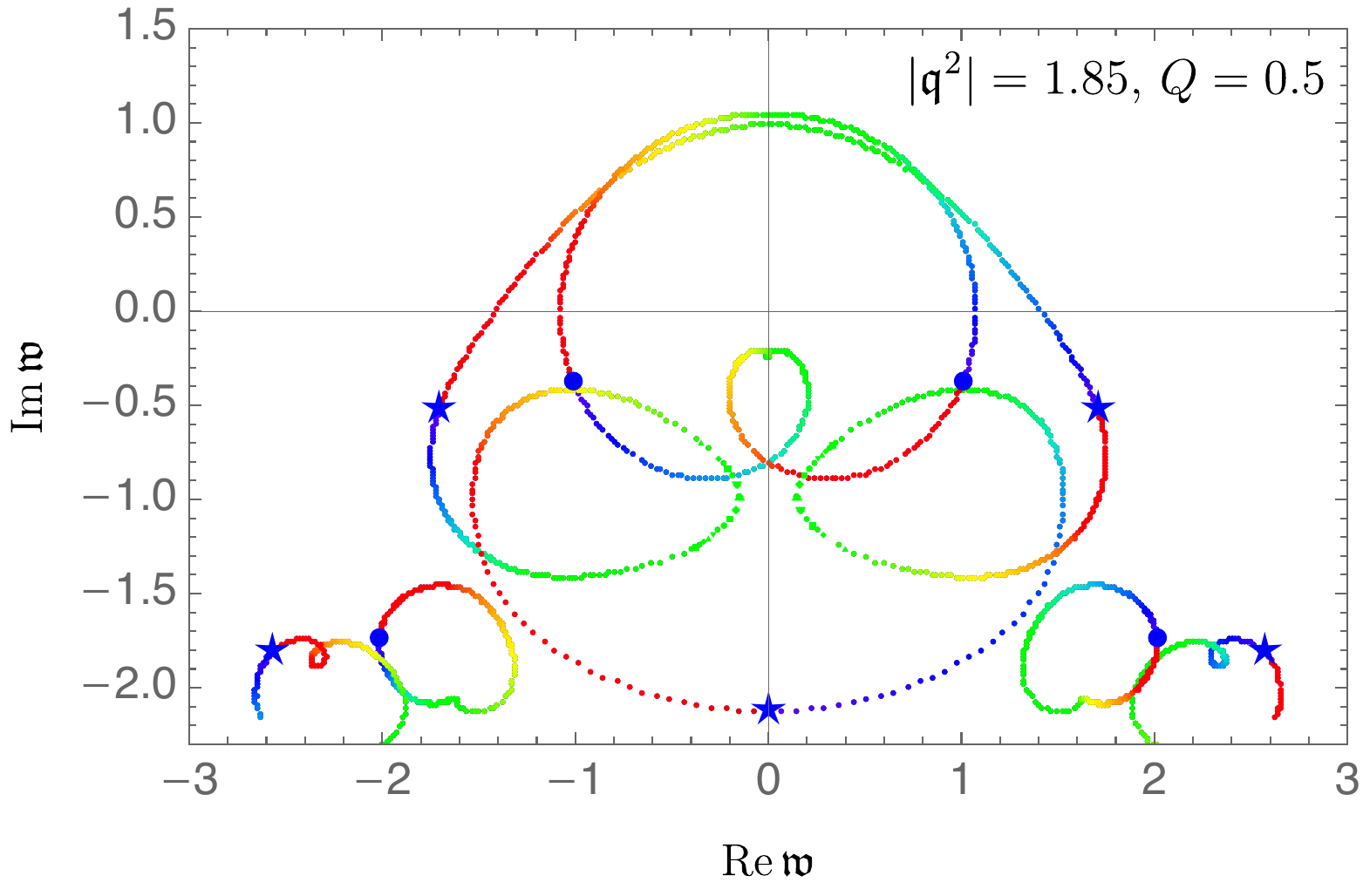}
	\includegraphics[width=0.42\textwidth]{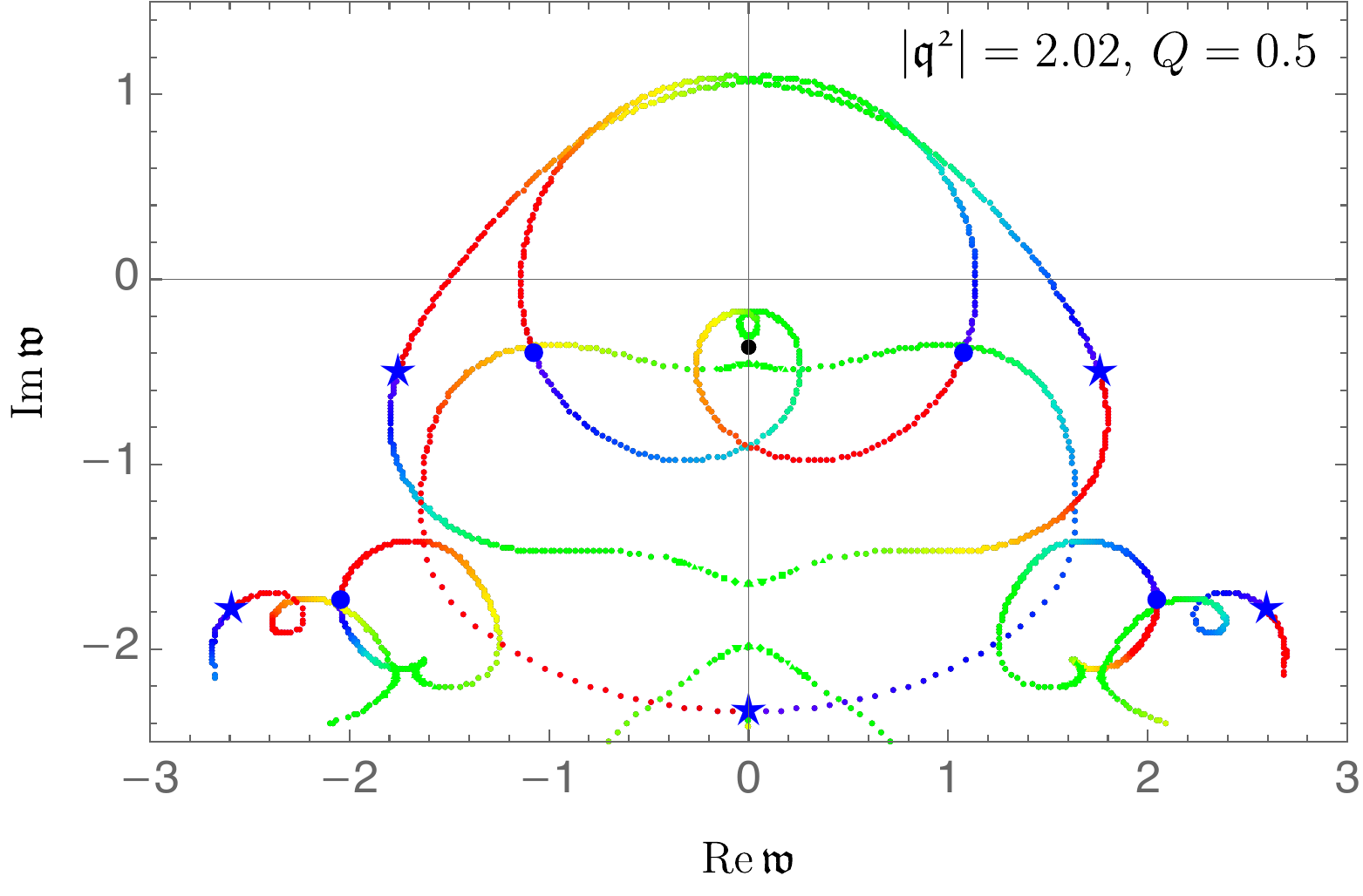}\,\,\,\,\includegraphics[width=0.42\textwidth]{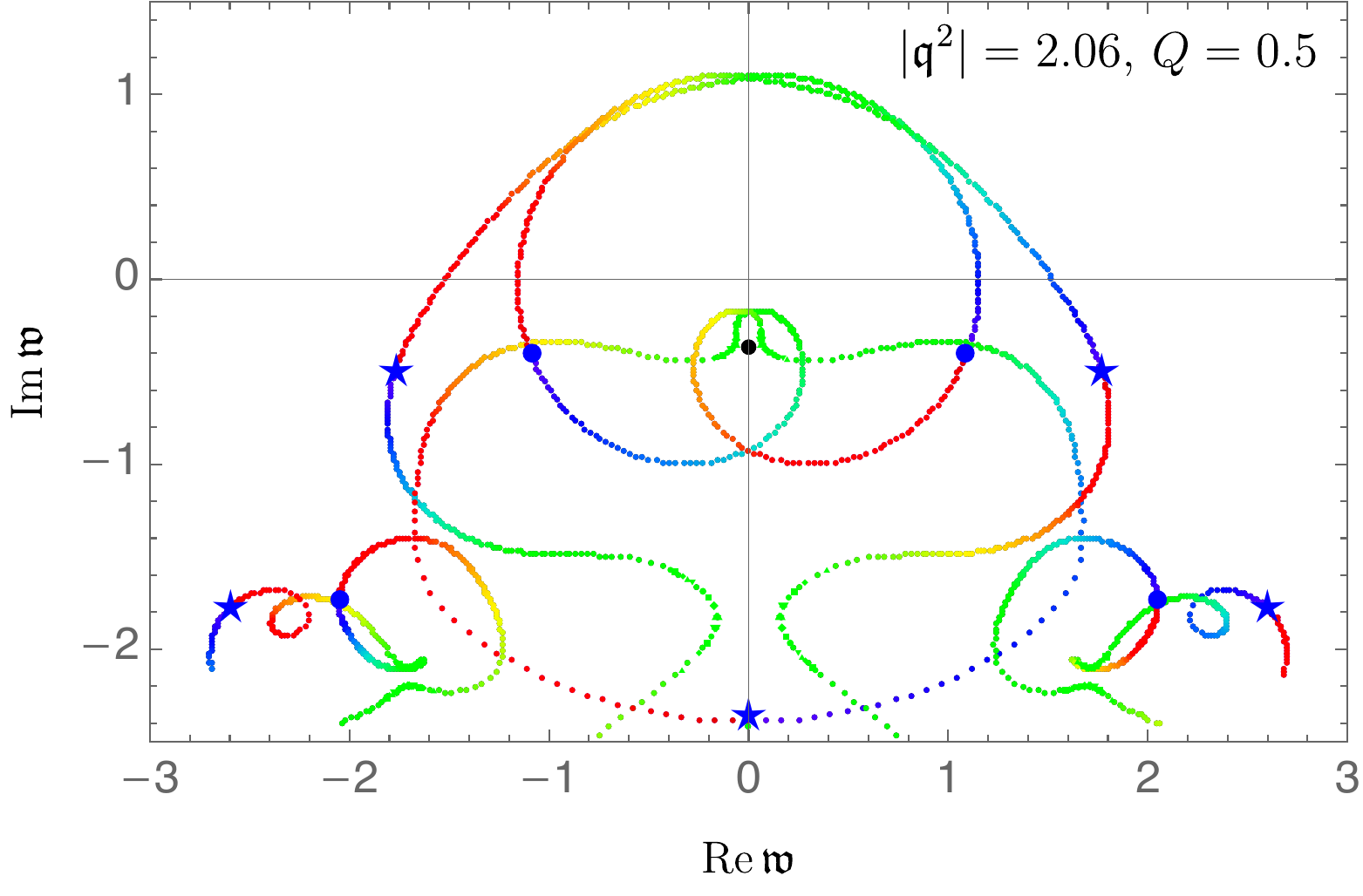}
	\caption{Poles of the retarded two-point function in the spin 0 channel at $Q=0.5$, in the complex $\wn-$plane, at various values of	the complexified momentum $\qn^2=|\qn^2|e^{i \theta}$.  Large dots and large stars correspond to the poles with  purely real momentum (i.e. at $\theta=0$). As $\theta$ increases from $0$ to $2\pi$, each pole moves counter-clockwise	following the trajectory whose color changes continuously from blue to red. While at $|\qn^2|=0.75$ (top left panel) all poles follow closed orbits, at $|\qn^2|=1.50$ (top middle panel), the hydrodynamic diffusion pole and the two lowest star gapped poles follow open orbits. It simply means that the dispersion relation of $\wn_{\text{diffusion}}(\qn)$ has branch point singularities  in the complex momentum squared plane at $(0.75)^{1/2}<|\qn_c|<(1.50)^{1/2}$. It is clear that the point $(Q=0.5, |\qn_c|)$ lies on the red curve in Fig.\ref{q_c_Sound_diiusion}. At $|\qn^2|=1.85$ (top right panel) the orbits of the hydrodynamic sound pole and the two nearest dot poles are still closed. But interestingly, the trajectory of the diffusion pole seems going to become closed again. At $|\qn^2|=2.02$ (bottom left panel), the orbit of diffusion pole is already closed, however it nearly collide the left sound pole at a position marked by black dot on the vertical axis. The collision point is identified with critical value of momentum $|\qn_c^2|=2.04$. As expected, the point $(Q=0.5, |\qn_c|=(2.04)^{1/2})$ lies on the blue curve in Fig.\ref{q_c_Sound_diiusion}. After the collision, for instance at $|\qn^2|=2.06$, the orbits of sound poles and the diffusion pole are no longer closed. They exchange their positions cyclically as the phase  increases from $0$ to $2\pi$. This is the manifestation of the $\boldsymbol{\bar{T}\bar{J}-}$\textbf{crossing}.}
	\label{Complex_Sound_Chennel_1_2}
\end{figure}
\par\bigskip 
\noindent

$\boldsymbol{(ii)\,\, 0.386\le Q \le 0.633}$: In this interval we choose to show the results associated with $Q=0.5$. See Fig.\ref{Complex_Sound_Chennel_1_2}. By increasing $|\qn|^2$, the poles tend to collide. Interestingly, it turns out that the first collision of the sound poles (the highest dots in the figure) is with the other hydro pole in this channel, namely the diffusion pole denoted by a star on imaginary axis.  The collision point is identified with critical value of momentum  $|\qn_c|^2\approx 2.04$. One then concludes that at $Q=0.5$ the radius of convergence of the derivative expansion for the sound branch of Puiseux series is $|\qn^{\text{sound}}_c|\approx(2.04)^{1/2}\approx 1.43$.
Let us emphasize that while one of the two colliding poles comes from the \textit{master} energy density spectrum, the other one belongs to the spectrum of \textit{master} charge density.  Therefore
it is reasonable to call such crossing of trajectories the $\boldsymbol{\bar{T}\bar{J}-}$\textbf{crossing}.  

\begin{figure}[tb]
	\centering
	\includegraphics[width=0.35\textwidth]{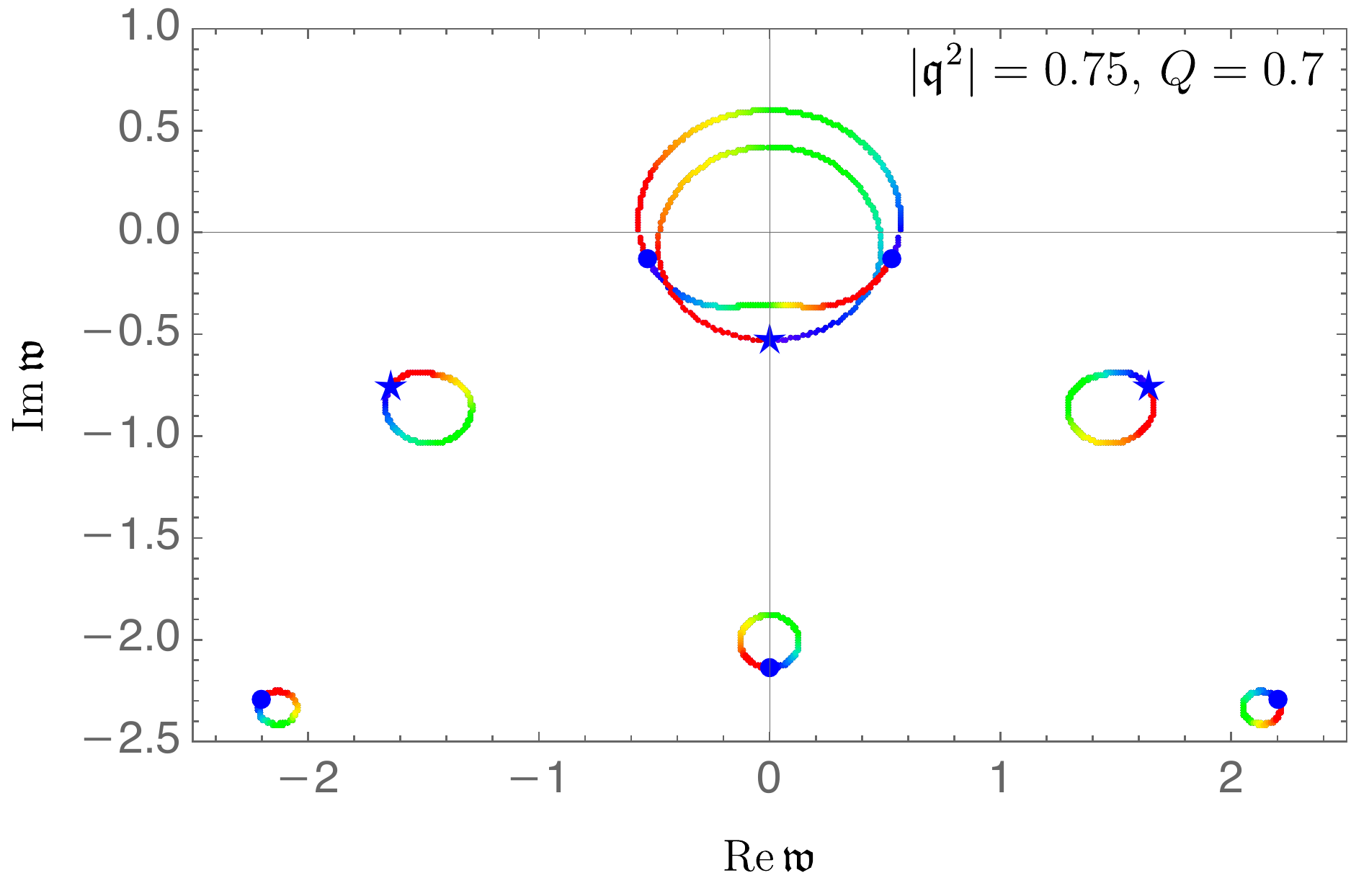}\includegraphics[width=0.35\textwidth]{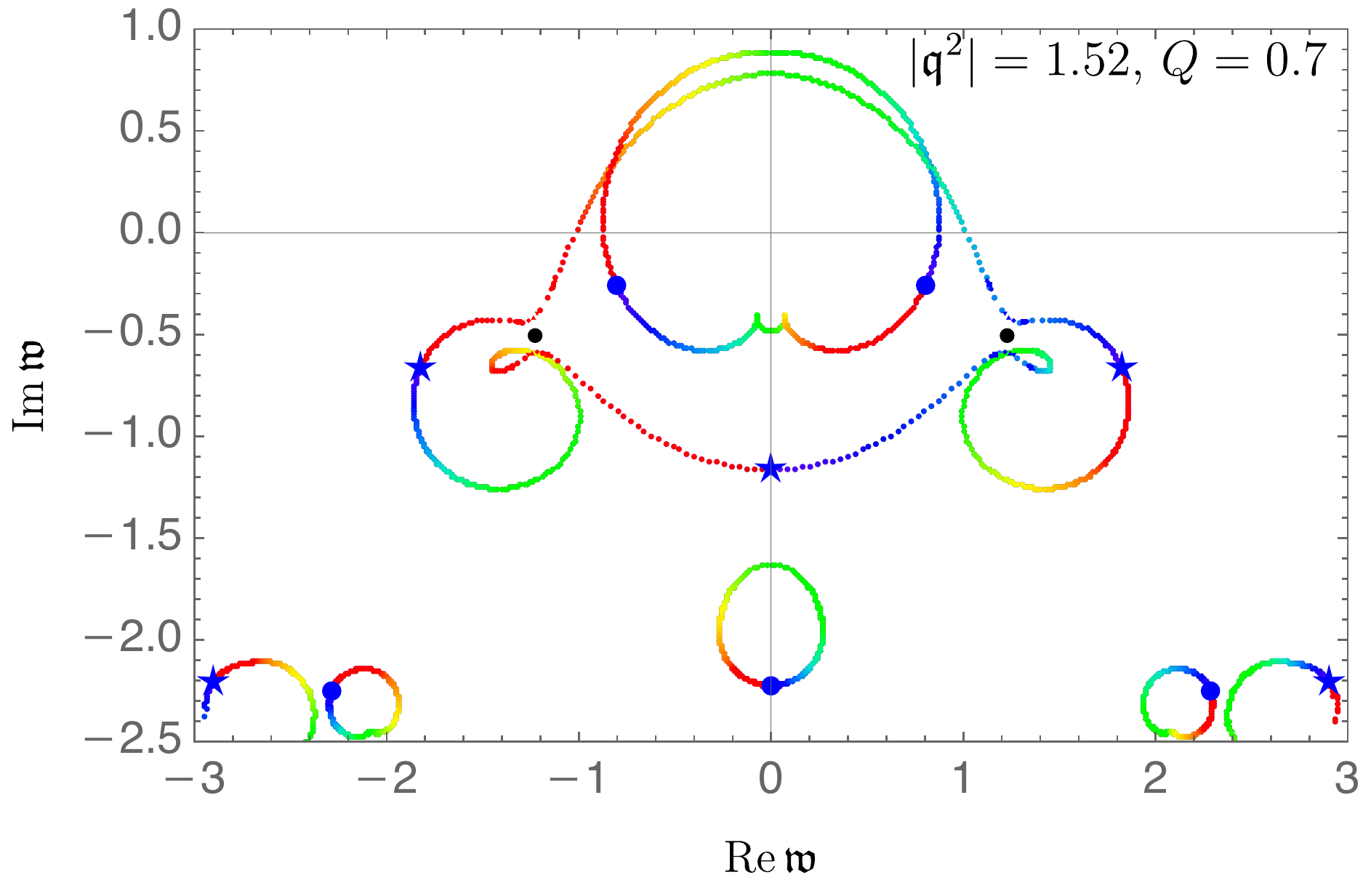}\includegraphics[width=0.35\textwidth]{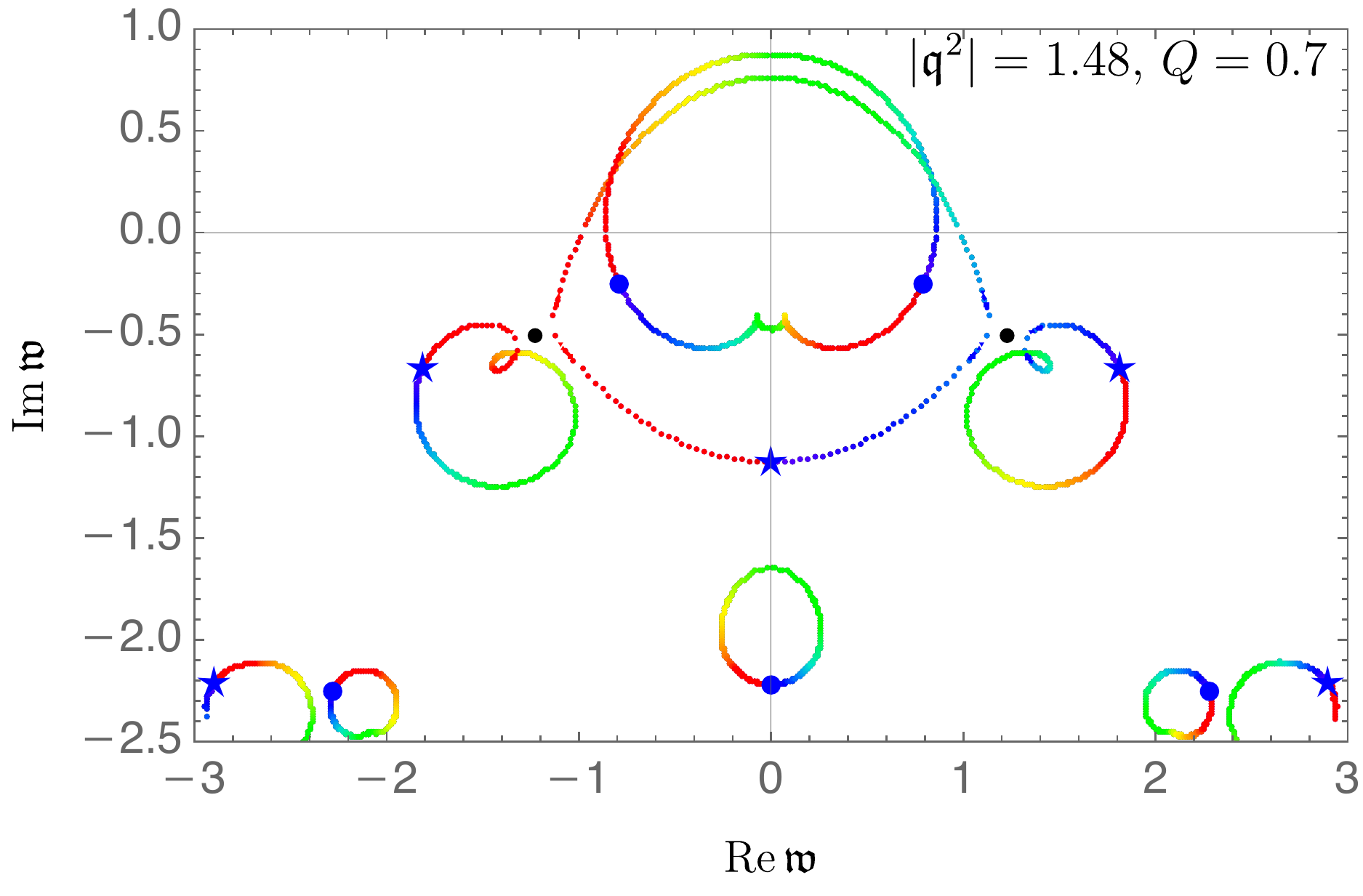}
	\includegraphics[width=0.42\textwidth]{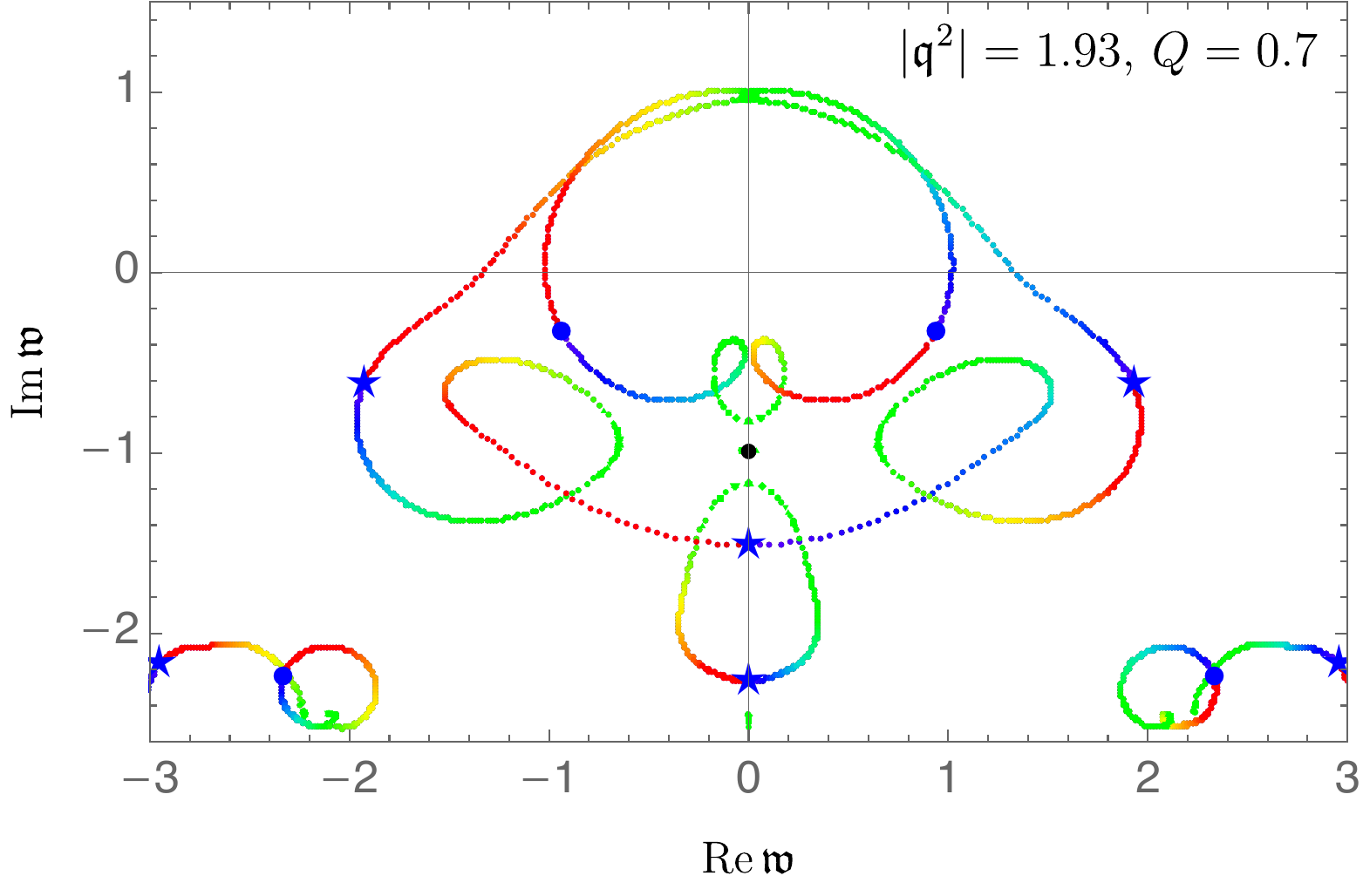}\,\,\,\,\includegraphics[width=0.42\textwidth]{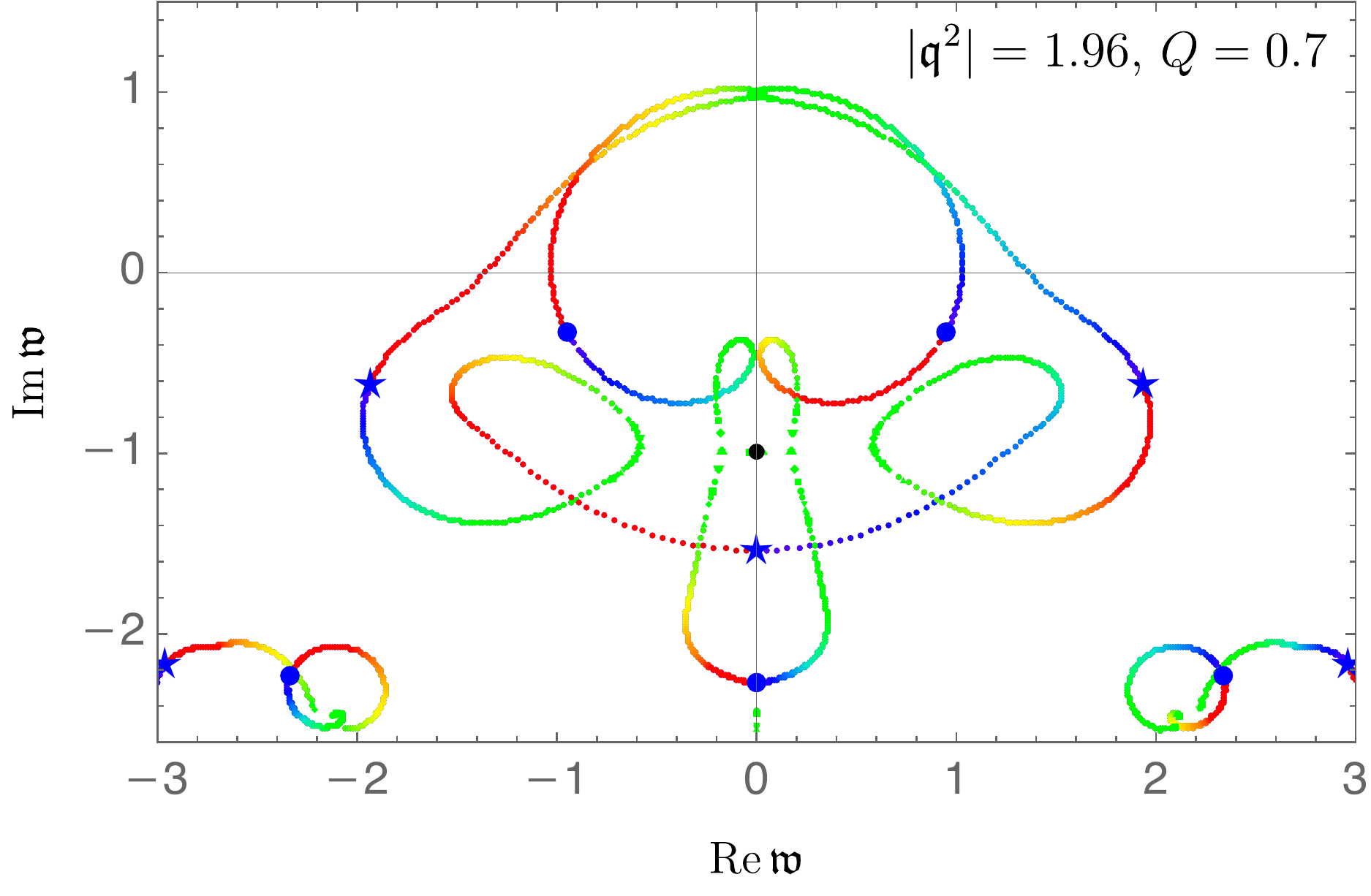}
	\caption{Poles of the retarded two-point function in the spin 0 channel at $Q=0.7$, in the complex $\wn-$plane, at various values of	the complexified momentum $\qn^2=|\qn^2|e^{i \theta}$.  Large dots and large stars correspond to the poles with  purely real momentum (i.e. at $\theta=0$). As $\theta$ increases from $0$ to $2\pi$, each pole moves counter-clockwise	following the trajectory whose color changes continuously from blue to red. At $|\qn^2|=0.75$ (top left panel) all poles follow closed orbits. By increasing $|\qn^2|$, the trajectory of the hydrodynamic diffusion pole comes close to those of the nearest star gapped poles.  At $|\qn^2|=1.48$ (top middle panel) they nearly collide at the positions marked by black dots. The actual collision point is identified with the critical value of the momentum $|\qn_{c}^{2}|\approx 1.50$. It is clear that the point $(Q=0.7, |\qn_c|=(1.50)^{1/2})$ lies on the red curve in Fig.\ref{q_c_Sound_diiusion}. At $|\qn^2|=1.52$ (top right panel), the orbits of the diffusion pole and the two lowest star gapped poles are no longer closed; these poles cyclically exchanges their positions as the phase $\theta$ increases from $0$ to $2\pi$. This is actually the manifestation of $\boldsymbol{\bar{J}\bar{J}-}$\textbf{crossing}. By further increasing $|\qn^2|$, the trajectory of sound poles come close to the that of the nearest gapped pole on the imaginary axis. At $|\qn^2|=1.93$ (bottom left panel) they nearly collide at the positions marked by black dots. The actual collision point is identified with the critical value of the momentum $|\qn_{c}^{2}|\approx 1.95$. As expected, the point $(Q=0.7, |\qn_c|=(1.95)^{1/2})$ lies on the blue curve in Fig.\ref{q_c_Sound_diiusion}.  At $|\qn^2|=1.96$ (bottom right panel) they exchange their positions cyclically as the phase  increases from $0$ to $2\pi$. This is the manifestation of the $\boldsymbol{\bar{T}\bar{T}-}$\textbf{crossing}.}
	\label{Complex_Sound_Chennel_7_10}
\end{figure}
\par\bigskip 
\noindent

It should be noted that at the same $|\qn|$ and $\theta=\pi$ that such collision occurs, another collision takes place in the upper half plane.  The latter is between the sound pole and one of the two lowest gapped poles of the \textit{master} charge density spectrum.  Since the geometry of collision in the latter $\bar{T}\bar{J}-$crossing point is finer than what can be clearly observed in the Fig.\ref{Complex_Sound_Chennel_1_2}, we have shown it by two zoom-in plots in the Appendix.\ref{zoom_in}.  In fact, after the two mentioned collisions occur, five poles join and form a very complicated closed trajectory. They are actually the two sound poles, the two lowest poles of the diffusion spectrum and the lowest purely imaginary pole of the sound spectrum. These five poles exchange their positions cyclically when $\theta$ increases from $0$ to $2\pi$.

$\boldsymbol{(iii)\,\, 0.633\le Q \le 0.850}$: In this interval we choose to show the results associated with $Q=0.7$. See Fig.\ref{Complex_Sound_Chennel_7_10}. 
At this value of $Q$, specifically,  we  illustrate crossings associated with critical points of both sound and diffusion branches of Puiseux series. It turns out that by increasing $|\qn^2|$, the first collision of diffusion pole would be with the nearest star gapped poles (see the black dots in top middle and top right panels). This occurs at $|\qn_c^2|\approx 1.50$. Since both colliding poles are star poles, we call such crossing of trajectories the $\boldsymbol{\bar{J}\bar{J}-}$\textbf{crossing}.\
The first collision of the sound poles, however, will occur at a larger value   of momentum. As can be seen in the bottom panels of Fig.\ref{Complex_Sound_Chennel_7_10}, at $|\qn_c^2|\approx 1.95$ sound poles collide with the lowest purely imaginary pole of the \textit{master} energy density spectrum.
Since  a dot pole is colliding with another dot pole, such crossing of trajectories is a $\boldsymbol{\bar{T}\bar{T}-}$\textbf{crossing}.

As a result one concludes that at $Q=0.7$, the diffusion dispersion relation $\wn_{\text{diffusion}}(\qn^2)$ converges for $|\qn|<(1.50)^{1/2}$ while the convergence of $\wn_{\text{sound}}(\qn)$ is for $|\qn|<(1.95)^{1/2}$.

Before ending this subsection let us give two comments regarding our results. \textit{Firstly},
as mentioned implicitly earlier and also can be seen by the behavior of red curve in Fig.\ref{q_c_Sound_diiusion}, the critical  point of the diffusion branch, not only at $Q=0.7$ but also in the whole range of $0\le Q\le 0.85$ is of $\bar{J}\bar{J}-$crossing type. \textit{Secondly}, there is an interesting point with the lower half plane $\bar{T}\bar{J}-$crossing in the interval (ii).  In fact in the latter point, the sound pole collides with the other hydro pole, namely the diffusion pole. This is quite different from the other collision points in the sense that it implies convergence of the derivative expansion for $ 0.386\le Q \le 0.610$ is solely determined by the hydrodynamic poles. This is a novel aspect of level-crossing specific to the systems at finite chemical potential. In previous studies at vanishing $\mu$, level-crossing was found exclusively as the result of interplay between hydrodynamic and non-hydrodynamic poles. 
 
 \subsection{Spin 1 channel}
 The typical spectrum of quasinormal modes in this channel was already shown in Fig.\ref{quasi_ExZ1_Q12_fig}. The spectrum includes poles associated with transverse \textit{master} momentum density (represented by dots) as well as those associated with transverse charge current (represented by stars). We argued that there would exist only one branch of Puiseux series  passing through $(0,0)$, namely the dispersion relation of the shear mode.  
 
\begin{figure}
	\centering
	\includegraphics[width=0.7\textwidth]{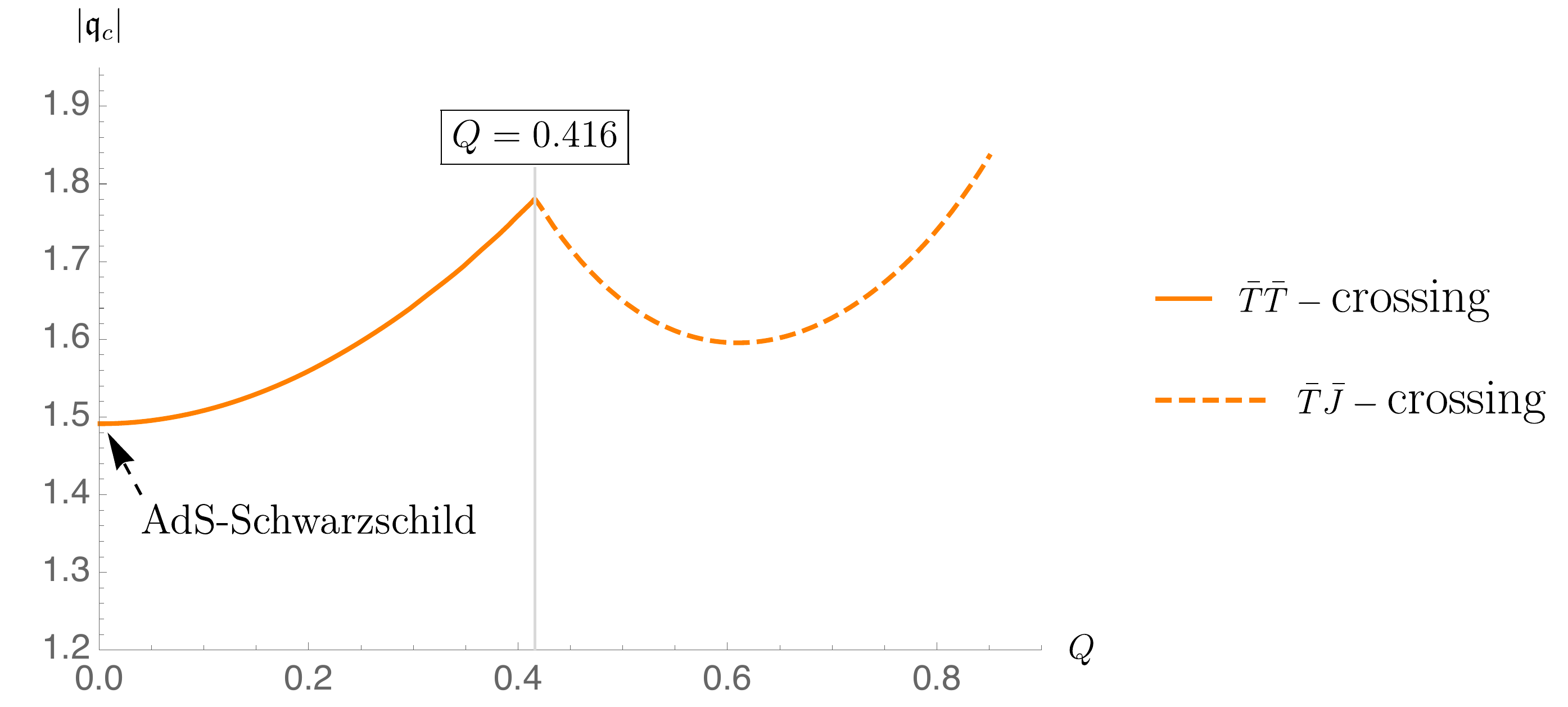}	
	\caption{Radius of convergence of the derivative expansion associated with $\wn_{\text{shear}}$ versus $Q$ in the spin 1 channel. At small values of $Q$ spectrum of transverse \textit{master} current is fully decoupled. We find that when $Q$ is lower than $0.416$, a $\bar{T}\bar{T}-$crossing determines the convergence radius. As $Q$ exceeds the latter value, before than $\bar{T}\bar{T}-$crossing occurs, the shear pole collides with a gapped pole of $\bar{J}^{L}$ spectrum. Thus in this interval  a $\bar{T}\bar{J}-$crossing determines the convergence radius of $\wn_{\text{shear}}$. The intersection point with the vertical axis, related to the neutral fluid case, was found in \cite{Grozdanov:2019kge}.}
	\label{converegence_shear}
\end{figure}

In this subsection we investigate how the radius of convergence of the derivative expansion for shear mode changes with $Q$. 
To this end, we numerically find  the spin 1 spectrum of quasinormal modes at complex momenta.
The result has been given in Fig.\ref{converegence_shear}\footnote{In the previous version of the paper,  we had missed to put the results associated with the whole range of $Q$ advertised in the Introduction. In fact we have extended the domain of $Q$  in Fig.\ref{converegence_shear}  from $0-0.5$ in the previous version  to $0-0.85$ in the current one.}. It is clear that within the range $0\le Q\le 0.85$, two different types of collisions correspond to the convergence radius of $\wn_{\text{shear}}(\qn^2)$. Thus we study the collision  of complexified quasinormal modes in two intervals:
$(i)\, 0\le Q \le 0.418$ and $(ii) \,0.418 \le Q \le 0.850$.

\begin{figure}[tb]
	\centering
	\includegraphics[width=0.35\textwidth]{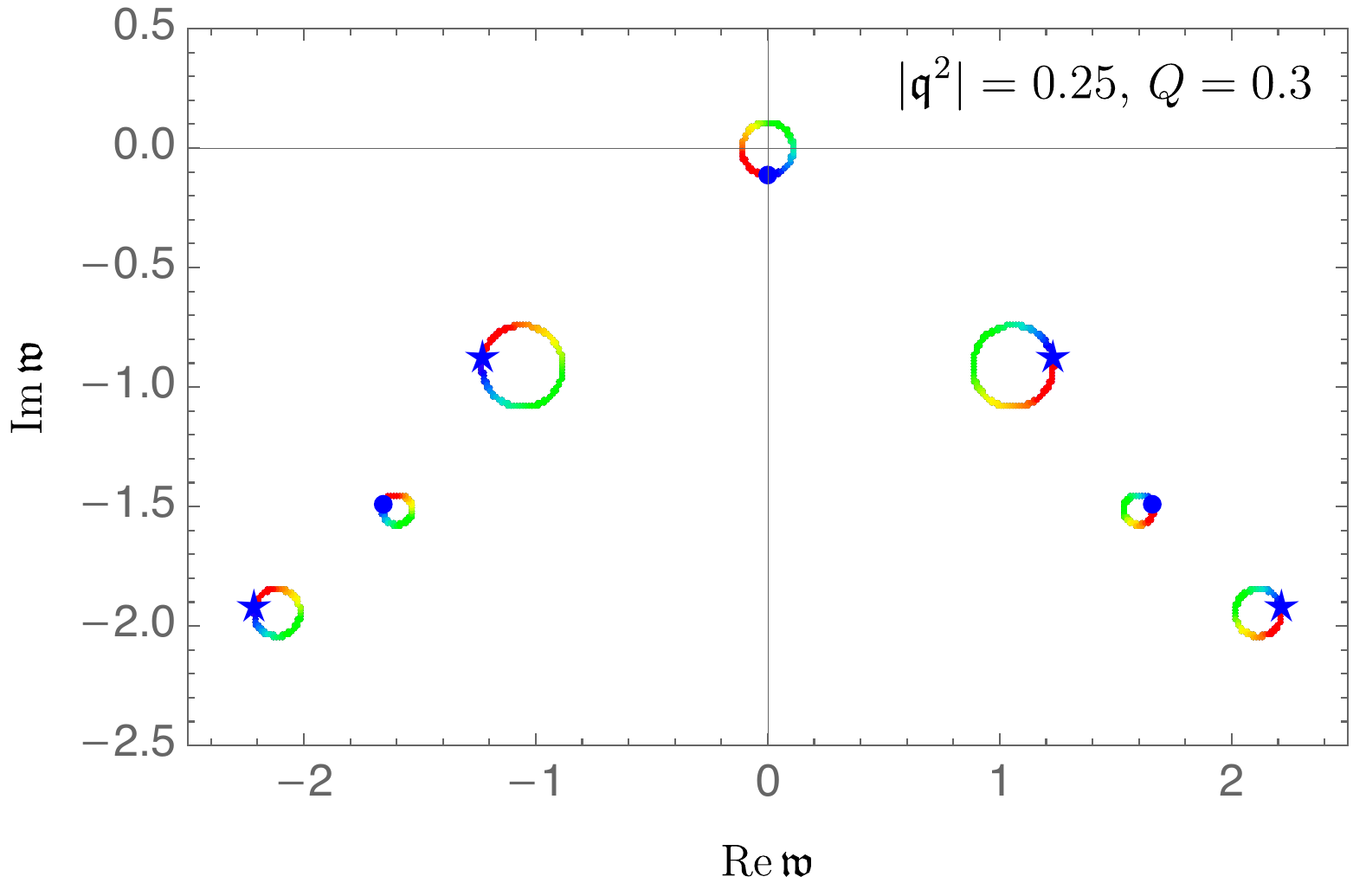}\includegraphics[width=0.35\textwidth]{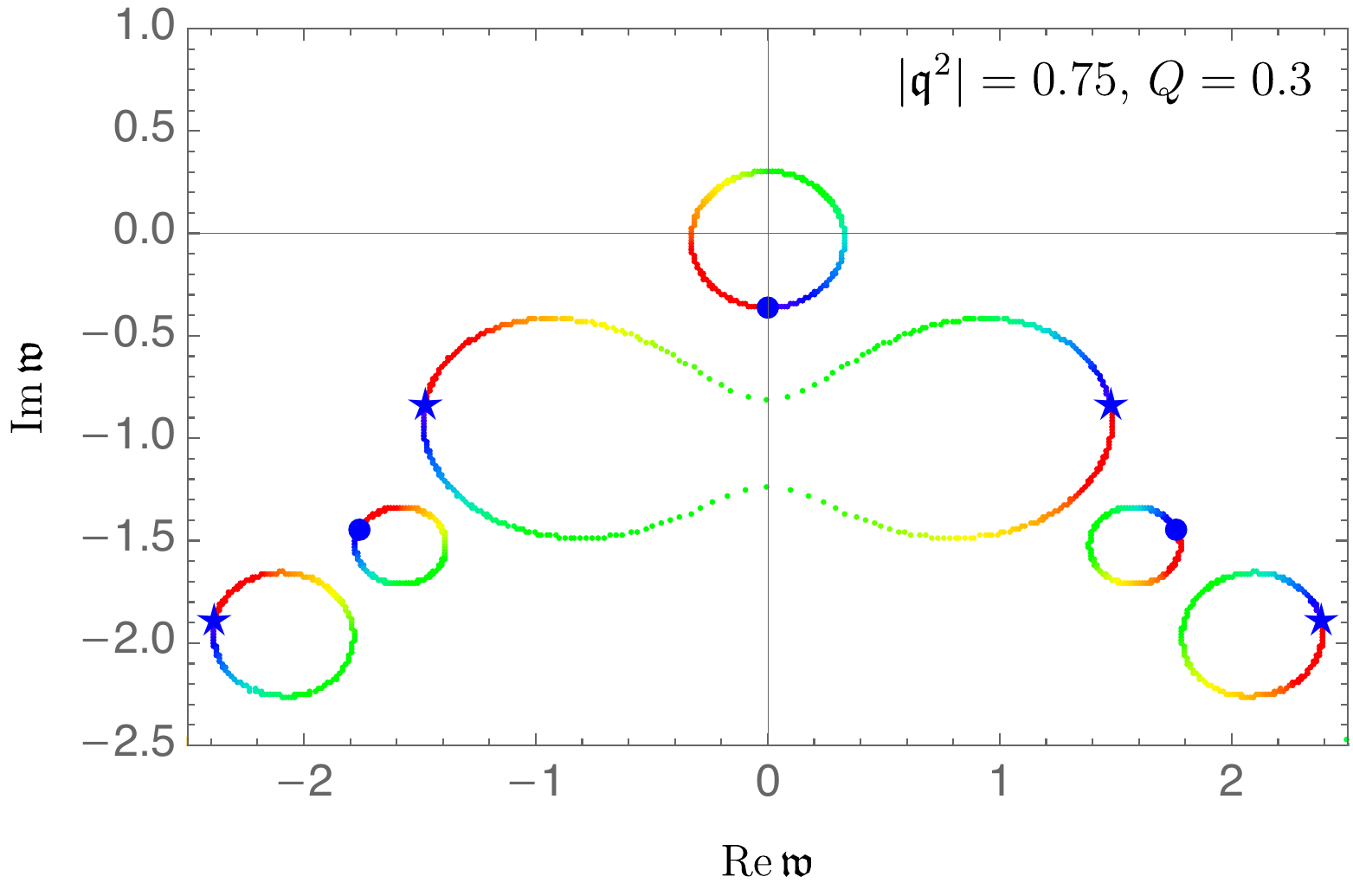}\includegraphics[width=0.35\textwidth]{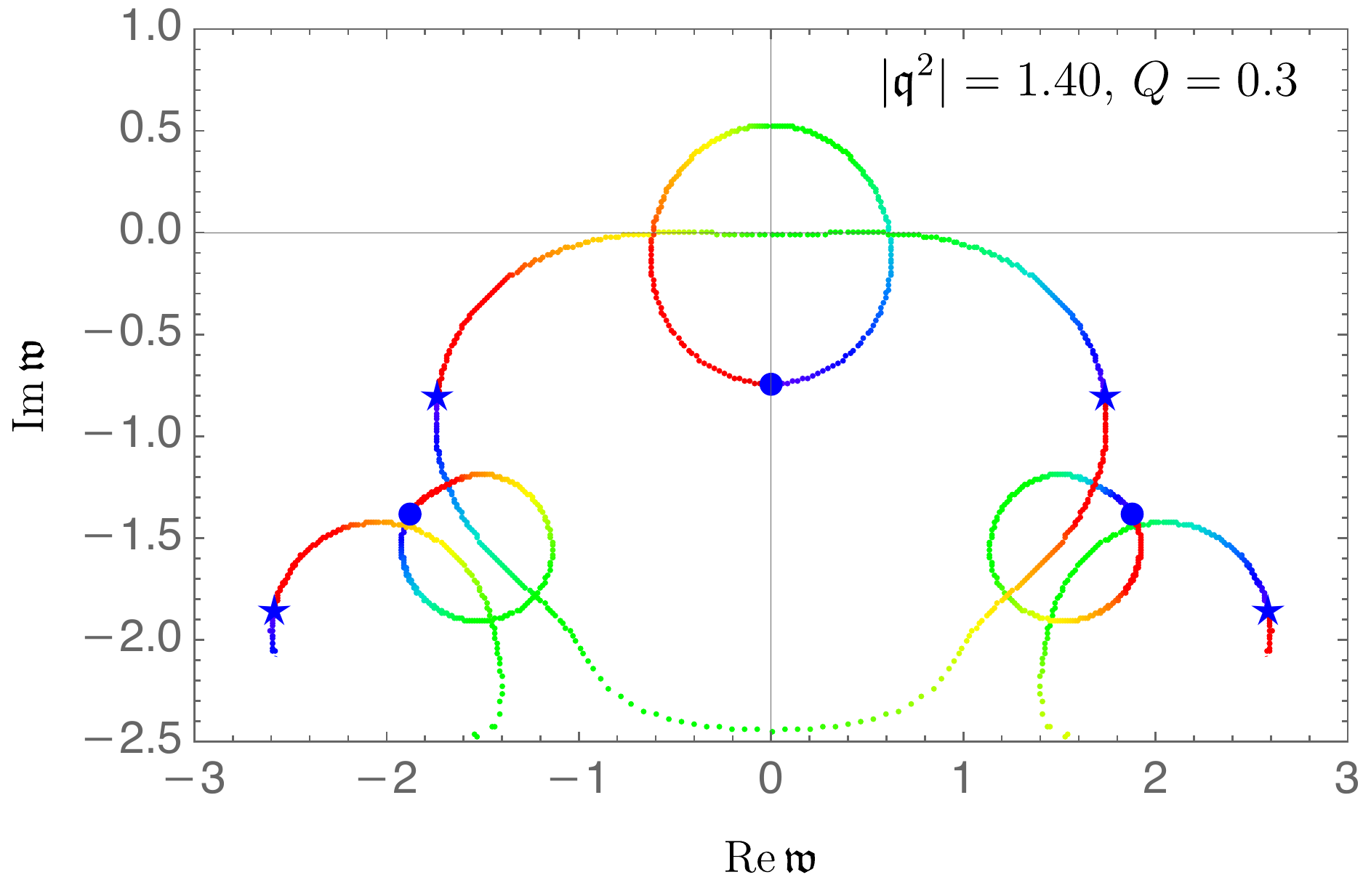}
\includegraphics[width=0.35\textwidth]{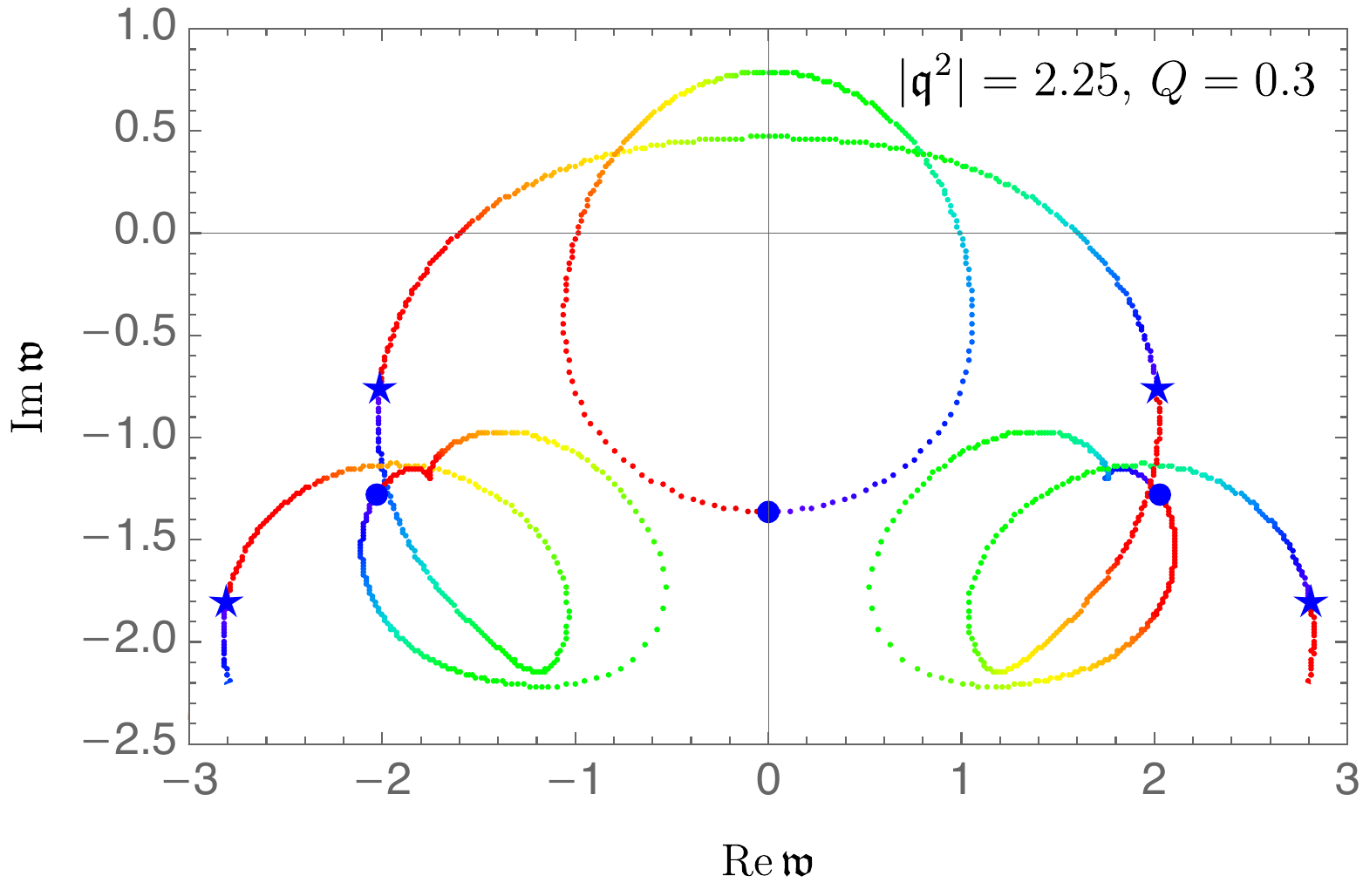}\includegraphics[width=0.35\textwidth]{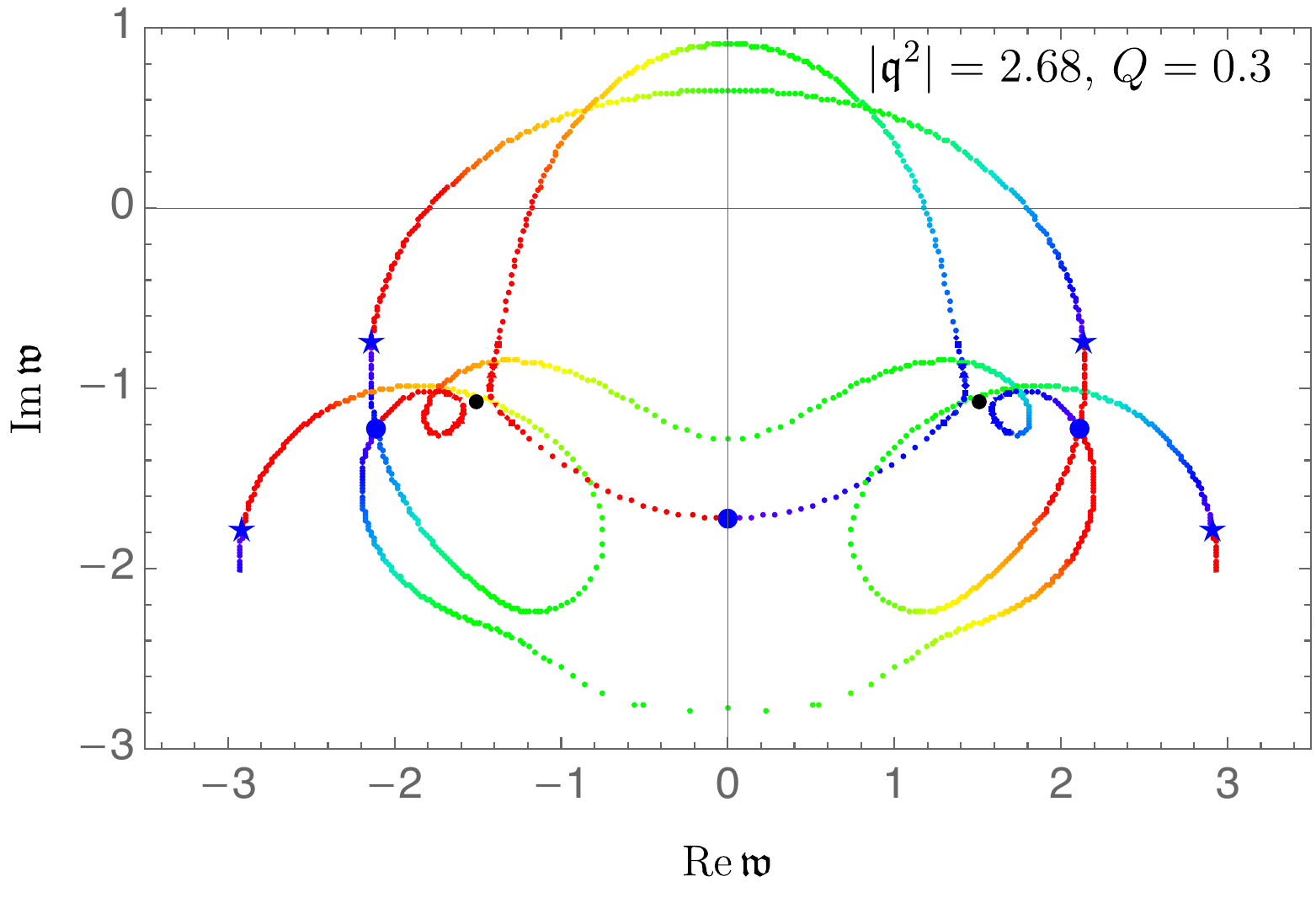}\includegraphics[width=0.35\textwidth]{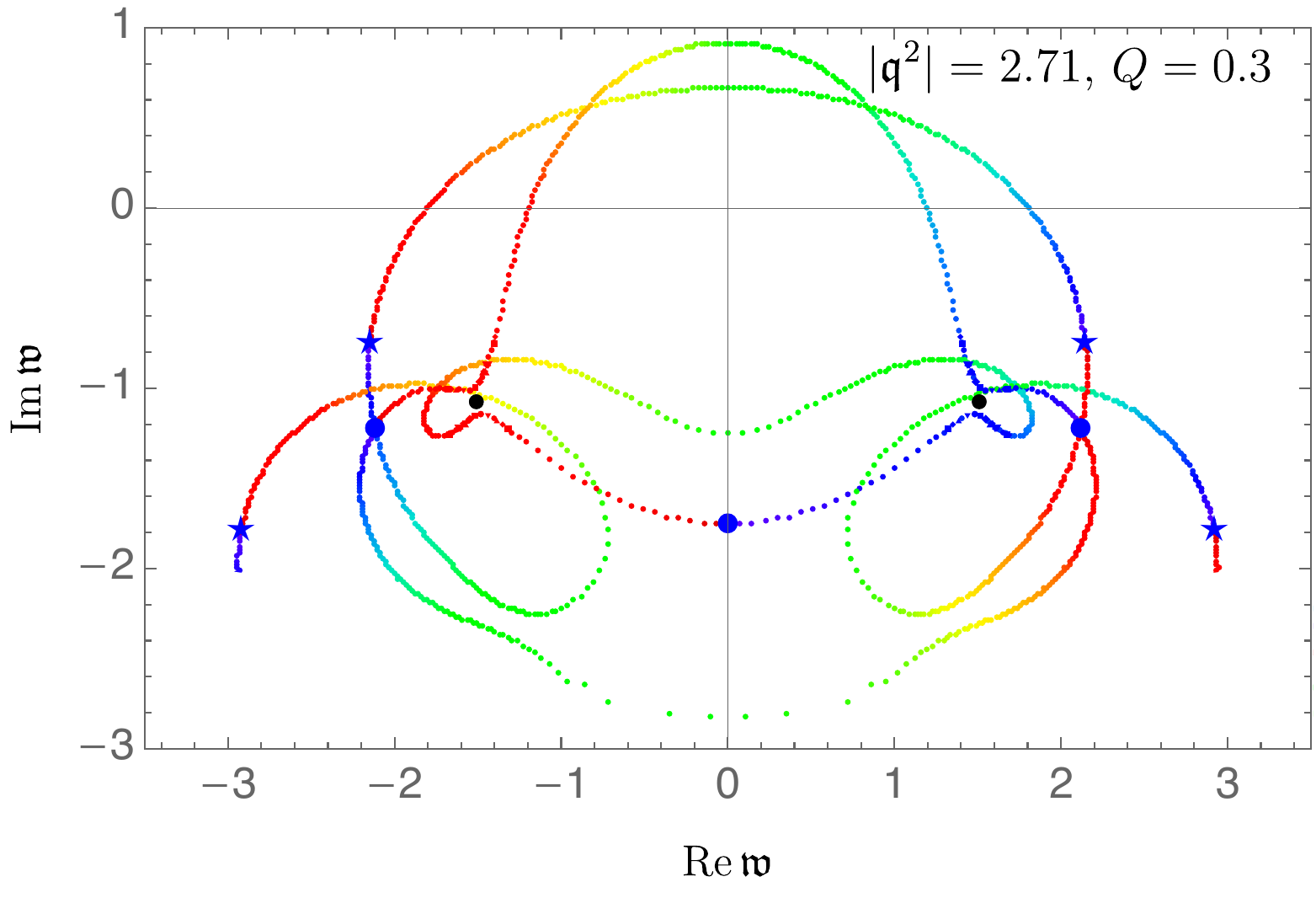}	\caption{Poles of the retarded two-point function in the spin 1 channel at $Q=0.3$, in the complex $\wn-$plane, at various values of	the complexified momentum $\qn^2=|\qn^2|e^{i \theta}$.  Large dots and large stars correspond to the poles with  purely real momentum (i.e. at $\theta=0$). As $\theta$ increases from $0$ to $2\pi$, each pole moves counter-clockwise	following the trajectory whose color changes continuously from blue to red. At $|\qn^2|=0.25$ (top left panel) all poles follow closed orbits. By increasing $|\qn^2|$, firstly the two lowest star gapped poles collide. After the collision, for instance at $|\qn^2|=0.75$ (top middle panel) their orbits are no longer closed; they exchange their positions cyclically as  the phase $\theta$ increases from $0$ to $2\pi$.  At $|\qn^2|=1.4$ (top right panel), the orbits of dot poles are still closed. By further increasing $|\qn^2|$, the trajectory of shear pole comes close to those of nearest dot gapped poles; at $|\qn^2|=2.69$, they nearly collide at the positions marked by black dots (bottom middle panel). Just after the collision, at $|\qn^2|=2.71$ (bottom right panel) the orbits of shear pole and the two lowest dot gapped pole are no longer closed; three of them exchange their positions cyclically  as $\theta$ increases from $0$ to $2\pi$.
 This is the manifestation of $\bar{T}\bar{T}-$crossing-crossing.}
	\label{Complex_Shear_Chennel_3_10}
\end{figure}
\par\bigskip 
\noindent

$\boldsymbol{(i)\,\, 0\le Q \le 0.418}$: 
In this interval, we have  shown the situation of complexified poles at several values of $|\qn^2|$ associated with $Q=0.3$, in Fig.\ref{Complex_Shear_Chennel_3_10}. 
As it is seen, by increasing $|\qn^2|$, firstly the lowest star gapped poles collide. It does actually occur at some critical value of $0.25<|\qn_c^2|<0.75$. By further increasing (the top right panel and bottom left panel), trajectories of the lowest dot gapped poles, which are still closed, come close to that of the diffusion pole, namely the single dot pole on the imaginary axis. The first critical point of the spectral curve on the shear branch of Puiseux series  turns out to be at $|\qn_{c}^2|\sim2.70$. We have demonstrated the situation of poles just before and just after the collision in the bottom middle and bottom right panels, respectively. We then conclude that at $Q=0.3$,  the dispersion relation of shear mode converges for  $|\qn|<|\qn_{c}|\approx(2.70)^{1/2}\approx 1.64$.
 Since colliding poles both belong to spectrum of transverse \textit{master} momentum density, we call such type of crossing point as the $\bar{T}\bar{T}-$crossing.  

\begin{figure}[tb]
	\centering
	\includegraphics[width=0.35\textwidth]{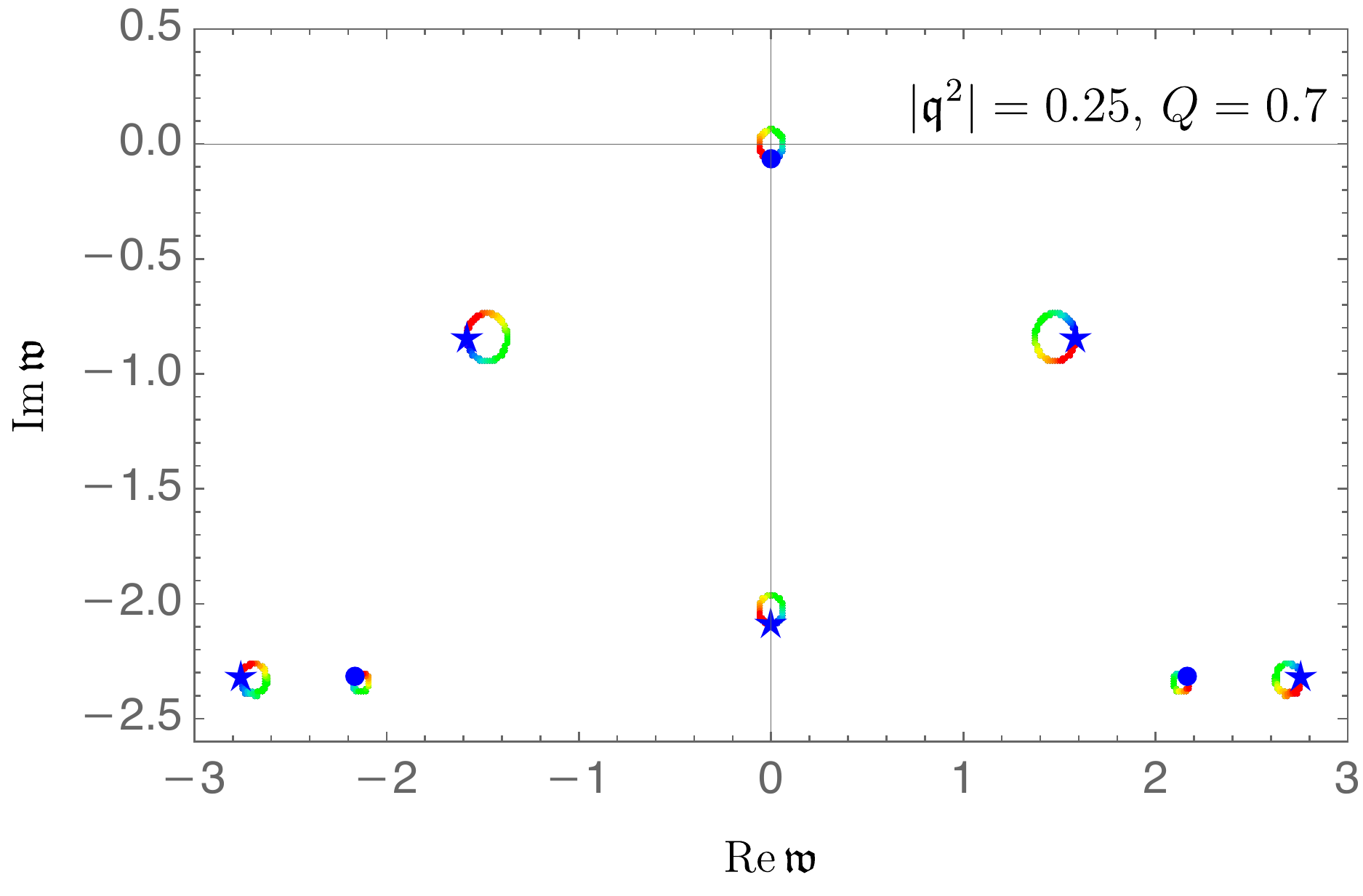}\includegraphics[width=0.35\textwidth]{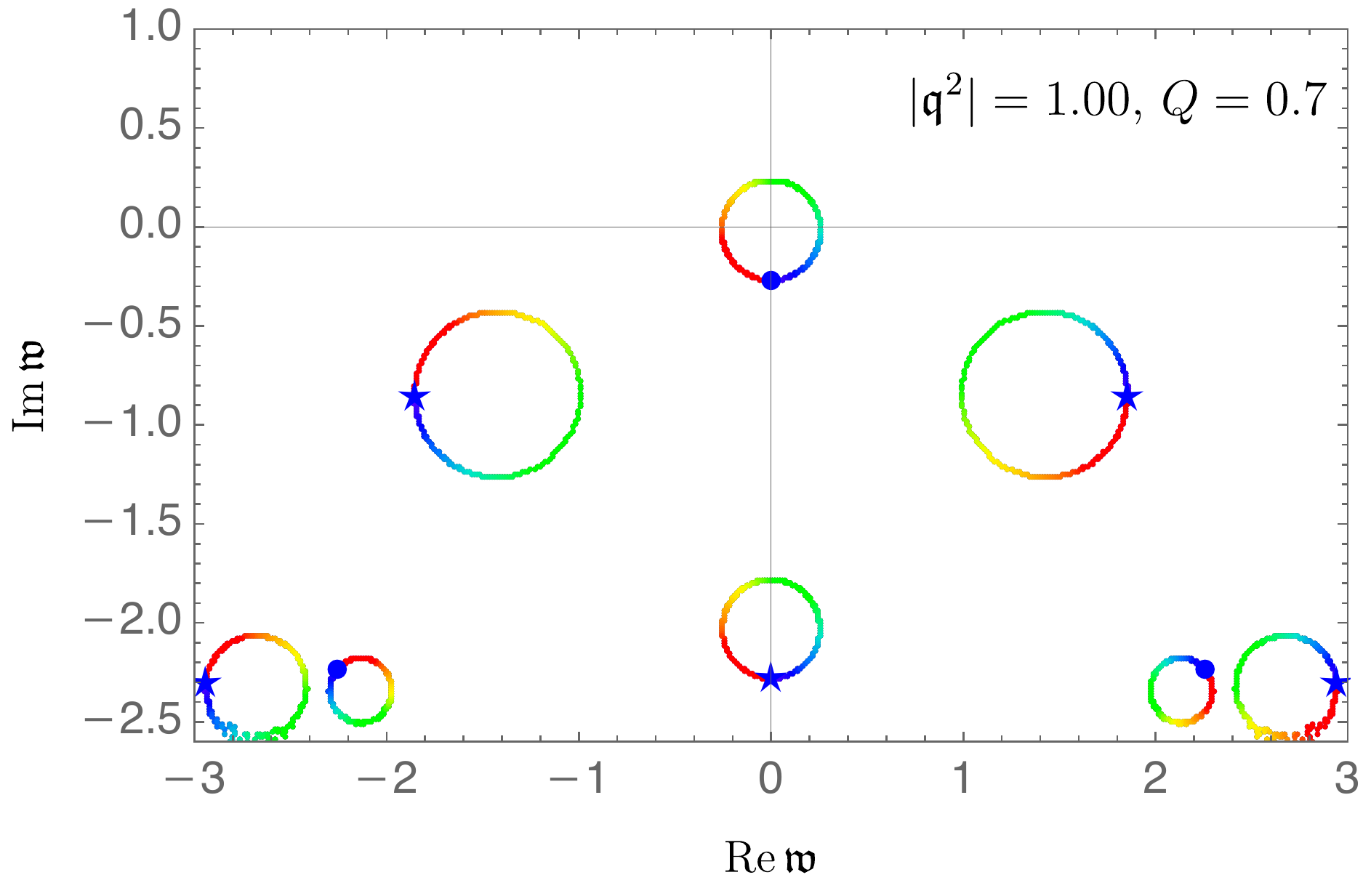}\includegraphics[width=0.35\textwidth]{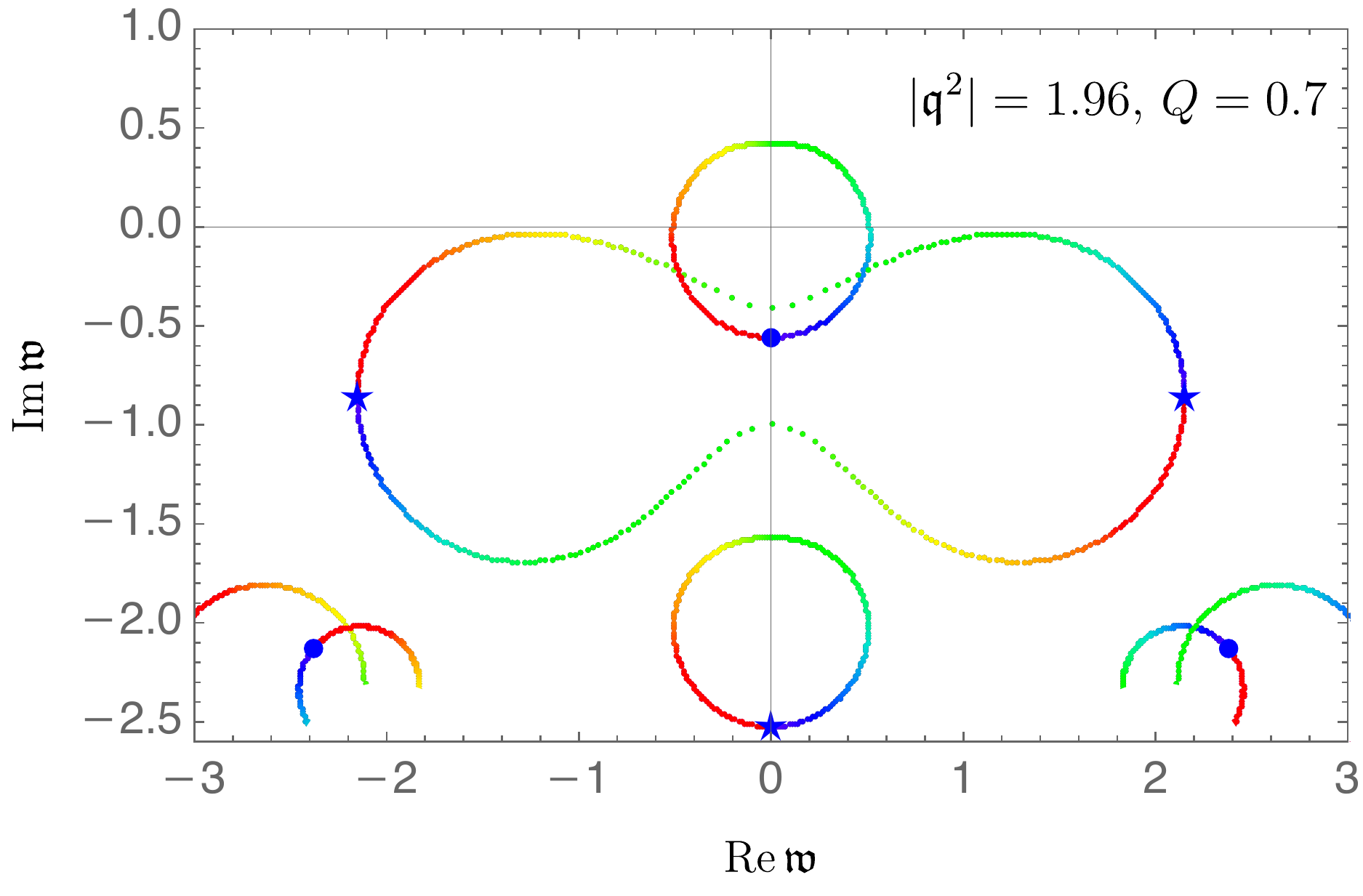}
	\includegraphics[width=0.35\textwidth]{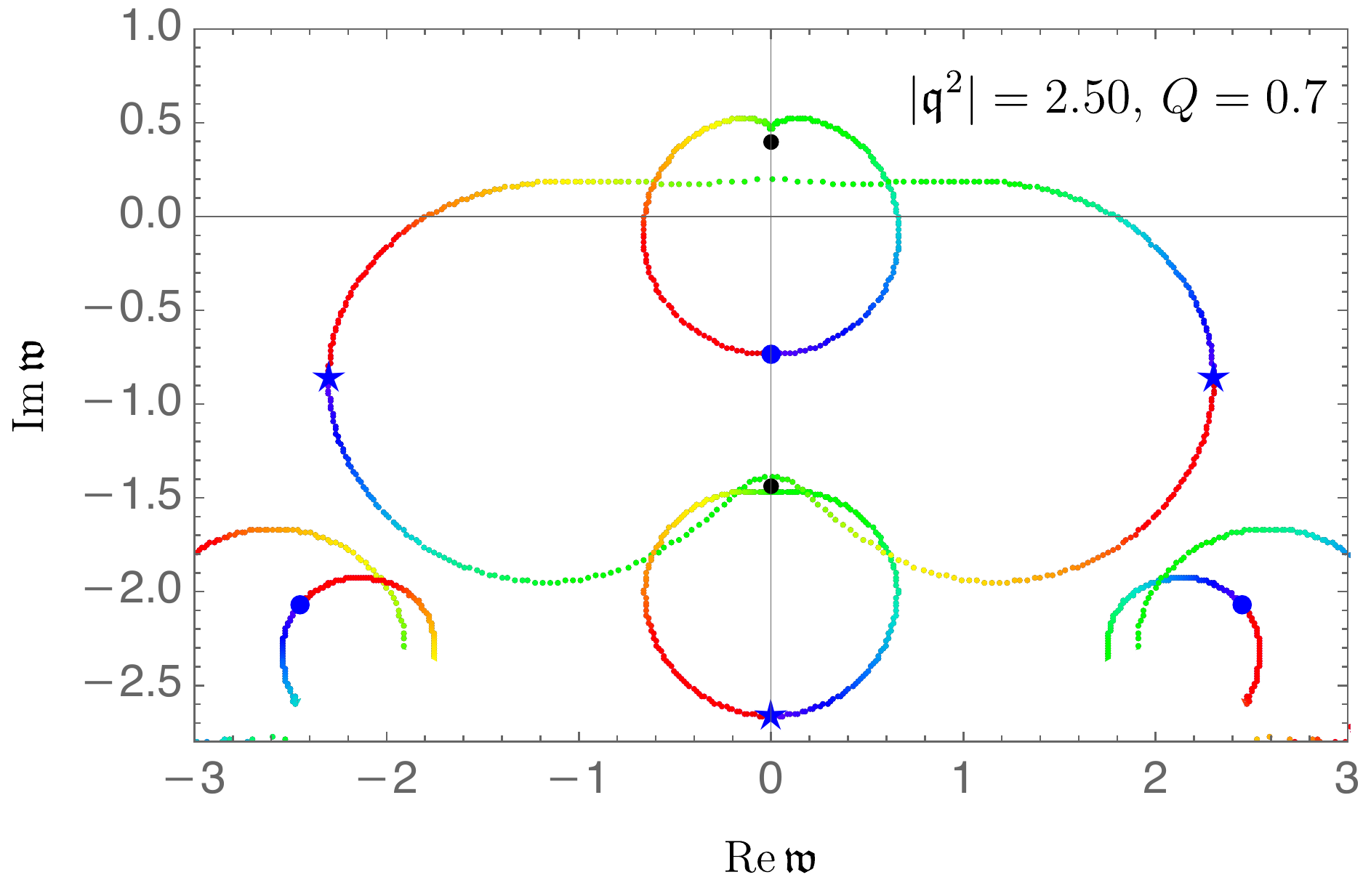}\includegraphics[width=0.35\textwidth]{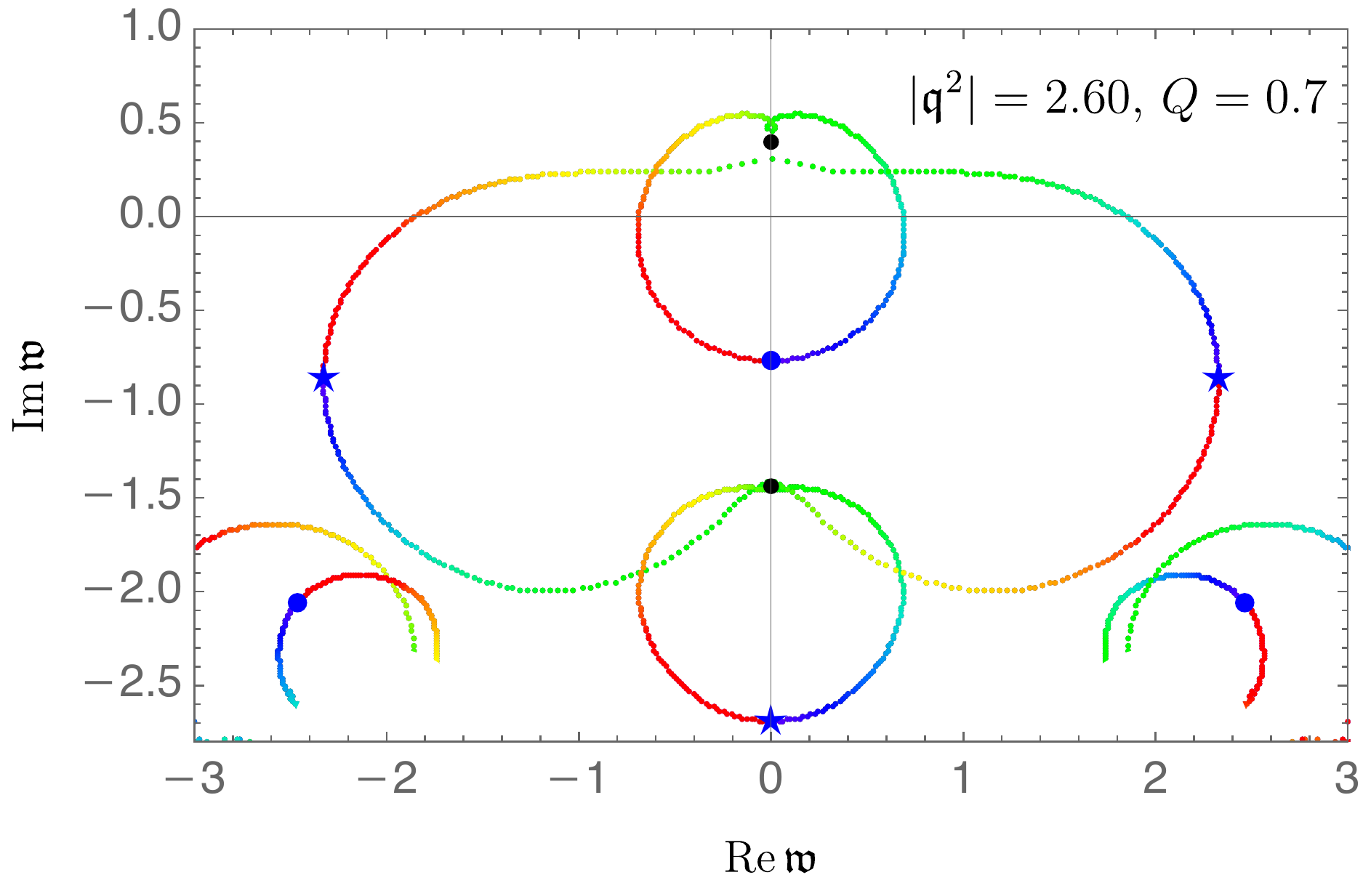}\includegraphics[width=0.35\textwidth]{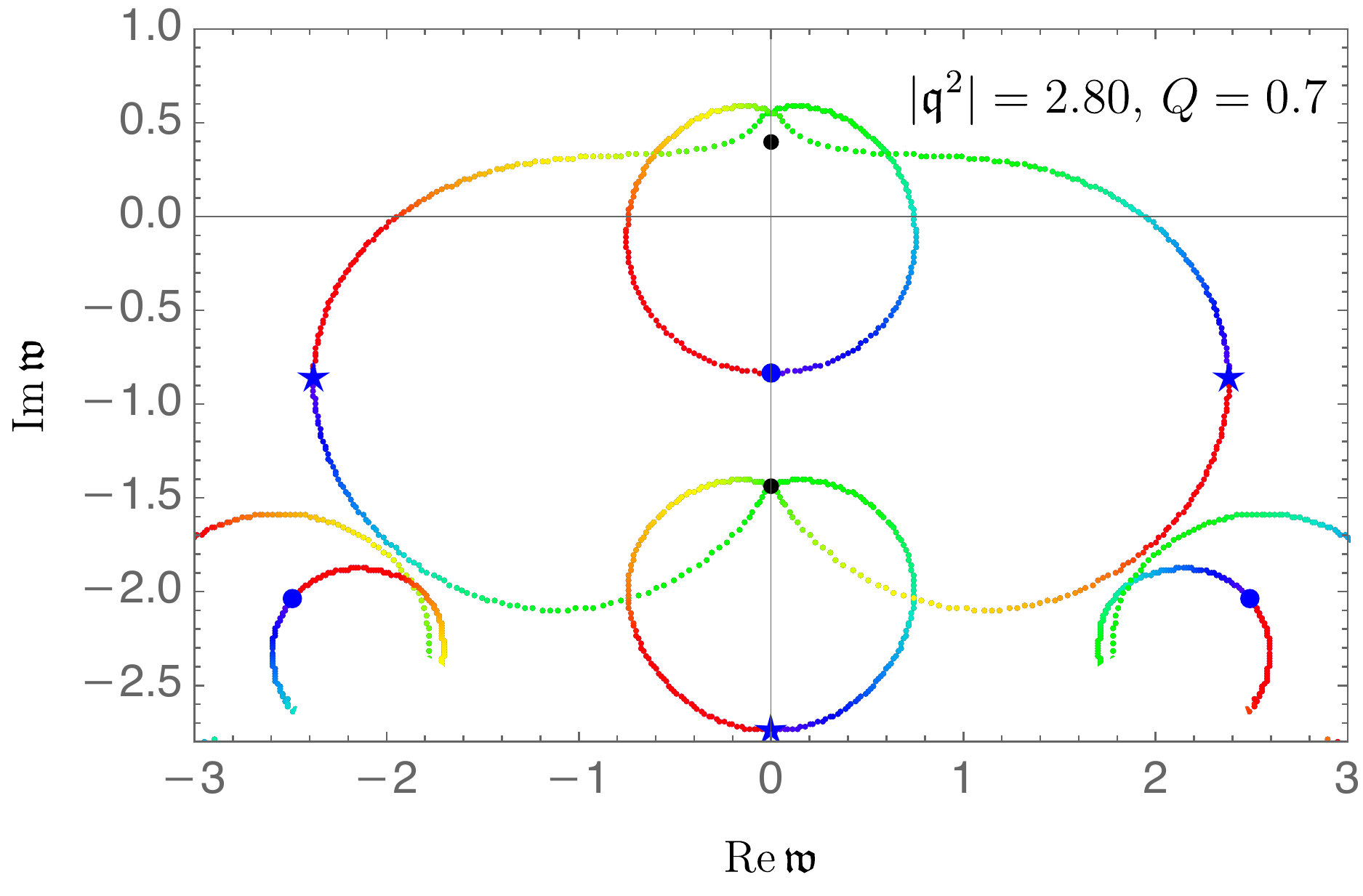}	\caption{Poles of the retarded two-point function in the spin 1 channel at $Q=0.7$, in the complex $\wn-$plane, at various values of	the complexified momentum $\qn^2=|\qn^2|e^{i \theta}$.  Large dots and large stars correspond to the poles with  purely real momentum (i.e. at $\theta=0$). As $\theta$ increases from $0$ to $2\pi$, each pole moves counter-clockwise	following the trajectory whose color changes continuously from blue to red. At $|\qn^2|=0.25$ (top left panel) all poles follow closed orbits. By increasing $|\qn^2|$, firstly the two lowest star gapped poles collide. After the collision, for instance at $|\qn^2|=1.96$ (top right panel) their orbits are no longer closed; they exchange their positions cyclically as  the phase $\theta$ increases from $0$ to $2\pi$.  At $|\qn^2|=2.50$ (bottom left panel), the orbits of dot poles are still closed. By further increasing $|\qn^2|$, the trajectory of shear pole comes close to those of the nearest dot gapped poles; at $|\qn^2|=2.60$, they nearly collide at the positions marked by black dots (bottom middle panel). Just after the collision, at $|\qn^2|=2.80$ (bottom right panel) the orbits of shear pole and the star pole lying on the imaginary axis are no longer closed; they join to the lowest star poles and form a very complicated orbit. Four of them exchange their positions cyclically  as $\theta$ increases from $0$ to $2\pi$. However, since the shear mode collides with a gapped pole of $\bar{J}^T$ spectrum, the transition between the last two plots is the manifestation of  $\bar{T}\bar{J}-$crossing.}
	\label{Complex_Shear_Chennel_7_10}
\end{figure}
\par\bigskip 
\noindent
 
$\boldsymbol{(ii)\,\, 0.418\le Q \le 0.850}$: In this interval we choose to show the results associated with $Q=0.7$. See Fig.\ref{Complex_Shear_Chennel_7_10}. 
 It turns out that by increasing $|\qn^2|$, the first collision of the shear pole would be with the nearest star gapped poles (see the black dots in bottom panels). This occurs at $|\qn_c^2|\approx 2.64$. Since one dot pole is colliding with a star one, it is actually a  $\boldsymbol{\bar{T}\bar{J}-}$\textbf{crossing}.
It should be noted that at the same $|\qn|$ and $\theta=\pi$ that such collision occurs, another collision takes place in the lower half plane.

Let us recall one of the reasons for which we started the present work. As mentioned in the Introduction, ref. \cite{Withers:2018srf} finds the radius of convergence of the shear mode at a fixed value of $\mu/T$ in a $2+1$ dimensional system. Our result shows that at least within the range of $\mu/T$ studied in the current paper, there is an interval within which the convergence radius of the shear mode decreases with $Q$ (and $\mu/T$), similar to the analytic formula found for AdS$_4$ RN model in ref. \cite{Withers:2018srf}.  The idea that the analytic formula of ref. \cite{Withers:2018srf} would probably get change by taking into account the behavior of other gapped poles,  with $Q$ decreasing, was firstly predicted in \cite{Grozdanov:2019uhi}. Our results given in Fig.\ref{converegence_shear} confirm this idea in four dimensions. Additionally as has been pointed out to in the figure, at $\mu=0$ we find a finite radius of convergence in complete agreement with ref. \cite{Grozdanov:2019kge}.

\subsection{Comment about the radius of convergence in the spin 0 channel and review of the results}
As it was shown in Fig.\ref{q_c_Sound_diiusion}, at each value of $Q$ there are two radii of convergence associated with spin 0 hydro modes. One is related to the dispersion of sound modes and the other to that of the diffusion mode. On the other hand since the corresponding boundary operators couple, each of the correlators contain all the above-mentioned modes. Then one may ask  what would essentially the radius of convergence be; would that be uniquely given by the smaller one between the two radii?

The important point is, one should not look for a unique radius of convergence at each specific value of $Q$. To understand why this is so, one can explore the small $\qn$ limit of the spectral function:
\begin{equation} \label{}
\boldsymbol{F}_0(\qn^2\rightarrow 0, \wn, Q)=0.
\end{equation}
The left side of this equation, namely the spectral curve, is actually  the common denominator of all coupled correlators.
From \eqref{hydro_mode_holog_spin_0} it is clear that the above equation has the following solutions:
\begin{equation} \label{sol_leading}
\wn_{\text{sound}}^{\pm}=\pm\frac{1}{\sqrt{3}}\qn+\,\mathcal{O}(\qn^2),\,\,\,\,\,\wn_{\text{diffusion}}=-i\qn^2+\mathcal{O}(\qn^4).
\end{equation}
These are actually different branches of the spectral curve in the vicinity of the origin and passing through it. Let us emphasize that their corresponding slops at the origin are different, too: $-1/\sqrt{3}$, $+1/\sqrt{3}$ and $0$. 

At larger values of $\qn$, the solutions \eqref{sol_leading} can be improved by taking into account more number of terms. The latter is the subject of Puiseux series which was given by  \eqref{Puiseux}. We immediately find
\begin{equation} \label{}
a_1=\,1/\sqrt{3},\,\,\,\,c_1=1.
\end{equation}
Now one should decide whether to study $\wn^{\pm}_{\text{sound}}$ or  $\wn_{\text{diffusion}}$. If the former is desired,  it is needed to substitute
\begin{equation} \label{}
\wn^{\pm}_{\text{sound}}=\,\pm \frac{1}{\sqrt{3}} \qn+\, i a_2\qn^2\pm\,a_3\qn^2+\,\cdots
\end{equation}
in the spectral curve equation and iteratively find the coefficients. Then the convergence radius of the hydrodynamic derivative expansion associated with the sound mode is given by
\begin{equation} \label{}
|\qn_c|_{\text{sound}}=\left(\lim_{n\rightarrow\infty}\bigg|\frac{a_{n+1}}{a_n}\bigg|\right)^{-1}.
\end{equation}
For the case of diffusion pole, one substitutes 
\begin{equation} \label{}
\wn_{\text{diffusion}}=\,-i\, \qn^2-\,i c_2 \qn^4-\,ic_3\qn^6+\,\cdots
\end{equation}
in the spectral curve equation and by finding the coefficients, convergence radius of the hydrodynamic derivative expansion associated with this mode is then determined by
\begin{equation} \label{}
|\qn_c|_{\text{diffusion}}=\left(\lim_{n\rightarrow\infty}\bigg|\frac{c_{n+1}}{c_n}\bigg|\right)^{-1/2}.
\end{equation}
\textit{The above discussion makes it clear that $|\qn_c|_{\text{sound}}$ and $|\qn_c|_{\text{diffusion}}$ do not have anything to do with each other. }In another words, the spectral curve is not required to have  a unique radius of convergence at a specific $Q$. It has actually three branches passing through the origin (see Fig.\ref{imaginary_dispersion_Spin_0} as illustration of this point at complex momenta). \textit{Depending on the slop by which we start to get away from the origin, i.e. $-1/\sqrt{3}$, $1/\sqrt{3}$ or $0$, we will probe different branches of Puiseux series given in \eqref{Puiseux}}. Correspondingly, the radius of convergence on the first two branches (sounds) will be different from that of the third one (diffusion).

Before ending this subsection we briefly review our results concerning the spin 0 and 1 channels in  table.\ref{Table}. Let us also denote that we did not study spin 2 channel in this section. The reason is that its corresponding spectral curve does not have any branch passing through the origin (see Fig.\ref{quasi_Exy}). The latter is equivalent to say that there is no any spin 2 hydrodynamic mode. 
\begin{table}[!htb]
	\label{table one}
	\begin{center}
		\begin{tabular}{|c|c|c|c|c|}
			\hline
			\hline
			Spin&	Type of mode & Range of $Q$ & Type of level-crossing & Type of colliding poles \\
			\hline
			\hline
			\multirow{8}{*}{0}
			&		&&&\\
			&		& $0\le Q \le 0.386$ & $\bar{T}\bar{T}$  & hydro with non-hydro\\ 
			&		&&&\\
			&	Sound    & $0.386\le Q \le 0.633$ &  $\bar{T}\bar{J}$ & hydro with hydro\\
			&		&&&\\
			
			&		&  $0.633\le Q \le 0.850$ & $\bar{T}\bar{T}$ &hydro with non-hydro\\
			&		&&&\\
			\hhline{~----}
			&	&  & & \\ 
			
			&	diffusion     & $0\le Q\le 0.850$ &  $\bar{J}\bar{J}$ &hydro with non-hydro \\
			
			&	&   &  &  \\
			\hline
			\hline
			&		&  & & \\ 
			
			&	      & $0\le Q\le 0.416$ &  $\bar{T}\bar{T}$ & hydro with non-hydro\\
			1	&	Shear	&   & &  \\
			&	      & $0.416\le Q\le 0.850$ &  $\bar{T}\bar{J}$ & hydro with non-hydro\\
			&		&   & &  \\
			\hline
			\hline
		\end{tabular}
	\end{center}
	\caption{We have classified all different types of level-crossing in terms of spin, the value of $Q$ and type of colliding poles. For example, the second row of the table is saying that sound mode is an eigen frequency for the spectral curve of spin 0 fluctuations. When $0.386\le Q \le 0.633$, the derivative expansion associated with dispersion relation of sound converges; the radius of convergence is identified with the collision of a hydrodynamic mode, actually the sound pole itself, belonging to the spectrum of \textit{master} energy density $\bar{T}^L$, and a non-hydrodynamic mode, which is actually gapped and comes from the  spectrum of $\bar{J}^L$. }
	\label{Table}
\end{table}

\section{Quasinormal modes and the chaos point}
\label{chaos}
Earlier than establishing the relation of quasinormal modes at complex momenta with convergence of the derivative expansion, they were found to be related with quantum chaos.
The first observation of such relation turns to ref. \cite{Grozdanov:2017ajz}. Before recalling the main idea of the latter reference, let us recall that  the exponential decrease of the out-of-time-order-correlator (OTOC) 
\begin{equation} \label{OTOC_0}
\langle \,V(t,\vec{x})W(0,0)V(t,\vec{x})W(0,0)\,\rangle_{\beta}=1-\frac{1}{N} \,e^{\lambda\left( (t-t_*)-\frac{x}{v_B}\right)}+\cdots
\end{equation}
 in a thermal large-$N$ system is the manifestation of quantum chaos \cite{LArkin, Shenker:2013pqb,kitaev:2014bwb,Roberts:2014isa,Shenker:2014cwa,Maldacena:2015waa,Davison:2016ngz,Kukuljan:2017xag,Ling:2016ibq,Blake:2017qgd}.  Here $V$ and $W$ are  two generic few-body operators,  $\lambda$ is the quantum Lyapunov exponent and $t_*$ is scrambling time. $v_{B}$ is the butterfly velocity.
 Equation \eqref{OTOC_0} can be also written in the form of a plane wave
\begin{equation} \label{OTOC}
\langle \,V(t,\vec{x})W(0,0)V(t,\vec{x})W(0,0)\,\rangle_{\beta}=1-\frac{1}{N} \,e^{-i \omega (t-t_*)+ i q |\vec{x}|}+\cdots
\end{equation}
with purely imaginary values  for both momentum and frequency
\begin{equation} \label{chaos_p}
\omega_{ch}= i \lambda,\,\,\,\,\,\,\,\,\,\,\,\,\,\,\,q_{ch}=i \frac{\lambda}{v_{B}}.
\end{equation}
The point $(\omega_{ch}, q_{ch})$ is called the chaos point\footnote{Since throughout this paper  subscript $c$ is used for frequencies and momenta associated with critical points  of  spectral curves, we then unconventionally use the subscript $ch$ to label the chaos point. }.

In holography, \eqref{OTOC_0} is mapped onto the back reaction of an small amount of energy thrown towards the horizon of a two-sided black hole \cite{Shenker:2013pqb}.  The resultant deformed geometry is described by a Dray-'t  Hooft shock wave \cite{Dry}. But as has been explained in \cite{Dry2}, the same geometry can be found from linearized Einstein equations too.  In addition, shock wave solution deals with dynamics of energy-momentum in the boundary theory. Then one may conclude that the same information about quantum chaos and scrambling can be extracted from studying the linear perturbations of metric around the horizon. It was actually shown to be the case firstly in \cite{Grozdanov:2017ajz} and then in \cite{Blake:2018leo}.  The statement is that at point $(q_*, \omega_*)=(i \lambda/v_B, i \lambda)$, one specific component of Einstein equations at the horizon becomes trivial. As a result the energy density response function in the boundary theory is infinitely multivalued at exactly $(q_*, \omega_*)$. This point is called the pole-skipping point \cite{Blake:2018leo}. More precisely, this point lies on the analytically continued dispersion relation of the sound mode in the boundary theory \cite{Grozdanov:2017ajz}. The coincidence of $(q_{ch}, \omega_{ch})$ with $(q_*, \omega_*)$ can be regraded as the existence of a direct link between hydrodynamics and quantum chaos, at least in holographic systems\footnote{Pole-skipping has been also derived as a general prediction of effective field theory in maximally chaotic systems in ref. \cite{Blake:2017ris}. This phenomenon has been explicitly shown to happen in 2-dim CFT at large central charge \cite{Haehl:2018izb} and recently in higher dimensions \cite{Haehl:2019eae}. See \cite{Grozdanov:2018kkt}  for considering the stringy corrections. }. 

In the following two subsections we investigate on the relation between hydrodynamics and quantum chaos in our holographic model.

\subsection{Chaos point from shock wave computations}
In this subsection we analytically compute the chaos point in a system dual to a AdS$_5$ RN black brane. We exploit the result of ref.  \cite{Blake:2017qgd}.
In this reference, based on the shock wave propagation picture in the bulk of AdS$_5$ \cite{Shenker:2013pqb},  the butterfly velocities for an anisotropic Q-lattice have been computed. Considering the background as 
\begin{equation}\label{Blake_metric}
ds^2=\,-F(r)dt^2+\frac{dr^2}{F(r)}+h_{T}(r)\,(dx^2+dy^2)+h_{L}(r) \,dz^2,
\end{equation}
it was shown that butterfly velocities in longitudinal, $L$, and transverses, $T$, directions  are given as the following 
\begin{equation}\label{Blake_butterfly}
v_{L}=\frac{2 \pi T}{\sqrt{h_L} m}\bigg|_{r_h},\,\,\,\,\,\,\,\,v_{T}=\frac{2 \pi T}{\sqrt{h_T} m}\bigg|_{r_h},\,\,\,\,\,\,\,\text{with}\,\,\,m^2=\,\pi T\left(\frac{2h'_T h_L+h'_l h_T}{h_Th_L}\right)\bigg|_{r_h}.
\end{equation}
In our case $h_{T}(r)=h_L(r)=\,r^2$ and so $m=6\pi T/r_h$. The isotropic butterfly velocity then reads
\begin{equation}\boxed{
v_{B}=\sqrt{\frac{2-Q^2}{3}}}.
\end{equation}
This result is an exact expression in the whole range of $Q$, including $Q=0$ case corresponding to AdS$_5$ Schwarzschild  black brane \cite{Shenker:2013pqb}, as well as  $Q=\sqrt{2}$ case corresponding to the extremal AdS$_5$ RN black brane. Considering the fact that all systems with gravity dual saturate the chaos bound \cite{Maldacena:2015waa}, the chaos point in our system is found to be given by
\begin{equation}\label{chaos_point}
(\qn_{ch}, \wn_{ch})=\,\left(\frac{q_{ch}}{2\pi T}, \frac{\omega_{ch}}{2\pi T}\right)=\,\left(\pm i \,\sqrt{\frac{3}{2-Q^2}}\, ,\,i\right).
\end{equation}
In next subsections, we find pole-skipping points and discuss their relation with  \eqref{chaos_point}.
\subsection{Pole-skipping}
As discussed earlier, pole-skipping points of energy density response function can be found from a near horizon analysis of perturbations associated with bulk fields. To this end, it is convenient to work with the ingoing Eddington-Finkelstein coordinates. In this system of coordinates, bulk solutions \eqref{back_ground} take the following general form
\begin{eqnarray}
ds^2&=&2 dv dr- f(r)dv^2+ h(r)\left(dx^2+dy^2+\frac{}{}dz^2\right),\\
A&=&-\frac{\sqrt{3}q_b}{2 r^2}\left(dv-\frac{dr}{f(r)}\right),
\end{eqnarray}
where $v$ is the ingoing time coordinate. Because of the relation between energy dynamics and quantum chaos in maximally chaotic systems \cite{Grozdanov:2017ajz,Blake:2018leo},   it is natural to look for pole-skipping points in spin 0 channel of fluctuations. But as was recently shown,  the same information about chaos can be  extracted from both spin 1 and spin 2 channels too, at least in a holographic system with vanishing $\mu$  \cite{Grozdanov:2019uhi}.  Thus for completeness, we study pole-skipping in the spin 1 channel as well. 

To clarify the notation, let us mention that we take metric and gauge field perturbations as $\delta g_{\mu\nu}(r,v,\vec{x})=\delta g_{\mu\nu}(r)e^{-i\omega v+i q z }$ and $\delta A_{\mu}(r,v,\vec{x})=\delta A_{\mu}(r)e^{-i\omega v+i q  z}$. The near horizon expansions of these perturbations are given by
\begin{equation}\label{}
\delta g_{\mu\nu}(r)=\,\sum_{n=0}^{\infty}\delta g_{\mu \nu}^{(0)}\,(r-r_h)^n,\,\,\,\,\,\,\delta A_{\mu}(r)=\,\sum_{n=0}^{\infty}\delta A_{\mu}^{(0)}\,(r-r_h)^n.
\end{equation}

\subsubsection*{Spin 0 Channel}
By taking the momentum  along the third boundary direction, spin 0 perturbations in the bulk are $\delta g_{vv}$, $\delta g_{vr}$, $\delta g_{rr}$, $\delta g_{vz}$, $\delta g_{rz}$, $\delta g_{xx}+\,\delta g_{yy}$, $\delta g_{zz}$, $\delta A_{r}$, $\delta A_v$ and $\delta A_z$. Following \cite{Blake:2018leo}, we expand $E_{vv}$ component of Einstein equations about the horizon $r=r_h$; to linear order in perturbations, we arrive at
\begin{equation}\label{E_vv}
\left(q^2-\frac{3}{2}ih'(r_h)\omega\right)\delta g_{vv}^{(0)}- i\left (i \omega +\frac{1}{2}f'(r_h)\right)\left( 2q \delta g_{vz}^{(0)}+\frac{}{}\omega \delta g_{ii}^{(0)}\right)=\,0.
\end{equation}
Using metric functions given around \eqref{back_ground}, equation \eqref{E_vv} at the horizon $r_h=R$ becomes
\begin{equation}
E^{(0)}_{vv}\equiv\,\left(q^2-3 i R\,\omega\right)\delta g_{vv}^{(0)}- i \left(i \omega +\frac{}{}R(2-Q^2)\right)\left( 2q \delta g_{vz}^{(0)}+\frac{}{}\omega \delta g_{ii}^{(0)}\right)=\,0.
\end{equation}
It is clear that this equation trivially holds at 
\begin{equation}\label{pole_skipping_point}
(\qn_*, \wn_*)=\,\left(\frac{q_*}{2\pi T}, \frac{\omega_*}{2\pi T}\right)=\,\left(\pm i \,\sqrt{\frac{3}{2-Q^2}}\, ,\,i\right).
\end{equation}
Therefore, at exactly the above points the rank of matrix $E^{(0)}_{\mu \nu}$ decreases by one.  The latter is equivalent to say that line of poles of energy density response function in the boundary suddenly skips at \eqref{pole_skipping_point}. Thus the points given by \eqref{pole_skipping_point} are nothing but the pole-skipping points of energy density response function \footnote{It has been recently shown that  such behavior may also happen when studying perturbations of probe bulk fields other than metric itself \cite{Blake:2019otz}. However the resultant pole-skipping points in such cases lie totally in the lower half of $\text{Im} \wn- \text{Im} \qn$ plane. 
 	See also \cite{Natsuume:2019vcv,Wu:2019esr,Ahn:2019rnq, Li:2019bgc,Natsuume:2019sfp,Natsuume:2019xcy,Ceplak:2019ymw,Das:2019tga,Abbasi:2019rhy,Liu:2020yaf,Ahn:2020bks}.}.
The latter will be confirmed numerically in the following. 

\begin{figure}
	\centering
	\includegraphics[width=0.48\textwidth]{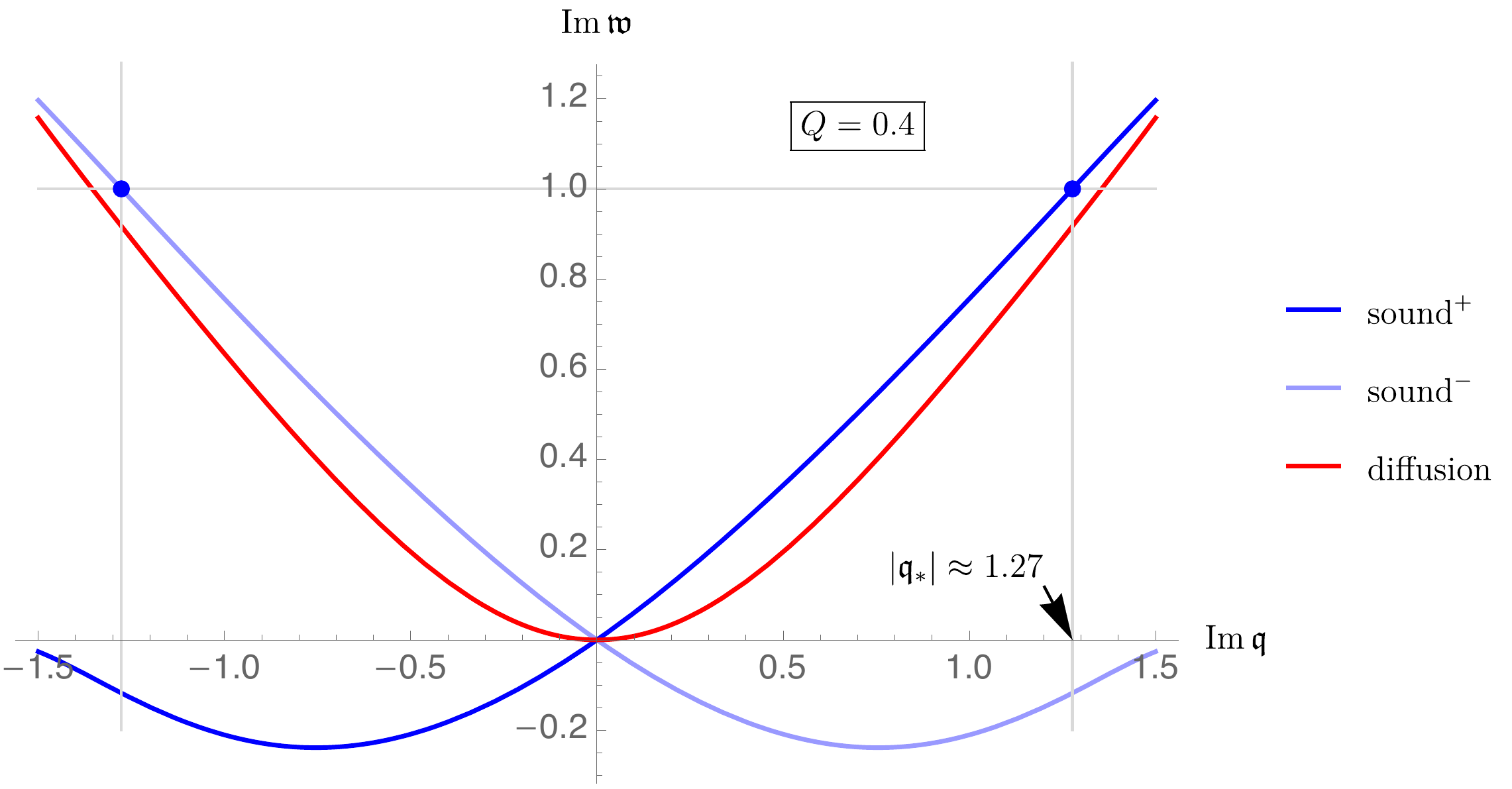}\,\,\,\,\,\,\,\,\includegraphics[width=0.42\textwidth]{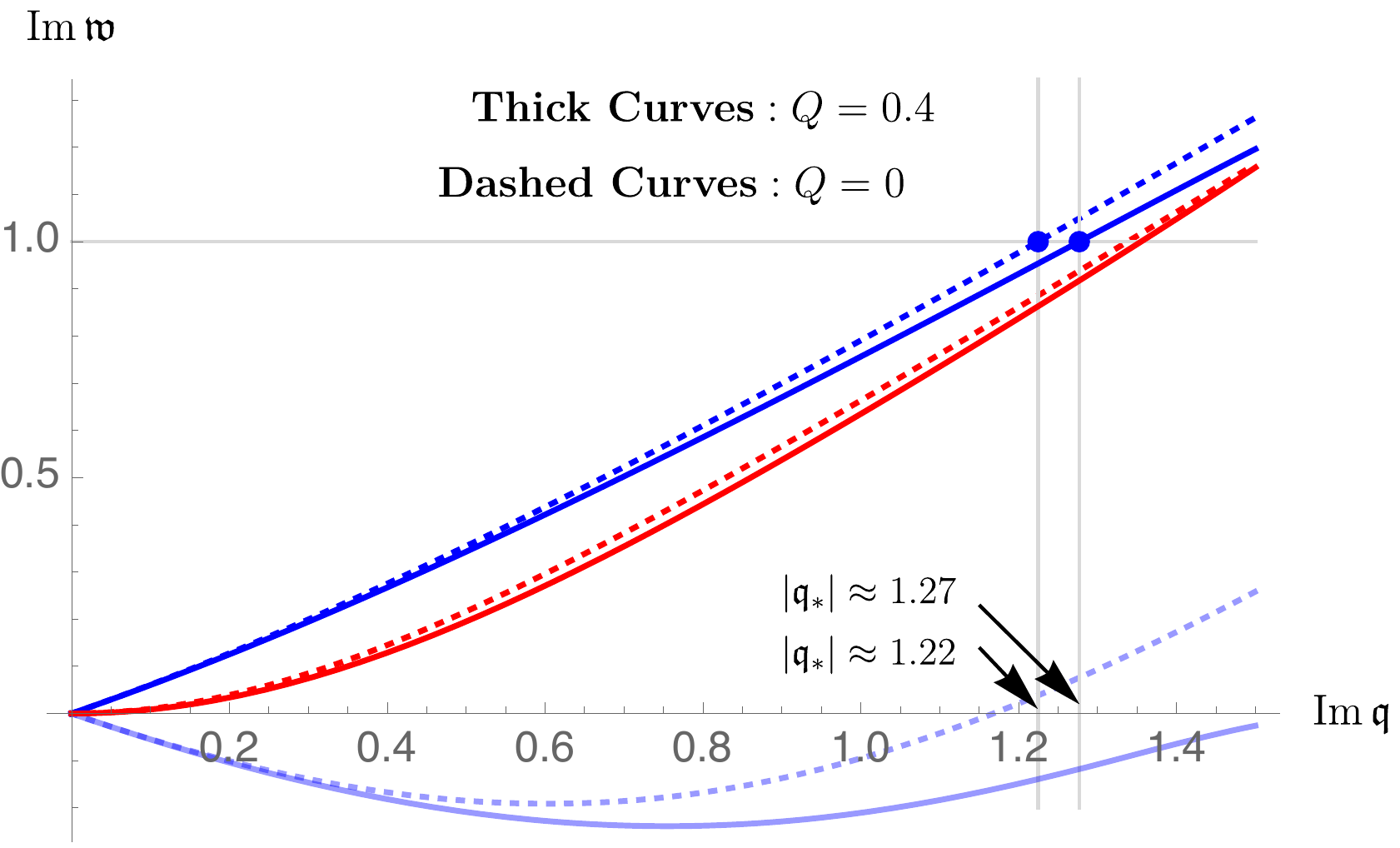}
	\caption{Left panel: Three branches of Puiseux series at purely imaginary momenta, associated with $Q=0.4$ in spin 0 channel. Obviously the points \eqref{pole_skipping_4_10} lie on the sound branches, manifesting the hydrodynamic origin of quantum chaos. Right panel: Comparison between the left panel dispersion relations with those of associated with a AdS$_5$ Schwarzschild black brane ($Q=0$).    }
	\label{imaginary_dispersion_Spin_0}
\end{figure}

More concretely,  the above-mentioned pole-skipping points may lie on a hydrodynamic dispersion relation associated with the energy density response function \cite{Blake:2017ris}. Let us recall that energy dynamics is related to the spin 0 channel. Hydrodynamic poles in this channel are the sound and the diffusion modes. On the other hand, according to \eqref{Puiseux}, these modes become purely imaginary at purely imaginary values of $\qn$. One then takes a fixed value of $Q$ and numerically finds frequencies of  these modes at several  purely imaginary momenta other than the momentum of pole-skipping points. Then, interpolation between the resultant points gives the dispersion relations we look for.

In Fig.\ref{imaginary_dispersion_Spin_0}, we have demonstrated the results related to $Q=0.4$. The left panel shows how pole-skipping points 
\begin{equation}\label{pole_skipping_4_10}
(\qn_*, \wn_*)=\,\left(\pm i \,\sqrt{\frac{3}{2-Q^2}}\, ,\,i\right)\bigg|_{Q=0.4}\approx\,(\pm1.27\,i,\,i).
\end{equation}
lie on the sound branches. The same can be  done for other values of $Q$. As a result, we understand that each of the chaos points \eqref{chaos_point} in our system  lies on the analytically continued dispersion relation  of one of the sound modes. This result completes our discussion about the hydrodynamic origin of quantum chaos in a holographic system at finite chemical potential. 
In the right panel of Fig.\ref{imaginary_dispersion_Spin_0}, we have compared our results with  those  associated with AdS$_5$ Schwarzschild case with $Q=0$ \cite{Grozdanov:2019uhi}.

\begin{figure}
	\centering
	\includegraphics[width=0.6\textwidth]{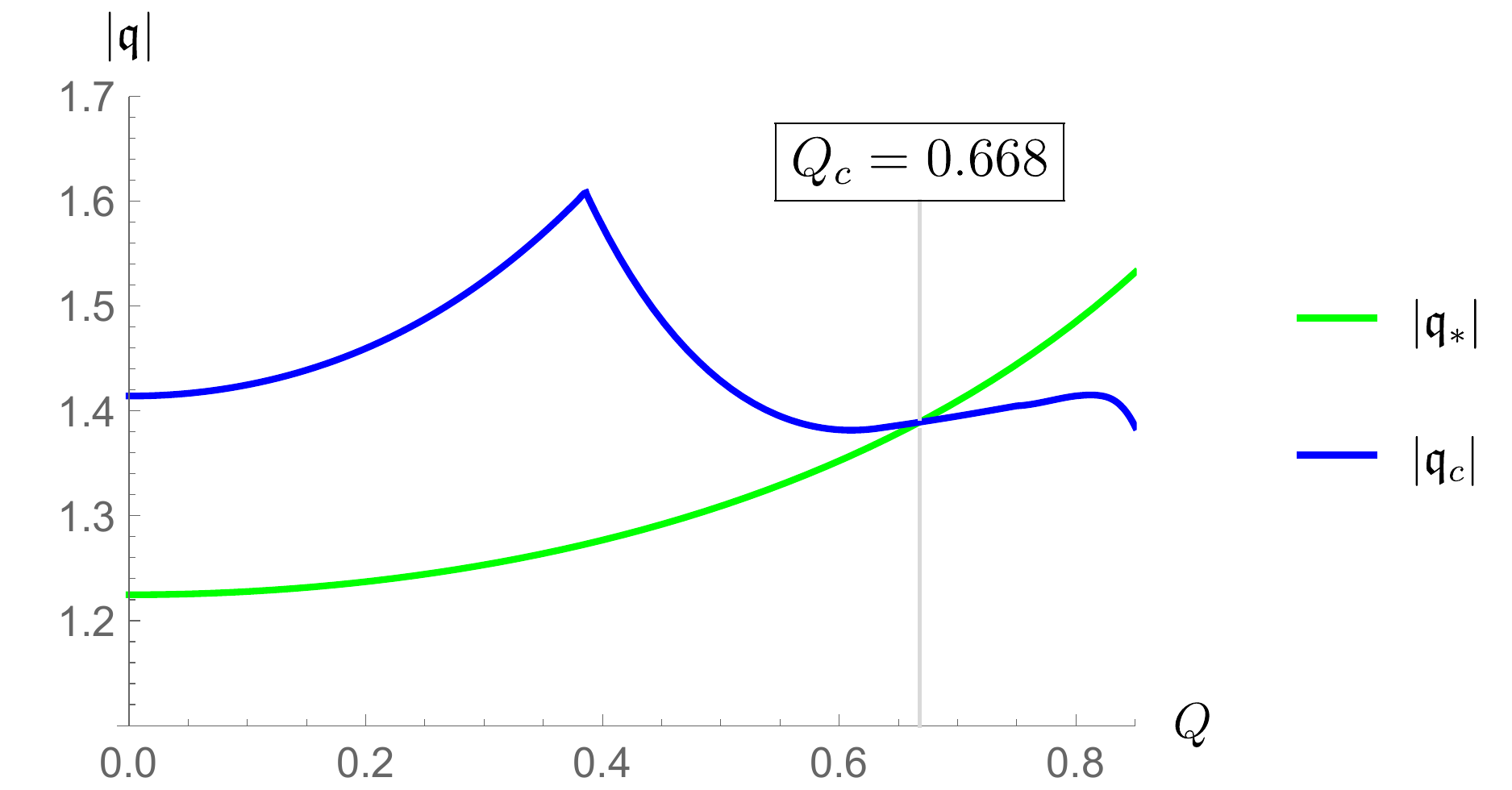}
	\caption{Comparison between the convergence radius  of the derivative expansion associated with  hydrodynamic sound mode (denoted by blue curve, already given in Fig.\ref{q_c_Sound_diiusion}) and the absolute value of momentum corresponding to pole-skipping points (denoted by green curve).  When $Q<Q_c$, pole-skipping and the chaos points lie within the domain of convergence of classical hydrodynamics. Thus they can be found perturbatively  in a derivative expansion from energy density response function. However, when $Q>Q_c$, extracting any information about quantum chaos from energy density response function requires non-perturbative  computations \cite{Blake:2017ris}.}
	\label{copara_chaos_critical}
\end{figure}

At this point one may ask: Does pole-skipping happen within the domain of validity of the hydrodynamic derivative expansion? To answer this question, in Fig.\ref{copara_chaos_critical}, we compare the absolute value of momentum at the pole-skipping point, namely $|\qn_{*}|$, with the radius of convergence of the hydrodynamic dispersion relation $|\qn_c|$.  
It is seen that when $Q$ exceeds the critical value $Q_c\approx0.668$, $|\qn_{*}|$ lies outside of the  domain of convergence of the derivative expansion. Is it physically reasonable?

 Recalling  $|\qn_{*}|=|\qn_{ch}|$,  one notices that when $Q>Q_c$ the chaos point lies outside the regime of validity of the hydrodynamic derivative expansion, too.  This is exactly one of  the reasons for which a quantum theory of hydrodynamics was constructed in ref. \cite{Blake:2017ris}.
 As explicitly mentioned by its authors, to capture the exponential behavior  of  \eqref{OTOC_0}, they constructed an effective hydrodynamic theory non-perturbatively in derivatives\footnote{ At this point it is worth mentioning that by "hydrodynamic origin of quantum chaos" posed in \cite{Blake:2017ris}, they actually mean a quantum hydrodynamic origin. However, what we have dealt with in this paper have been about classical hydrodynamics.}. Then our above mentioned results for $Q>Q_c$ can be interpreted as  it follows. When $Q<Q_c$, the derivative expansion is sufficient to see pole-skipping while for the range $Q>Q_c$ a non-perturbative treatment is required to find it from energy density response function \cite{Blake:2017ris}.

 A very explicit example of  pole-skipping outside the regime of convergence of hydrodynamics was found in the self-dual graviton-axion model  \cite{Grozdanov:2019uhi}. While the derivative expansion associated with dispersion relation of  gapless mode in this theory diverges at the pole-skipping point, it was shown that the Borel resummation  of the series converges at the same point. Such Borel resummation is in fact one kind of non-perturbative methods which we already argued should be used in the range $Q>Q_c$ \footnote{We thank Sa\v{s}o Grozdanov for bringing this example to our attention.}.  

\subsubsection*{Spin 1 Channel}
 Spin 1 perturbations in the bulk are $\delta g_{vx}$, $\delta g_{rx}$, $\delta g_{rr}$, $\delta g_{zx}$,  $\delta A_{r}$ and  $\delta A_x$.  Now the question is if there exists a combination of spin 1 components of dynamical equations at the horizon, namely $E^{(0)}_{\mu x}=0$ and $E^{(0)}_{x}=0$, which trivially holds at \eqref{pole_skipping_point}.  
 In the vanishing $\mu$ limit, Maxwell's equations decouple and one finds $E^{(0)}_{vx}+i \sqrt{2/3} E^{(0)}_{xz}$ to be the desired combination \cite{Grozdanov:2019uhi}. This expression becomes trivial at $(\mp i \qn_*, -\wn_*)$.

\begin{figure}
	\centering
	\includegraphics[width=0.6\textwidth]{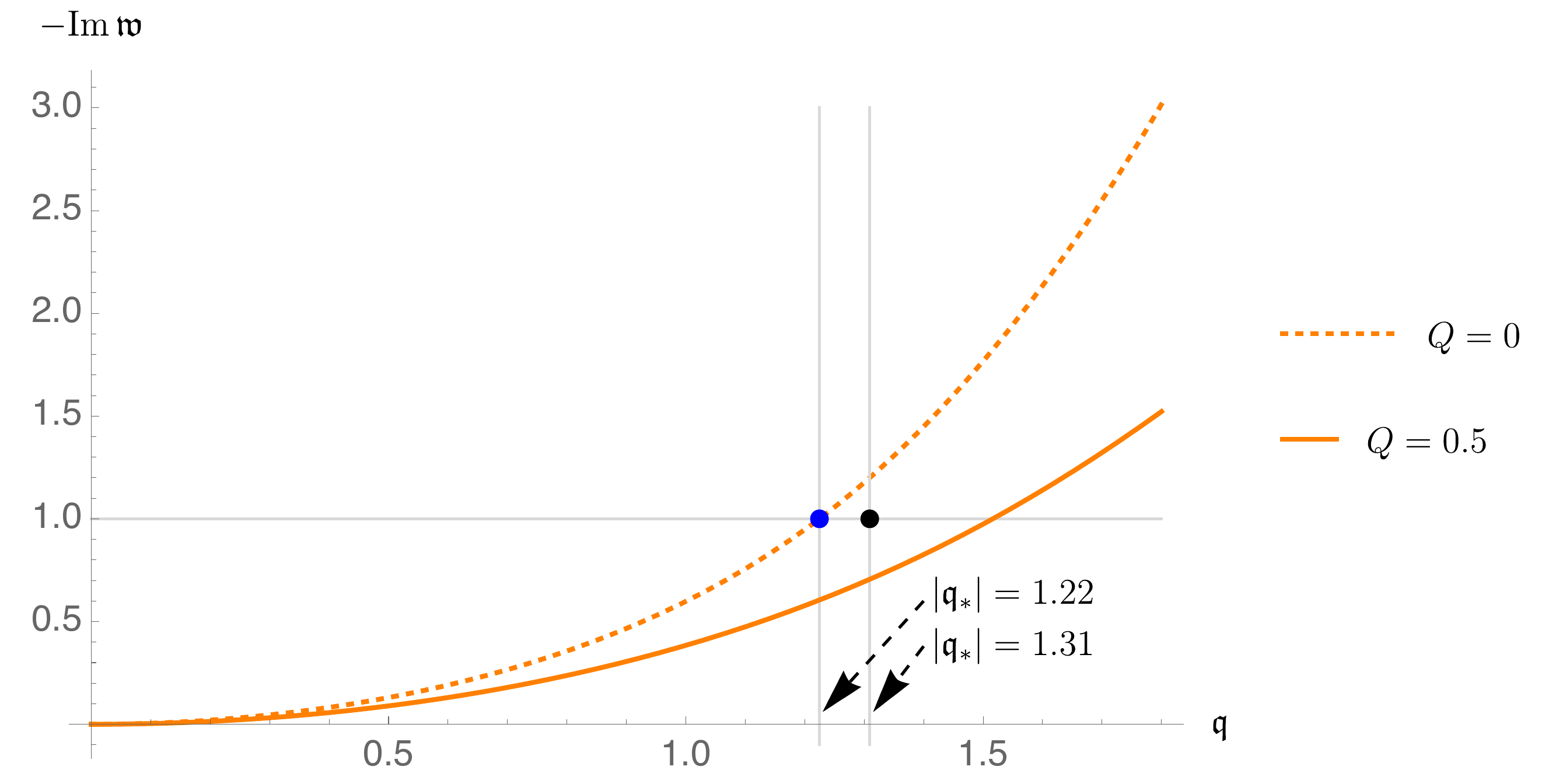}
	\caption{In a neutral fluid, $Q=0$, hydrodynamic branch of poles in spin 1 channel has a pole-skipping point denoted by blue dot. This point turns out to be $(\mp i\qn_*, - \wn_*)$ and so includes the same information about the chaos  point $(\qn_*, \wn_*)$. However, when $Q\ne0$, it will no longer be the case. As an example, black dot denotes the point $(\mp i\qn_*, - \wn_*)$ evaluated at $Q=0.5$; it obviously does not lie on the line of poles associated with $Q=0.5$.}
	\label{Shear_pole_skipping}
\end{figure}

In our case with $\mu\ne0$, however, Maxwell's equations are coupled to Einstein equations. One finds that the simplest possible combination close to what we want is
\begin{equation}\label{pole_skipping_eq_shear}
\left(E^{(0)}_{vx}+i\sqrt{\frac{2}{3}}\,\frac{1+Q^2}{\sqrt{1-Q^2/2}}\,\,E^{(0)}_{xz}+\frac{\sqrt{3}}{2}R\,Q\, E^{(0)}_{x}\right)\bigg|_{(-\wn_*, \mp i \qn_*)}=\,\frac{3}{2}Q^2\,\left(-3 \delta g_{vx}^{(0)}+i \sqrt{6-3Q^2} \delta g_{xz}^{(0)}\right).
\end{equation}
Although at $Q=0$ this  gives the decoupled equation $E^{(0)}_{vx}+i \sqrt{2/3} E^{(0)}_{xz}=0$, at a general value of $Q$ the left hand side combination is still coupled to metric perturbations of the right hand side \footnote{One another decoupling case is $Q=\sqrt{2}$, i.e. the extremal case, in which $E_{xz}^{(0)}=0$ decouples from the rest of equations.}. 
Thus at $Q\ne0$ we do not expect to see any pole-skipping in the shear channel which is related to the upper half plane pole-skipping in the sound channel \footnote{ In other words, there still might be some other pole-skipping points in the lower half of complex frequency plane associated with spin 1 channel, unrelated to quantum chaos (see for instance section (5.1) in ref. \cite{Blake:2019otz} for similar issues). We thank Sa$\check{\text{s}}$o Grozdanov  for pointing this out to us.}. We do not prove the latter statement explicitly but we just show that $(\mp i \qn_*, -\wn_*)$ do not correspond to  pole-skipping of the shear dispersion relation. In Fig.\ref{Shear_pole_skipping} we have demonstrated it for $Q=0.5$.  The points
\begin{equation}\label{pole_skipping_Shear}
( \mp i\qn_*, - \wn_*)=\,\left(\sqrt{\frac{3}{2-Q^2}}\, , \,-i\right)\bigg|_{Q=0.5}\approx\,(1.31,\,-i)
\end{equation}
does not lie on the dispersion relation of the shear mode associated with $Q=0.5$.
\textit{The above observation shows that in holographic systems, the chaos points \eqref{pole_skipping_point} are not related in general to pole-skipping in any channel other than the spin 0 one. }

Again, we emphasize that the above-mentioned statement is not in contradiction with those of ref. \cite{Blake:2017ris}. Even if the points \eqref{pole_skipping_Shear} were really pole-skipping points, they would not have anything to do with the "hydrodynamic origin of quantum chaos". The reason is that the points \eqref{pole_skipping_Shear}  would lie in the lower half of the complex frequency plane while those being responsible for exponential growth of OTOC have $\text{Im}\, \wn>0$ and are located in the upper half  plane.

Before ending this section let us denote that we intentionally do not study possible pole-skipping points in the spin 2 channel. Looking at eq. (4.53) in ref. \cite{Grozdanov:2019uhi}, we understand that pole-skipping in this channel cannot happen in the upper half plane. As a result, existence or non-existence of such points is not related to quantum chaos at all and so will be unrelated to the scope of our discussion.

\section{Review, Conclusion and Outlook}
\label{conclusion}
In this  paper we explored three aspects of  quasinormal modes in a holographic system at finite chemical potential: 
\newline $\textcircled{1}$ The dependence of the quasinormal mode spectra and also the hydrodynamic excitations on the value of $\mu/T$ (or equivalently on $Q$), in different channels of spin.
\newline $\textcircled{2}$ The collisions of hydrodynamic and non-hydrodynamic poles at complex momenta, and specifically changes in the pattern of collisions when $\mu/T$ varies.
\newline $\textcircled{3}$ The relation between pole-skipping of energy density response function and hydrodynamics in the system
and also extraction of chaos information from energy dynamics by use of the derivative expansion. 

Let us emphasize that the necessary tool for studying all above aspects was initially provided in \sec{method}. In fact besides all analytical and numerical methods and techniques used in the paper, it was equation \eqref{det_quasi} by use of which we were able to produce most of our results, i.e. those associated with spin 0 and 1 channels. 

In part $\textcircled{1}$, among all other results we think that finding analytic expressions for hydrodynamic excitations is of more importance. To best of our knowledge, hydrodynamic modes \eqref{hydro_mode_holog_spin_0} and \eqref{hydro_mode_holog_spin_1}  were never found in the literature.

In part $\textcircled{2}$, our main results are definitely related to spin 0 channel. We have shown that convergence radius of the derivative expansion for the sound mode non-trivially depend on the value of $\mu/T$. In the vanishing $\mu$ limit, it has been shown that convergence of the derivative expansion both in momentum \cite{Grozdanov:2019kge,Withers:2018srf} and in position space \cite{Heller:2020uuy} corresponds to the collision of hydrodynamic and non-hydrodynamic modes. However, for $\mu\ne0$ we have found that within the range $0.386\le Q\le0.633$ or equivalently $1.1\lesssim \mu/T\lesssim 2$, this is the collision between two hydrodynamic modes, namely sound and  diffusion modes, which determines the radius of convergence for the sound dispersion relation.  Then it would be interesting to investigate how to extract information about the non-hydrodynamic sectors from the large order behavior of derivative expansion, in the mentioned range. We leave study on this issue to a future work.

In part $\textcircled{3}$, we provided a new evidence for the hydrodynamic origin of quantum chaos. As we showed, for all values of $\mu/T$, the chaos points found from shock wave computations precisely coincide with the pole-skipping points of energy density response function. However  comparison between momentum of the pole-skipping points with the radius of converegnece of sound mode leads to an interesting result. Let us recall that quantum chaos seems not to be captured by classical hydrodynamics. The reason is that hydrodynamics is applicable for time intervals $\Delta t$ much larger than $1/T$, while in a maximally chaotic system the exponential behavior in \eqref{OTOC} \cite{Maldacena:2015waa} can be captured when $\Delta t\sim 1/T$  \cite{Blake:2017ris}. It shows why a to-all-order non-perturbative theory of hydrodynamics is needed to describe the chaos. On the other hand, we have found that in the range $0\le \mu/T\lesssim 2.1$, chaos information can be extracted perturbatively in the standard hydrodynamic derivative expansion. This results is simply the consequence of our another result saying that in the mentioned range, pole-skipping happens in the domain of convergence of the hydrodynamic derivative expansion.

It should be noted that the discussion of previous paragraph associated with part $\textcircled{3}$, might implicitly confirm the result shown in Fig.\ref{q_c_Sound_diiusion} of part $\textcircled{2}$.   Let us assume that the blue curve in that figure was monotonically increasing by $Q$ and always lay higher than the green curve in Fig.\ref{copara_chaos_critical}. Since an increase in $Q$ at a fixed $\mu$ is equivalent to a decrease in $T$, then one would conclude that  even as $T\rightarrow 0$, where the derivative expansion definitely breaks down, 
the chaos point could be still found perturbatively in a derivative expansion. This is absolutely wrong.
Consequently, the quantum nature of chaos forces the blue curve in Fig.\ref{copara_chaos_critical} not to be  always higher than the green one.
 
Based on the above discussion, it becomes more important to study the quasinormal modes at higher values of $Q$ than those studied in this paper \cite{navid:2020}, specifically in low temperature regime which is in correspondence with a near extremal black hole \cite{Moitra:2020dal}. 
\section*{Acknowledgment}
N.A. would like to thank Javad Tabatabaei for collaboration in the early stages of this work.
N.A. would like to thank Casey Cartwright, Ali Davody and  Matthias Kaminski for useful discussions and specially appreciate Matthias Kaminski's suggestions on the draft. We are particularly grateful to Sa$\check{\text{s}}$o  Grozdanov for reading the draft and providing valuable comments. 
We would also like to thank Marco Ruggieri, Gianluca Giuliani, Bonan Zhang and Shen-Song Wan  for discussions on related topics. N.A. acknowledges Hong-Fei Zhang  and Wen-Hui Long for their  supports.
The work of S.T. was supported in part by National Natural Science Foundation of China, under Grant No. 11575254. We thank Benjamin Withers for correcting our 
over-interpretation of his results in the first version of this paper. We thank Christiana Pantelidou for correspondence after the second version of the paper appeared on arXiv.
\appendix
 \section{Diffeomorphism and gauge transformations}
 \label{diff_gauge_App}
In the spin $0$ channel,  all diff-gauge transformations are given as the following
 \begin{equation}\label{gauge_trans_Spin_0}
 \begin{split}
 h_{tt}&\rightarrow h_{tt}+ 2 i \omega \,\xi_t(r)+\frac{\big(f(r)a(r)\big)'}{b(r)}\xi_r(r),\\
 h_{tz}&\rightarrow h_{tz}-i q \, \xi_t(r)+ i \omega\, \xi_z(r),\\
 h_{zz}&\rightarrow h_{zz}-2 i q \, \xi_z(r) -\frac{a'(r)}{b(r)} \xi_r(r),\\
 h&\rightarrow h -\frac{a'(r)}{b(r)} \xi_r(r),\\
 a_{t}&\rightarrow a_t+ i \omega\, \phi(r)-i \omega\, \frac{\tilde{q}\, c(r)}{a(r)f(r)} \xi_t(r)-\frac{\tilde{q}\,c'(r)}{b(r)} \xi_r(r),\\
 a_{z}&\rightarrow a_z- i q\, \phi(r)+i q\, \frac{ \tilde{q}\,c(r)}{a(r)f(r)} \xi_t(r),
 \end{split}
 \end{equation}
 where $h=\sum_{\alpha}h_{\alpha \alpha}$ and $\alpha\in \{x,y\}$.
 The transformations in the spin $1$ channel  are
 \begin{equation}\label{gauge_trans_Spin_1}
 \begin{split}
 h_{t\alpha}&\rightarrow h_{t\alpha}+  i \omega \,\xi_{\alpha}(r),\\
 h_{z\alpha}&\rightarrow h_{z\alpha}-  i q \,\xi_{\alpha}(r),\\
 a_{\alpha}&\rightarrow a_{\alpha}.
 \end{split}
 \end{equation}
 Finally in the tensor channel:
 \begin{equation}\label{gauge_trans_Spin_2}
 h_{\alpha\beta}-\frac{1}{2}h \delta_{\alpha \beta}\rightarrow h_{\alpha\beta}-\frac{1}{2}h \delta_{\alpha \beta}.
 \end{equation}
 
\section{ Frobenius solution}
\label{Frobenius}
In order  to find the quasinormal modes in channel of spin 0 and 1, one has to find $Z$ and $E$ from  two specific coupled second order differential equations. Following the explanations given in \cite{Kovtun:2005ev}, the solutions obeying the ingoing boundary condition at the horizon can be represented as 
\begin{equation}
\begin{split}
Z(u)=&(1-(1+Q^2)u^2+Q^2 u^3)^{-i \wn/2}\sum_{n=0}^{\infty} a_n(\wn, \qn)(1-u)^n,\\
E(u)=&(1-(1+Q^2)u^2+Q^2 u^3)^{-i \wn/2}\sum_{n=0}^{\infty} b_n(\wn, \qn)(1-u)^n.
\end{split}
\end{equation}
Substituting the above two solutions into the corresponding coupled differential equations, one finds the coefficients $a_n$ and $b_n$ of the series expansion, all in terms of $a_0$ and $b_0$. Then by imposing the Dirichlet boundary condition at $u=0$, 
\begin{equation}
Z(0)=\sum_{n=0}^{\infty} a_n(\wn, \qn)=\,0,\,\,\,\,\,\,\,\,\,
E(0)=\sum_{n=0}^{\infty} b_n(\wn, \qn)=\,0
\end{equation}
one arrives at a system of coupled algebraic equations which can be formally written as 
\begin{equation}
\begin{split}
\mathcal{C}_{11}(\wn, \qn) a_0+\,\mathcal{C}_{12}(\wn, \qn) b_0=&\,0,\\
\mathcal{C}_{21}(\wn, \qn) a_0+\,\mathcal{C}_{22}(\wn, \qn) b_0=&\,0.
\end{split}
\end{equation}
Quasinormal spectrum then can be determined by demanding the above  equations have non-trivial solutions for $(\wn, \qn)$:
\begin{equation}\label{detreminant_numeric}
\mathcal{C}_{11}(\wn, \qn)\,\,\mathcal{C}_{22}(\wn, \qn)-\,\mathcal{C}_{12}(\wn, \qn) ,\mathcal{C}_{21}(\wn, \qn)=\,0
\end{equation}
and solving equation \eqref{detreminant_numeric} numerically; the latter is done by taking a sufficiently large but finite number of terms, $n$, in the sum.
\section{ Near-boundary behavior of gauge invariant quantities}
\label{bdy_bahavior}
In this section we use the results of ref. \cite{Witten:1998qj} (see also ref. \cite{Son:2007vk} for a review of the related stuff)  to find the near boundary behavior of $Z(u)$ and $E(u)$ fields. We consider the AdS$_5$ space time given by
\begin{equation}
ds^2=\,\frac{1}{z^2}\big(\eta_{\mu\nu}dx^{\mu}dx^{\nu}+dz^2\big).
\end{equation}

Let us start by considering $Z(u)$ (note that $u\sim z^2$). As we already showed in the paper, in every channel of spin, $Z$ was basically constructed out of $H_{\mu\nu}$ components.
 The advantage of working with $H_{\mu\nu}$ instead of $h_{\mu\nu}$ is that the former obeys a scalar field-like equation of motion in the bulk  \cite{Son:2007vk}. From ref.  \cite{Witten:1998qj}, the near-boundary fall-off of a scalar field of mass $m$ in the five dimensional bulk is given by
\begin{equation}
z\rightarrow0:\,\,\,\,\,\,\,\,\,\phi(z)\,\approx\, A \,z^{4-\Delta}+\,B \,z^{\Delta},\,\,\,\,\,\,\,\,\,\,\,\,\,\Delta=2+\sqrt{2+m^2}.
\end{equation}
Since $Z(u)$ is associated with massless perturbations ($m=0$) in the bulk, one takes $\Delta=4$ and then writes
\begin{equation}\label{Z_bdy}
z\rightarrow0:\,\,\,\,\,\,\,\,\,Z(z)\,\approx\, (c_1+\cdots) +\,(c_2+\cdots) \,z^{4}
\end{equation}

On the other hand, $E(u)$ is constructed out of $a_{\mu}$, i.e. the one form bulk field perturbations. The near-boundary fall-off of a $p-$form field of mass $m$ is given by \cite{Witten:1998qj} 
\begin{equation}
z\rightarrow0:\,\,\,\,\,\,\,\,\,a_{\mu_1\cdots\mu_p}(z)\approx A_{\mu_1\cdots\mu_p}z^{4-p-\Delta}+\,B_{\mu_1\cdots\mu_p}z^{\Delta-p}
\end{equation}
where $\Delta$ is the larger root of the equation $m^2=\,(\Delta-p)(\Delta+p-4)$. Since $a_\mu$ is a massless ($m=0$) $1-$form field ($p=1$),  we find $\Delta=3$ and then write
\begin{equation}\label{E_bdy}
z\rightarrow0:\,\,\,\,\,\,\,\,\,E(z)\,\approx\, (c_3+\cdots) +\,(c_4+\cdots)\,z^{2}.
\end{equation}
In the present paper we take the bulk radial coordinate being $u\sim\frac{1}{r^2}\sim z^2$. Thus we rewrite  \eqref{Z_bdy} and \eqref{E_bdy} in terms of $u$:
\begin{equation}\label{Z_E_bdy}
\begin{split}
u\rightarrow0:&\,\,\,\,\,\,\,\,\,Z(u)\,\approx\,  (c_1+\cdots) +\,(c_2+\cdots)\,u^2\\
u\rightarrow0:&\,\,\,\,\,\,\,\,\,E(u)\,\approx\,  (c_3+\cdots) +\,(c_4+\cdots)\,u.
\end{split}
\end{equation}
This is what was already given by \eqref{near_bdy_exp}.

\section{Comparing between the method developed in \sec{method} and that of Kaminski et al \cite{Kaminski:2009dh}}
\label{Comparison}
As mentioned in the paper, in ref. \cite{Kaminski:2009dh} a new method has been developed to find the quasinormal modes associated with the coupled bulk fields. In order for our method in \sec{method} to be comparable with that of ref. \cite{Kaminski:2009dh}, let us take the Fourier components of the  bulk fields as $\Phi_k^I$. This is actually the notation adopted in the mentioned reference. In our case  with two coupled fields $Z$ and $E$, one writes
\begin{equation}
\{\Phi_k^I\}\equiv\big(Z, E\big)=(1-u^2)^{-i\wn/2}\big(G(u), Y(u)\big)\,\,\,\,\,\,\,I=1,2.
\end{equation}
Both the ref. \cite{Kaminski:2009dh} and us impose the ingoing boundary condition on $\Phi_k^I$ by considering the coefficient function $(1-u)^{-i\wn/2}$. The difference between methods of two papers, however, arises when considering the second boundary condition at the horizon $u=1$.  The latter is actually fixing the values of $\{G(1), Y(1)\}$.

Since our computations in this paper are based on the analytic Frobenius expansion, we are able to work with two unknown parameters $\big(G(1), Y(1)\big)=\,\big(C_Z,C_E\big)$. On the other hand, because of their numerical method,  authors of  \cite{Kaminski:2009dh} need to know the numerical values of $\big(G(1), Y(1)\big)$. The idea of ref. \cite{Kaminski:2009dh} is to take two \textit{linearly independent} sets of values as
\begin{equation}\label{Basis_sol_numeric_bdy}
\big(G^{(1)}(1), Y^{(1)}(1)\big)=\big(1,1\big),\,\,\,\,\,\,\,\big(G^{(2)}(1), Y^{(2)}(1)\big)=\big(1,-1\big).
\end{equation}
Then for each of the above two sets,  they numerically find $G(u)$ and $Y(u)$ in the bulk. One may formally show the solutions as
\begin{equation}\label{Basis_sol_numeric}
\big(G^{(1)}(u), Y^{(1)}(u)\big),\,\,\,\,\,\,\big(G^{(2)}(u), Y^{(2)}(u)\big).
\end{equation}
The last step is to put the above solutions in a matrix:
\begin{equation}\label{H}
H(u)=\,\begin{pmatrix}
	G^{(1)}(u)&G^{(2)}(u)\\
	Y^{(1)}(u)& Y^{(2)}(u)
\end{pmatrix}.
\end{equation}
 Then they argue that  the spectrum of quasinormal modes associated  with the fields $\{\Phi_k^I\}$ is determined by
\begin{equation}\label{det_H}
\det\big[H( 0)\big]=\,0.
\end{equation}
This is actually eq. (2.35) in  ref. \cite{Kaminski:2009dh}.

Let us summarize:
\newline $\bullet$ Authors of \cite{Kaminski:2009dh} \textit{numerically} solve the coupled equations twice; first for $\big(G^{(1)}(1), Y^{(1)}(1)\big)=\big(1,1\big)$ and then for $\big(G^{(2)}(1), Y^{(2)}(1)\big)=\big(1,-1\big)$. Then they numerically solve \eqref{det_H} to find the quasinormal modes.
\newline $\bullet$ In our side, we \textit{analytically} solve the coupled equations for  $\big(G(1), Y(1)\big)=\,\big(C_Z,C_E\big)$, by use of the Frobenius expansion. The corresponding solutions are formally given by \eqref{ingoing_coupled_sol}. Then via numerically solving \eqref{det_quasi}, we find the quasinormal modes.

From the above discussions it is clear that the method developed in this paper differs from that of  \cite{Kaminski:2009dh}. Our method is based on the combination of analytic and numerical computations while their method is fully numeric-based. We can also find the quasinormal modes once we obtain the general analytic solutions in the bulk. But as discussed earlier, 
in the  ref.  \cite{Kaminski:2009dh},  \textit{two} numerical solutions are needed to find the quasinormal modes.

Apart from the differences discussed above, equations  \eqref{det_H}  and \eqref{det_quasi},  found in ref. \cite{Kaminski:2009dh} and in the present paper, respectively, seem to be similar. Although they are not the same, they carry the same information. In what follows, we show how those two equations can be transformed to each other. 

Since our coupled differential equations are ordinary, any two solutions of them can be linearly combined to produce another solution. Let us take the linear combination of solutions given in \eqref{Basis_sol_numeric} as a third solution
\begin{equation}\label{third_solution}
\big(\tilde{C}_1\,G^{(1)}(u)+\,\tilde{C}_2\,G^{(2)}(u), \tilde{C}_1\,Y^{(1)}(u)+\,\tilde{C}_2\,Y^{(2)}(u)\big)\xrightarrow[\text{}]{u=1}\big(\tilde{C}_1+\tilde{C}_2,\,\tilde{C}_1-\tilde{C}_2\big).
\end{equation}
This can be exactly the solution found in the present paper, given by \eqref{Our_solution} and \eqref{G_Y_bdy}, if one demands
\begin{equation}
\begin{array}{c}
\tilde{C}_1+\tilde{C}_2=\,C_Z \\
\tilde{C}_1-\tilde{C}_2=\,C_E \\
\end{array} \,\,\,\rightarrow\,\,\,\,\tilde{C}_1=\frac{C_Z+C_E}{2},\,\,\,\tilde{C}_2=\frac{C_Z-C_E}{2}.
\end{equation}
Substituting $\tilde{C}_1$ and $\tilde{C}_2$ back into the third solution \eqref{third_solution}, we can rewrite that in terms of $C_Z$ and $C_E$:
\begin{equation}\label{third_solution_2}\boxed{
\left(C_Z \frac{G^{(1)}(u)+G^{(2)}(u)}{2}+C_E\frac{G^{(1)}(u)-G^{(2)}(u)}{2},\,C_Z \frac{Y^{(1)}(u)+Y^{(2)}(u)}{2}+C_E\frac{Y^{(1)}(u)-Y^{(2)}(u)}{2}\right)}
\end{equation}
It is clear that at $u=1$ the above solution goes to $(C_Z, C_E)$ meaning that \eqref{third_solution_2} is exactly the same as \eqref{ingoing_coupled_sol}. Consequently, we can read off $g_Z(u)$, $g_E(u)$, $y_Z(u)$ and $y_E(u)$ in terms of $G^{(1)}(u)$, $G^{(2)}(u)$, $Y^{(1)}(u)$ and $Y^{(2)}(u)$. By defining the matrix $M$, we write 
\begin{equation}\label{M}
M(u):=\,\begin{pmatrix}
g_{Z}(u)&y_{Z}(u)\\
g_{E}(u)& y_{E}(u)
\end{pmatrix}=\,\frac{1}{2}\begin{pmatrix}
 G^{(1)}(u)+G^{(2)}(u)&G^{(1)}(u)-G^{(2)}(u)\\
Y^{(1)}(u)+Y^{(2)}(u)& Y^{(1)}(u)-Y^{(2)}(u)
\end{pmatrix}.
\end{equation}
This is actually the relation between the "general analytic solution found from Frobenius expansion  in the present paper" and the "two numerical solutions found in ref. \cite{Kaminski:2009dh}".
According to \eqref{det_quasi}, the quasinormal modes in our case are determined by  $\det\big[M(0)\big]=\,0$. But when considering both  \eqref{H} and \eqref{M}, we arrive at
\begin{equation}\label{final_comparison}\boxed{
\underbrace{\det\big[M(0)\big]}_{\text{eq.}\,\eqref{det_quasi}\,\text{in this paper}}=\,\,\,-\frac{1}{2}\underbrace{\det\big[H(0)\big]}_{\text{eq.}\,(2.35)\, 
	\text{in\,\cite{Kaminski:2009dh}}}=\,0} 
\end{equation}
This is the relation between two determinants appeared in ref. \cite{Kaminski:2009dh} and in the present paper.
The numerical factor $-1/2$ in \eqref{final_comparison} is indeed the Jacobi of transformation \eqref{M}.

Equation \eqref{final_comparison} emphasizes that equations \eqref{det_quasi} in the present paper and (2.35) in ref. \cite{Kaminski:2009dh} carry the same information, however as shown above, they are basically different from each other in the sense that they correspond to two different ways of finding the bulk solutions. 

\section{Numerical values of quasinormal modes associated with spin 0 channel}
\label{refrence_data}
As discussed in the text, our study in this paper is actually the first one about the spin 0 quasinormal modes on AdS$_5$ RN background. For this reason, in Table.\ref{Table_2} we have provided some reference data that might be useful in future studies of quasinormal modes. The numerical frequencies in the table are all related to $Q/Q_{\text{ext}}=0.5$ with $Q_{\text{ext}}=\sqrt{2}$.
\begin{table}[!htb]
	\label{table one}
	\begin{center}
		\begin{tabular}{|c||c|c||c|c||c|c|}
			\hline
			\hline
				& \multicolumn{6}{c|}{$Q/Q_{\text{ext}}=0.5$} \\
					\hhline{~------}
	& \multicolumn{2}{c||}{$\qn=1.0$} 
	& \multicolumn{2}{c||}{$\qn=0.5$} 
	& \multicolumn{2}{c|}{$\qn=0.0$} 
	\\
		\hhline{~------}
					$n$&	Re $\wn$ & Im $\wn$ & Re $\wn$ & Im $\wn$ &Re $\wn$&Im $\wn$\\
			\hline
			\multirow{8}{*}{}
	sound	&	$\pm0.6224006$	& $-0.1699176$ & $\pm0.2936770$  &$-0.0420687$&$0$& $0$\\ 
			diffusion	&	$0$	&$-0.7200600$ & $0$  & $-0.1624148$&$0$& $0$\\ 
				$1$	&	$\pm1.7098496$	& $-0.7324281$& $\pm1.5473713$  &$-0.8196519$ &$\pm1.4933083$&$-0.8426994$ \\ 
					$2$	&	$0$	&$-2.1357589$ &  $0$ & $-2.0385094$ &$0$& $-2.0000000$\\ 
						$3$	&$\pm2.8522545$	&$-2.2714762$ 	 &$\pm2.7470788$   &$-2.3268141$&$\pm2.7085986$& $-2.3437522$ \\ 
												$4$	&$\pm2.2455651$	&$-2.2996959$	& $\pm2.1734813$  &$-2.3461519$ &$\pm2.1524008$& $-2.3632861$\\ 
			\hline
			\hline
		\end{tabular}
	\end{center}
	\caption{ Numerical values of frequencies corresponding to the two hydrodynamic and the lowest four non-hydrodynamic  quasinormal modes in spin 0 channel at real momenta.}
	\label{Table_2}
\end{table}
\newline
We have shown the numerical values of hydrodynamic frequencies at $\qn=1$ and  $\qn=0.5$, in the first two rows of the table.  As expected, hydrodynamic frequencies vanish when $\qn=0$. 

In the next four rows, we have shown numerical values of frequencies at $\qn=1$, $\qn=0.5$ and $\qn=0$ associated with the lowest four non-hydrodynamic quasinormal modes in this channel. 

\section{Comparison with explicit  hydrodynamic computations}
In this section we reproduce the hydrodynamic excitations found from the study of perturbations, given in  \eqref{hydro_mode_holog_spin_0} and \eqref{hydro_mode_holog_spin_1}, via explicit hydrodynamic computations on the boundary.
The hydrodynamic regime of a holographic charged fluid has been studied in the context of fluid/gravity duality \cite{Banerjee:2008th,Erdmenger:2008rm} and also in 
\cite{Son:2006em}. Following \cite{Banerjee:2008th} we write the associated constitutive relations up to first order in the derivative expansion, in 4-dimensions, as the following
\begin{equation}
\begin{split}
T_{\mu\nu}=&\,p(\eta_{\mu\nu}+4u_{\mu}u_{\nu})-2 \eta \sigma_{\mu\nu},\\
j_{\mu}=&\,n \,u_{\mu}-\mathfrak{D} \left(P_{\mu}^{\nu}\partial_{\nu}n+ \frac{}{}3 (u^{\lambda}\partial_{\lambda}u_{\mu})n\right),
\end{split}
\end{equation}
with $P_{\mu\nu}=\eta_{\mu\nu}+u_{\mu}u_{\nu}$. The coefficients found in \cite{Banerjee:2008th,Erdmenger:2008rm} can be written as
\begin{equation}
p=\frac{(1+Q^2)}{16\pi G_5}\left(\frac{2\pi T}{2-Q^2}\right)^4,\,\,\,\,\,\,\,\,\,\,\eta=\frac{1}{16\pi G_5}\left(\frac{2\pi T}{2-Q^2}\right)^3
\end{equation}
and
\begin{equation}
n=\frac{\sqrt{3}Q}{16\pi G_5}\left(\frac{2\pi T}{2-Q^2}\right)^3,\,\,\,\,\,\,\,\,\,\,\mathfrak{D}=\frac{4-Q^4}{8\pi T(1+Q^2)}.
\end{equation}
Now by perturbing the equilibrium state of the fluid
\begin{equation}
u^{\mu}(x,t)=\big(1,\delta v_{x}(x,t),\delta v_{y}(x,t),\delta v_{z}(x,t)\big),\,\,\,\,T(x)=T+\delta T(x,t),\,\,\,\,\,Q(x,t)=Q+\delta Q(x,t)
\end{equation}
we search for the plane wave solutions $e^{-i \omega t+i q x_3}$ from the hydro equations of motion
\begin{equation}
\partial_{\nu}T^{\mu\nu}=\,0,\,\,\,\,\,\,\,\partial_{\mu} j^{\mu}=0.
\end{equation}
To first order in perturbations, these equations take the following form
\begin{equation}\label{linear_eq}
\begin{split}
0=&\left(\wn+\frac{i }{4 }\frac{2-Q^2}{1+Q^2}\qn^2\right)\delta v_{\perp}\,\,\,\,\,\,\,\,\,\,\,(\perp\in \{x,y\}),\\
0=&\,2(-2+Q^2)\bigg( T\qn  \delta v_z-3 \wn \delta T\bigg)+9T Q\frac{2+Q^2}{1+Q^2} \,\wn \,\delta Q,\\
0=&\,\frac{2}{9}(-2+Q^2)T \bigg(\wn +\frac{i }{3} \frac{2-Q^2}{1+Q^2}\qn^2\bigg)\delta v_z+2 (2-Q^2)\qn\,\delta T+3 T Q\frac{2+Q^2}{1+Q^2}\qn\,\delta Q,\\
0=&\,T Q (-2+Q^2) \left(1+\frac{3i}{4}\frac{4-Q^4}{1+Q^2}\wn\right) \qn\, \delta v_z+3 Q(2-Q^2)\left(\wn+\frac{i}{4}\frac{4-Q^4}{1+Q^2}\qn^2\right)\delta T\\
&\,\,\,\,\,\,\,+ T (2+5Q^2)\left(\wn+\frac{i}{4}\frac{4-Q^4}{1+Q^2}\qn^2\right)\delta Q.
\end{split}
\end{equation}
The corresponding eigen modes of these equations, namely the hydrodynamic modes, can be expressed in the  representations of earlierly discussed $SO(2)$  group. In the spin 0 channel, we find the following three modes 
\begin{equation}
\begin{split}
\wn_{1,2}=&\pm\frac{1}{\sqrt{3}}\qn-\frac{i}{6}\left(\frac{2-Q^2}{1+Q^2}\right)\,\qn^2,\,\,\,\,\,\,\,\\
\wn_3=&-\frac{i}{4}\left(\frac{4-Q^4}{1+Q^2}\right)\qn^2.
\end{split}
\end{equation}
The first two ones are the \textit{sound} modes and the third mode is the diffusive \textit{R-charge} mode.
These excitations are in fact the eigen modes of the three last equations in \eqref{linear_eq}.
It is obvious that at the limit $Q\rightarrow 0$, and up to the order $Q^2$,  the above expressions reduce to those in \eqref{hydro_mode_holog_spin_0}.
 
 From the first equation in \eqref{linear_eq}, we find the spin 1 modes as it follows 
\begin{equation}
\wn_{4,5}=-\frac{i}{4}\left(\frac{2-Q^2}{1+Q^2}\right)\qn^2
\end{equation}
which are the transverse \textit{shear} modes. At the limit $Q\rightarrow 0$, and up to the order $Q^2$, this expression becomes the one found in \eqref{hydro_mode_holog_spin_1}.

\section{Perturbative solution of equations \eqref{Sound_channel_diff} and \eqref{Shear_channel_diff}}
\label{App_sol_HYDRO_SOUND}
We list the solution functions $Z_{0}^{m.n}(u)$ and $E_{z}^{m,n}(u)$ according to ordering we find them through the perturbative solving the coupled equations \eqref{Sound_channel_diff}. Let us remind that in \eqref{Dirichlete_sound}, we have replaced $C_1$ with $(2-3\textswab{y}^2)C_1$, ($\textswab{y}=\wn/\qn$), and found the following expressions in terms of the rescaled $C_1$:
\begin{equation}\label{}
\begin{split}
Z_{0}^{0,0}(u)&=\,C_1\,(1- 3 \textswab{y}^2+ u^2),\,\,\,\,\,\,\,\,\,\,\,\,\,\,\,\,\,\,\,\,\,\,\,\,\,\,\,\,\,\,\,\,\,\,\,\,\,\,\,\,\,\,\,\,E_{z}^{0,0}(u)=C_2,\\
Z_{0}^{0,1}(u)&=\,-\frac{3 }{\textswab{y}}C_2\,(u^2-1),\,\,\,\,\,\,\,\,\,\,\,\,\,\,\,\,\,\,\,\,\,\,\,\,\,\,\,\,\,\,\,\,\,\,\,\,\,\,\,\,\,\,\,\,\,\,\,\,\,E_{z}^{0,1}(u)=\,\textswab{y}\,C_1(u-1),\\
Z_{0}^{0,2}(u)&=\,\frac{1}{3\textswab{y}^2-2}C_1\left[5-9u^2+4u^3-\frac{}{}6\textswab{y}^2(1-2u^2+u^3)\right],\\
E_{z}^{0,2}(u)&=-\frac{3}{2}C_2\,(u-1),\\
Z_{0}^{1,0}(u)&=\,2 i \,C_1\,(u^2-1),\,\,\,\,\,\,\,\,\,\,\,\,\,\,\,\,\,\,\,\,\,\,\,\,\,\,\,\,\,\,\,\,\,\,\,\,\,\,\,\,\,\,\,\,\,\,\,\,\,\,\,\,\,\,\,\,\,
E_{z}^{1,0}(u)=-\frac{i}{\textswab{y}^2}C_2\,\left[u-1+\textswab{y}^2\log\frac{u+1}{2}\right],\\\nonumber
Z_{0}^{1,1}(u)&=\,-\frac{i}{ \textswab{y}^3}C_2(u-1)\,\bigg[3\textswab{y}^2(u+1)\log \frac{u+1}{2}+2+3\textswab{y}^2-2u(u-1) \bigg], \,\,\,\,\,\,\,\,\,\,\,\,\,\\
E_{z}^{1,1}(u)&=\frac{i}{\textswab{y}}C_1\,\left[(\textswab{y}^2+1)(u-1)-\textswab{y}^2\log\frac{u+1}{2}\right],\nonumber
\end{split}
\end{equation}
	and
\begin{equation}\label{}
\begin{split}
Z_{0}^{1,2}(u)&=\,i C\bigg[-\frac{3}{2}(1+3\textswab{y}^2-3u^2)\log\frac{1+u}{2}\\\nonumber
& -\frac{u-1}{2(2\textswab{y}^2-3)}\bigg(8(u^2-u-1)+2\textswab{y}^2(u^2+4u+1)-3\textswab{y}^4(3+7u^2)+27\textswab{y}^6\bigg)\bigg],\\
E_{z}^{1,2}(u)&=-i C \left[(5+3 u) \log\frac{1+u}{2}+\frac{(u-1)(1+u+\textswab{y}^2 u)}{2\textswab{y}^2(1+u)}\right].
\end{split}
\end{equation}
The perturbative solutions to \eqref{Shear_channel_diff} are given by
\begin{equation}\label{}
\begin{split}
Z_{1}^{0,0}(u)&=\,C_3\,\left(1+\frac{i \qn^2}{2 \wn} (1-u^2)\right),\,\,\,\,\,\,\,\,\,\,\,\,\,\,\,\,\,\,\,\,\,\,\,E_{x}^{1,0}(u)=C_4,\\
Z_{1}^{0,1}(u)&=\,\frac{3 }{2}C_4\,(1-u^2),\,\,\,\,\,\,\,\,\,\,\,\,\,\,\,\,\,\,\,\,\,\,\,\,\,\,\,\,\,\,\,\,\,\,\,\,\,\,\,\,\,\,\,\,\,\,\,\,\,E_{x}^{1,1}(u)=\,i C_3\frac{\qn^2}{2 \wn}(1-u),\\
Z_{1}^{0,2}(u)&=\,iC_3 \frac{\qn^2}{2\wn}u^2(u-1),\,\,\,\,\,\,\,\,\,\,\,\,\,\,\,\,\,\,\,\,\,\,\,\,\,\,\,\,\,\,\,\,\,\,\,\,\,\,\,\,E_{x}^{1,2}(u)=\,\frac{3}{2}C_4(u-1).
\end{split}
\end{equation}

\section{Fine structure of the Fig.\ref{Complex_Sound_Chennel_1_2} in the vicinity of the second collision point }
\label{zoom_in}
\begin{figure}
	\centering
	\includegraphics[width=0.42\textwidth]{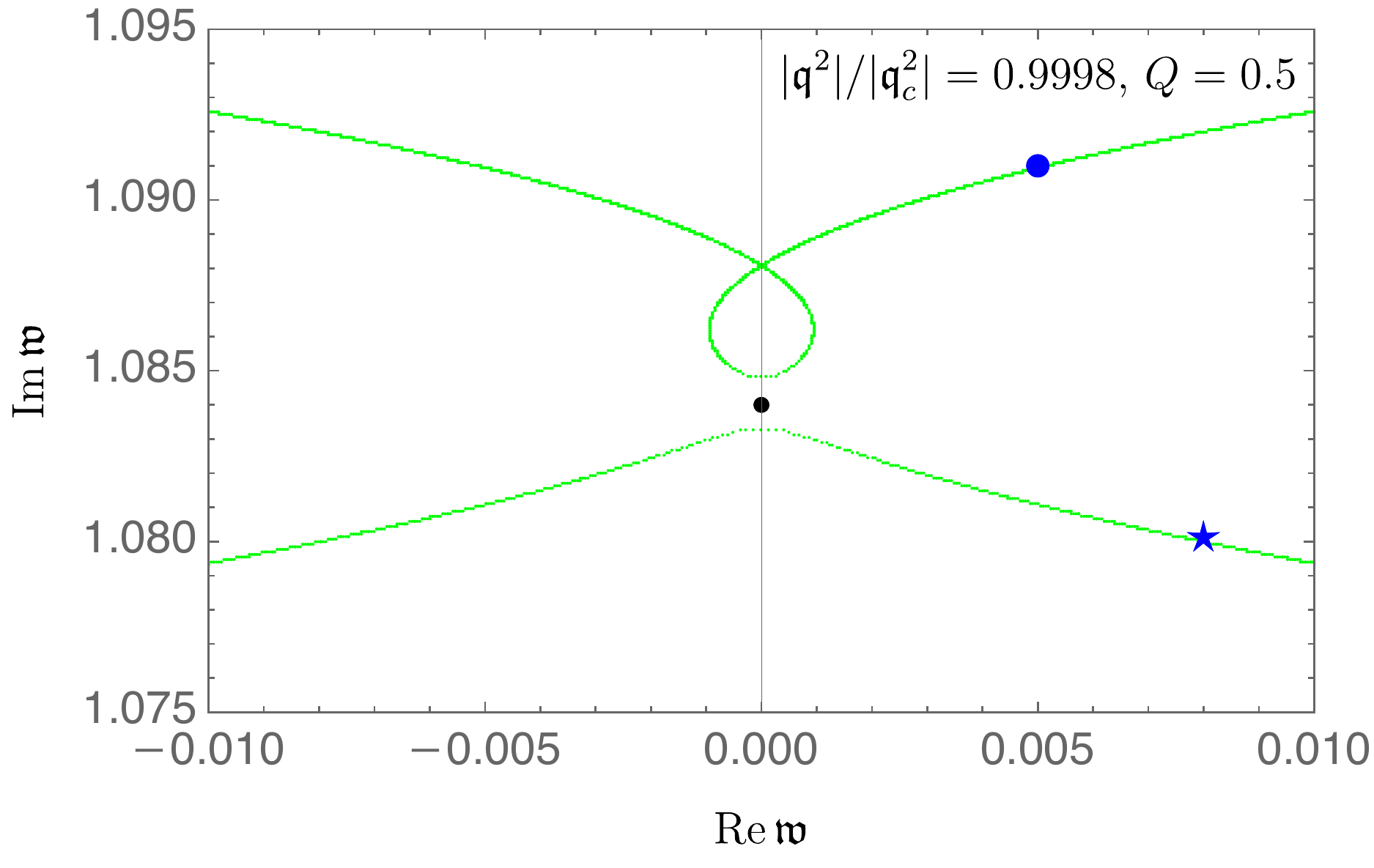}\,\,\,\,\,\,\,\,	\includegraphics[width=0.42\textwidth]{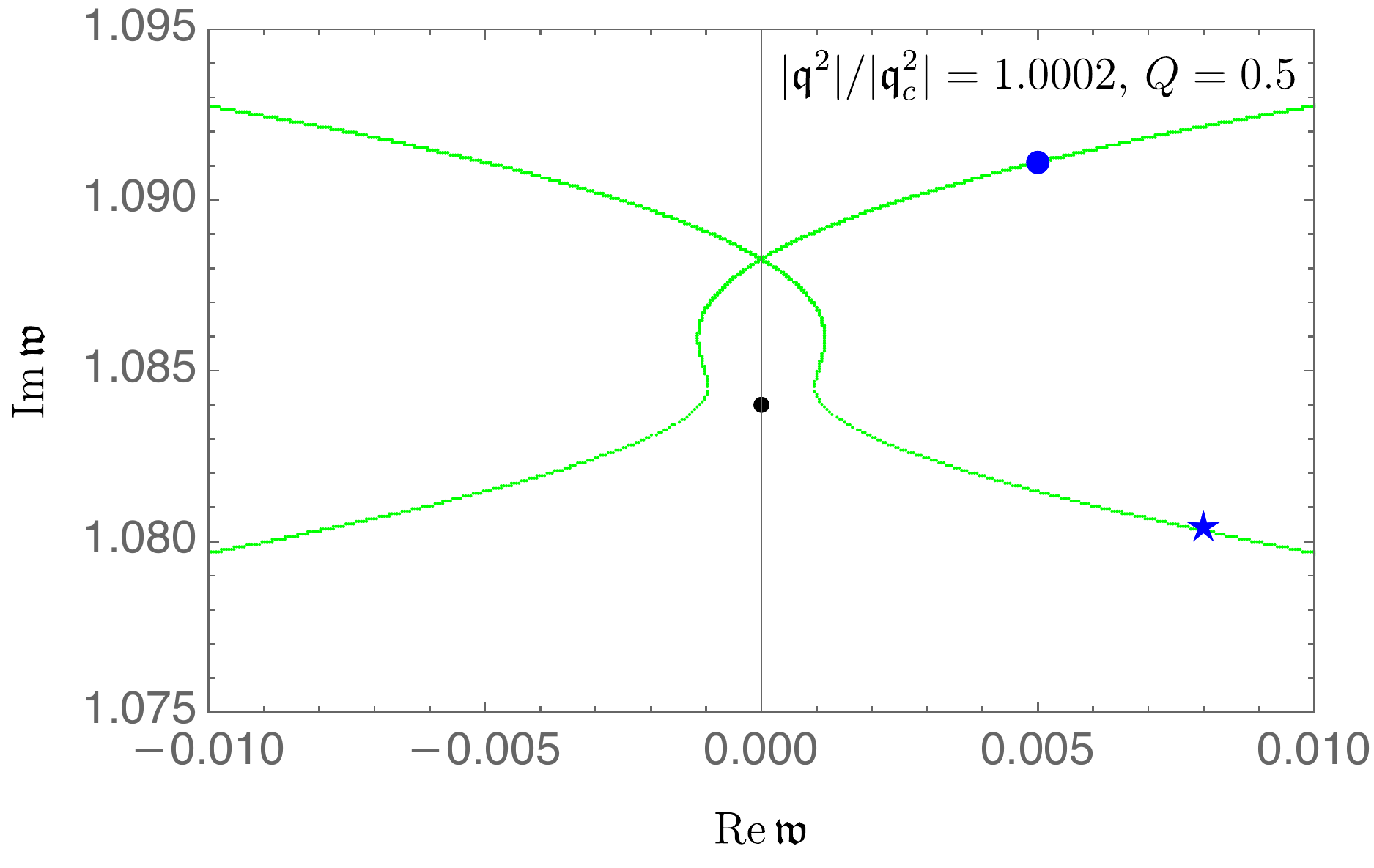}
	\caption{Fine structure geometry of the second collision point in the spin 0 channel at $Q=0.5$, in the complex $\wn-$plane, at various values of the complexified momentum $\qn^2=|\qn^2| e^{i \theta}$. When the phase $\theta$ increases from $\sim 0.495\pi$ to $\sim 0.505 \pi$,  blue dot and star poles lying on  the green trajectories move from right to left along them. The left panel shows the situation slightly before the collision occurs. The collision point is marked with a black dot.  The right panel corresponds to a value of $|\qn^2|$ slightly greater than that at which the two poles collide.}
	\label{zoom}
\end{figure}

As shown in the paper, within the range $0.386\le Q\le 0.633$, two types of collisions simultaneously determine the convergence radius of derivative expansion associated with the sound mode in the spin 0 channel. Of the latter two, the one which is more interesting was  discussed in details in Fig.\ref{Complex_Sound_Chennel_1_2}. The other collision takes place between the sound pole and one of the lowest gapped poles of $\bar{J}^L$ spectrum. The collision point is located in the upper half plane on the imaginary axis. Geometry of this collision is just recognized when one zooms it in. We have done so in the Fig.\ref{zoom}.

\bibliographystyle{utphys}
\providecommand{\href}[2]{#2}\begingroup\raggedright\endgroup

\end{document}